\documentclass[12pt,a4paper]{article}

\usepackage{geometry}
\usepackage{amsmath}
\usepackage{amssymb}
\usepackage[dvips]{graphicx}
\usepackage{cite}
\usepackage{pstricks}
\usepackage{bm}
\usepackage{pbox}
\usepackage{placeins}
\usepackage{graphicx}
\usepackage{caption}
\usepackage{subcaption}
\usepackage[T1]{fontenc}
\usepackage{footnote}
\usepackage{pdfpages}
\usepackage{hhline}
\usepackage{multirow}
\usepackage{enumitem}
\usepackage[toc,page]{appendix}
\allowdisplaybreaks

\definecolor{MyDarkBlue}{rgb}{0.1, 0.1, 0.8} 
\definecolor{MyLightBlue}{rgb}{0.22,0.51,0.9}
\definecolor{MyGreen}{rgb}{0.0, 0.5, 0.0}
\usepackage{times}
\usepackage[colorlinks=true,linkcolor=MyDarkBlue,citecolor=MyDarkBlue,
urlcolor=MyLightBlue,bookmarksnumbered=true,bookmarksopen]{hyperref}
\hypersetup{colorlinks, citecolor=red,linkcolor=MyDarkBlue, urlcolor=MyLightBlue}

\begin{document}
\vspace*{-0.2in}
\begin{flushright}
\end{flushright}
\vspace{0.5cm}
\begin{center}
{\Large\bf Fermion Masses and Mixings, Leptogenesis and Baryon Number Violation in Pati-Salam Model}\\
\end{center}

\vspace{0.5cm}
\renewcommand{\thefootnote}{\fnsymbol{footnote}}
\begin{center}
{\large
{}~\textbf{Shaikh Saad}\footnote{ E-mail: \textcolor{MyLightBlue}{shaikh.saad@okstate.edu}}
}
\vspace{0.5cm}

{\em Department of Physics, Oklahoma State University, Stillwater, OK 74078, USA }
\end{center}

\renewcommand{\thefootnote}{\arabic{footnote}}
\setcounter{footnote}{0}
\thispagestyle{empty}

\begin{abstract}
In this work we study a 
 predictive model based on a partially unified theory possessing the gauge symmetry of the Pati-Salam group, $SU(2)_L\times SU(2)_R\times SU(4)_C$ supplemented by a global  Peccei-Quinn   symmetry,  $U(1)_{PQ}$. A comprehensive  analysis of the Higgs potential  is carried out in a minimal set-up. The assumed Peccei-Quinn   symmetry  along with solving the strong CP problem, can provide axion as the dark matter candidate. This minimal set-up with limited number of Yukawa parameters can successfully incorporate the hierarchies in the charged fermion masses and mixings. The automatic existence of the heavy Majorana neutrinos generate the extremely small  light neutrino masses through the  seesaw mechanism, which is also responsible for producing the observed cosmological matter-antimatter asymmetry of the universe. We find interesting correlation between the low scale neutrino observables and the baryon asymmetry in this model.  Baryon number violating nucleon decay  processes  mediated by  the scalar diquarks and leptoquarks  in this framework are found to be, $n,p\to \ell+m, \ell^c + m$ ($m=$ meson, $\ell=$ lepton, $\ell^c=$ antilepton) and $n,p\to \ell+\ell^c+\ell^c$.  For some choice of the parameters of the theory, these decay rates  can be within the observable range. Another baryon number violating process, the neutron-antineutron oscillation  can also be in the observable range. 
\end{abstract}

\newpage
\section{Introduction}\label{intro}
Despite  being a very successful theory, the Standard Model (SM) of particle physics has many shortcomings. Such as, the SM does not provide any insights for understanding the hierarchical pattern of the masses and mixings of the charged fermions. Also the origin of the  neutrino oscillations is unexplained in the SM. The observed quantization of electric charge in the SM is also not obvious.  To explain these shortcoming of the SM, extensive search for finding new physics beyond the SM has been carried out in the literature. One of the most attractive extensions of the SM  proposed  in Refs. \cite{Pati:1973uk, Pati:1973rp, Pati:1974yy, Pati:2017ysg}  are based on  partial unification with non-Abelian gauge group $G_{224}=SU(2)_L\times SU(2)_R\times SU(4)_C$.  This Pati-Salam (PS) group is the most minimal quark-lepton symmetric model based on the $SU(4)_C$ group with the lepton number as the fourth color \cite{Pati:1974yy}. The minimal gauge group respecting  symmetry between the left-handed and right-handed  representations along with the  $SU(4)$-color symmetry and ensures electric charge quantization is the PS gauge group. Due to quark-lepton unification, one can hope to understand the flavor puzzle in the PS model.  The fermion multiplets of this theory automatically contain the right-handed neutrinos which are SM singlets, this is why  seesaw mechanism~\cite{Minkowski:1977sc} is a natural candidate in the PS model to explain the tiny  masses of the SM light neutrinos. Furthermore, our universe does not show symmetry between matter and antimatter. The origin of this matter-antimatter asymmetry may have link with the origin of neutrino mass. In the seesaw scenario, the Majorana mass term violates the lepton number conservation, so employing  the seesaw mechanism in the PS framework, the observed baryon asymmetry of the universe can be incorporated by the Baryogenesis via  Leptogenesis mechanism. In such a framework, the lepton asymmetry that is generated dynamically, later converted into the baryon asymmetry by the $(B+L)$-violating sphaleron interactions that exist in the SM. In the SM,  conservation of baryon  number and lepton number are accidental, however, violation of these  quantum numbers are natural in the PS model and baryon number violation induces interesting processes like nucleon decay and neutron-antineutron ($n-\overline{n}$) oscillation.

In this paper, we construct a minimal realistic model based on the PS gauge group augmented by a global $U(1)_{PQ}$ Peccei-Quinn (PQ) symmetry  in the non-supersymetric framework. Such an extension of the PS model by the PQ symmetry is not studied in the literature before and we show the possible implications of imposing this global symmetry into the theory.  Assuming an economical Higgs sector, we construct the complete Higgs potential and analyze it.
A complete analysis of the Higgs potential is also lacking in the literature  due to a large number of gauge invariant allowed terms in the scalar potential. In our framework, existence of the additional $U(1)_{PQ}$ symmetry forbids some of the terms that makes the analysis somewhat simpler.  
 The assumed minimal set of Higgs fields is  required not only to realize successful symmetry breaking of the PS group down to the SM and further down to $SU_{C}(3)\times U_{\rm{em}}(1)$, but also to reproduce realistic fermion masses and mixings. We discuss the possibility of baryon number violating processes such as nucleon decay and $n-\overline{n}$ oscillation in this set-up. Nucleon decay  processes in this framework are found to be, nucleon $\rightarrow$ lepton + meson, nucleon $\rightarrow$ antilepton + meson and  nucleon $\rightarrow$ lepton + antilepton + antilepton. In our set-up, we construct the dimension-9 and dimension-10 operators that mediate nucleon decay via the scalars within the minimal Higgs sector. Relative branching fractions of different modes of nucleon decay processes arising in this theory are computed on the dimensional ground.

We also analyze the predictions of this model for quark and lepton masses and mixings. Our numerical study shows full consistency with the experimental data. In addition to unifying quarks and leptons, seesaw mechanism arises naturally  in $G_{224}$ framework  due to the automatic presence of the right-handed neutrinos.  To solve the matter-antimatter asymmetry of the universe, we implement the novel idea of Baryogenesis via Leptogenesis. Utilizing the type-I seesaw scenario, the Baryogenesis via Leptogenesis mechanism links the  matter-antimatter asymmetry and the CP violation in the neutrino sector.   In search of successful baryon asymmetry,  we scan over the relevant parameter space and, present the predictions of our model of the neutrino observables.  In this work, on top of the PS gauge symmetry, we impose a global $U(1)_{PQ}$ PQ symmetry, that solves the strong CP problem. If the PQ symmetry is broken at the high scale $\sim 10^{11-12}$ GeV, then the pseudo-scalar Goldstone boson associated with this breaking can explain the observed dark matter relic density of the universe, so the dark matter candidate in this model is the axion. The presence of this global $U(1)$ symmetry in addition to  restricting some of the terms in  the Higgs potential it also forbids  few terms in  the Yukawa Lagrangian, hence helps to reduce the number of parameters in the theory significantly. We discuss the implications of both the high scale and low scale PS breaking scenarios.  With the economic choice of Higgs multiplets, we do a general study in $SU(2)_L\times SU(2)_R\times SU(4)_C \times U(1)_{PQ}$ set-up; a special case with the imposed discrete parity symmetry that demands $g_L=g_R$ at the PS symmetric phase is also considered  and additional restrictions due to the consequence of this  discrete symmetry are mentioned explicitly through out the text.  We also explore another interesting possibility, where with the absence of the discrete parity symmetry, $g_L=g_R$ unification can still be realized at the PQ breaking scale which however, requires extension of the   the minimal Higgs sector. 

The rest of the paper is organized as follows. In Sec. \ref{chapter01} we give the details of the model. In Sec. \ref{chapter02} we discuss the mass generation of the charged fermions as well as the neutrinos, then we briefly review the leptogenesis mechanism in Sec. \ref{chapter03}. Detailed numerical analysis of the charged fermion masses and mixings and also leptogenesis are  performed in Sec. \ref{chapter04}. Comprehensive analysis of the Higgs potential and computation of the Higgs boson mass spectrum are carried out in Sec. \ref{chapter05}. In Sec. \ref{chapter06} we find the baryon number violating processes within the model and construct the effective higher dimensional operators responsible for such processes and finally we  conclude in Sec. \ref{chapter07}.

\section{The model}\label{chapter01}

\subsection{The gauge group and spontaneous symmetry breaking chain} 

\noindent
\textbf{Breaking chain and particle content}

\vspace*{5pt}
\noindent
We work on a left-right symmetric partial unification theory based on the PS gauge group, $SU(2)_{L} \times SU(2)_{R} \times SU(4)_{C}$. $SU(4)_{C}$ is an extension of the QCD gauge group, $SU(3)_{C}$ with lepton as the fourth color and $SU(2)_{R}$ is right-handed gauge group similar to the SM $SU(2)_{L}$ weak interactions.
Starting from this gauge group, to break it down to  the SM group, several different breaking  chains are possible, but in this paper we assume the one step spontaneous symmetry breaking (SSB) of the PS group to that of the SM group,

\vspace{-10pt}
\begin{align}\label{eq:SSB}
G_{224} &\xrightarrow{M_X} SU(2)_{L} \times U(1)_{Y} \times SU(3)_{C}  \\
&\xrightarrow{M_{EW}} U(1)_{em} \times SU(3)_{C}.
\end{align}

\noindent   In our model, we assume the existence of the following  Higgs multiplets (under the PS group):
\vspace{-2pt}
\begin{align}\label{eq:Higgs}
\bm\Phi = (2,2,1), \;\;\; \bm\Sigma = (2,2,15),\;\;\; \bm\Delta_{R} = (1,3,10).
\end{align}

\noindent 
The breaking of the PS symmetry by employing the Higgs multiplet $(1,3,10)$
 was first discussed in \cite{Mohapatra:1980qe}. 
Instead of $G_{224}$, if left-right parity  symmetry is also preserved (in this case we denote the group as $G_{224P}$), the existence of the Higgs field $\bm\Delta_{L}=(3,1,10)$ is needed due to the presence of the  parity symmetry. This choice of the Higgs multiplets is the minimal set.
This one step breaking of PS group to the SM can be achieved by the VEV of the (1,3,10) multiplet, $v_{R}=\langle \bm\Delta_{R} \rangle$~\cite{Pati:1983zp}. If the group is $G_{224P}$ then, in general the  breaking of the parity scale may not coincide with the breaking of the PS symmetry. However, breaking the $G_{224P}$ group by the VEV of $(1,3,10)$ automatically breaks the parity symmetry. The multiplet $\bm\Delta_{R}$, breaking $SU(4)_{C}$, $B-L$ and left-right symmetry spontaneously also provides masses to the heavy right-handed neutrinos. In an alternative approach the parity symmetry can be broken before breaking the PS group   by a parity odd singlet Higgs and then the PS symmetry can be broken by the usual $(1,3,10)$ VEV. The SM group can be broken by the scalar field $\bm\Phi$ that contains the SM doublet. The  VEV of $\bm\Phi$ field,

\vspace{-2pt}
\begin{equation}
\langle \bm\Phi \rangle = \begin{bmatrix}
k_{1} & 0 \\
0 & k_{1}^{\prime}
\end{bmatrix} \otimes diag(1,1,1,1)
\end{equation}
\vspace{2pt}

\noindent 
is responsible for generating Dirac mass terms for the SM fermions. But if only $\bm\Phi$ is responsible for generating charged fermion masses, one gets the unacceptable relations, $m_{e}=m_{d}$, $m_{\mu}=m_{s}$ and $m_{\tau}=m_{b}$. These lead to $m_{e}/m_{\mu}=m_{d}/m_{s}$, which are certainly not in agreement with experimental measured values. These bad relations are the consequences  of the multiplet $\bm\Phi$ being color singlet (in the $SU(4)_{C}$ space the fourth entry is also 1) and cannot differentiate fermions with different colors. 
To cure these bad relations, the existence of the Higgs multiplet $\bm\Sigma$ is assumed which is not color blind, and by acquiring VEV of the form: 

\vspace{-2pt}
\begin{equation}
\langle \bm\Sigma \rangle = \begin{bmatrix}
k_{2} & 0 \\
0 & k_{2}^{\prime}
\end{bmatrix} \otimes diag(1,1,1,-3)
\end{equation}

\noindent 
can correct these bad mass relations~\cite{Pati:1973rp, Pati:1974yy, Pati:1983zp}, $m_{e}=m^{\Phi}_{e}-3 m^{\Sigma}_{e}$ , $m_{d}=m^{\Phi}_{d}+m^{\Sigma}_{d}$ and so on. Even though  
the field $\bm\Phi$ treats quarks and leptons on the same footing, $\bm\Sigma$ field being color non-singlet, distinguishes them and brings additional  Clebsch factor of $-3$ for the leptons.  \\

\noindent
\textbf{Renormalization group equations  and the $v_{R}$ scale} 

\vspace*{5pt}
\noindent
According to phenomenological considerations, the required  hierarchical pattern of the VEVs must obey the following hierarchy: 

\vspace{-11pt}
\begin{equation}
\langle \bm\Delta_{R}  \rangle >> \langle \bm\Phi  \rangle  \sim \langle \bm\Sigma  \rangle >> \langle \bm\Delta_{L}  \rangle .
\end{equation}

\noindent 
As previously mentioned, in the model without the parity symmetry, $\bm\Delta_{L}$ field need not to be present. Even when this field is present,  we assume that this field does not get any explicit VEV. However, this field does  get small induced VEV due to the presence of specific types of quartic  terms in the Higgs potential that are linear in $\bm\Delta_L$.  After the EW symmetry breaking such acquired VEV is of the form, $\langle \bm\Delta_{L}  \rangle \sim \lambda \; v^2_{ew}/v_R$ (where $\lambda$ is the relevant quartic coupling).  The fields $\Phi$ and $\Sigma$ containing the weak doublets acquire VEVs around the electro-weak scale.  

 If parity is assumed to be a good symmetry, $v_{R}$ can be fixed by the renormalization group equations (RGEs) running of the gauge coupling constants by using low energy data. This additional discrete symmetry on top of the PS symmetry demands $g_{L}=g_{R}$. The one-loop RGEs for the gauge couplings are given by \cite{Machacek:1983tz}:  
\vspace{-5pt}  
\begin{equation}\label{eq:RGE01}
\frac{d \alpha^{-1}_{i}(\mu)}{d ln\mu} = \frac{a_{i}}{2 \pi} .
\end{equation}

\noindent   For the SM group, $G_{321}$ these coefficients are found to be \cite{Jones:1981we}: $b_i=(-7,-19/6, 41/10)$. Applying proper matching conditions for the coupling constants, 
\vspace{-2pt}  
\begin{equation}
\alpha^{-1}_{1Y}(M_{X}) = \frac{3}{5}  \alpha^{-1}_{2R}(M_{X})   + \frac{2}{5} \alpha^{-1}_{4}(M_{X})  , \;\; \alpha^{-1}_{2R}(M_{X})=\alpha^{-1}_{2L}(M_{X}), \;\; \alpha^{-1}_{4C}(M_{X})=\alpha^{-1}_{3C}(M_{X}),
\end{equation}

\noindent and using the  low energy data, $\alpha_{s}(M_{Z})=0.1184$, $\alpha^{-1}(M_{Z})=127.944$ and $s^{2}_{\theta_{W}}=0.23116$ taken from Ref.~\cite{Antusch:2013jca} (only the central values are quoted here), we find $M_{X}= 10^{13.71}$ GeV.  From now on, for models with parity symmetry broken by the $\bm\Delta_R$ VEV, we  set $v_{R}=10^{14}$ GeV for the rest of the analysis. Specially when we will discuss the high scale leptogenesis, we stick to this value of $v_R$. On the other hand, if left-right parity symmetry is absent, then the scale $v_R$ is not fixed by the RGEs running. The differences in results for the cases with $G_{224}$ and $G_{224P}$ are mentioned explicitly through out the text when needed. \\

\noindent
\textbf{Left-right gauge coupling unification at the Peccei-Quinn scale}

\begin{figure}[th!]
\centering
\includegraphics[scale=1]{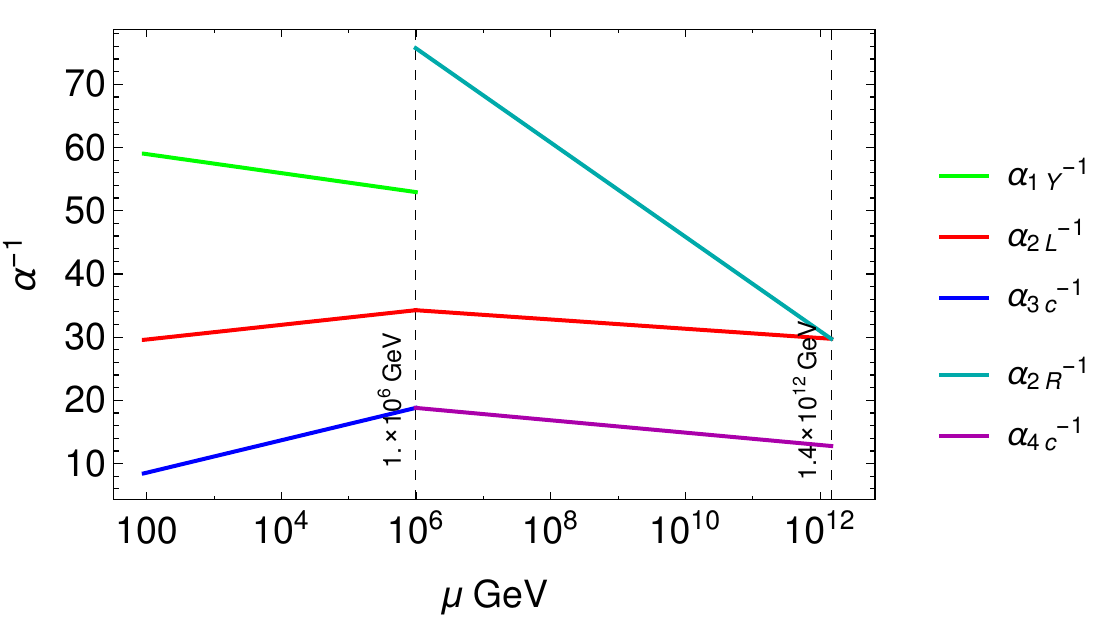}
\caption{One-loop gauge coupling running   of PS model without parity symmetry.  By including an extra $(1,3,10)$ multiplet and a real $(1,3,15)$ multiplet on the top of the minimal Higgs content that are a complex $(2,2,1)$, a complex $(2,2,15)$ and a $(1,3,10)$ multiplet, $g_L=g_R$ unification at the PQ scale  $\sim 10^{11-13}$ GeV can be realized. }\label{PQunification}
\end{figure}

\vspace*{5pt}
\noindent
In this subsection, we explore an alternative realization of $g_L=g_R$ unification without the presence of the left-right parity symmetry. As explained above, breaking the parity symmetry that demands $g_L=g_R$  along with the breaking of the PS symmetry by the $(1,3,10)$ multiplet restricts the PS breaking scale to be high $\sim 10^{14}$ GeV. If parity symmetry is absent, this scale is not determined by the RGEs running from the low energy experimental data and the PS breaking can happen at much higher or even at much lower scale. The experimental limits on the branching ratio for $K^0_L \rightarrow \mu^{\pm} e^{\mp}$ processes, mediated by the new gauge bosons $X_a$ ($a$ is the Lorentz index) with $(B-L)$ charge of $(4/3)$, implies that the $v_R$ scale that breaks the $SU(4)_C$ must be greater than about 1000 TeV \cite{Valencia:1994cj,Smirnov:2007hv}. Here  we explore the possibility of low scale PS scale breaking where $g_L=g_R$ unification can still be realized at the PQ scale $\sim 10^{11-13}$ GeV. However, this requires extension of the minimal Higgs sector. For example, by including an extra $(1,3,10)$ multiplet and a real $(1,3,15)$ multiplet on top of the minimal Higgs content that are a complex $(2,2,1)$, a complex $(2,2,15)$ and a $(1,3,10)$ multiplet, left-right gauge coupling unification can happen at the PQ scale as shown in Fig. \ref{PQunification}. For this plot, the PS breaking scale is fixed at $10^3$ TeV.  With this set of scalars, we find the RGE coefficients to be  $b_i=(2,61/3,8/3)$ for the group $G_{224}$.  
     
\FloatBarrier

\noindent
\textbf{Notation}

\vspace*{5pt}
\noindent
Our notation for indices is as follows: the indices for $SU(2)_{L}$ group are $\alpha, \beta, \gamma, \delta, \kappa = 1,2$, for $SU(2)_{R}$ group  
$\dot{\alpha}, \dot{\beta}, \dot{\gamma}, \dot{\delta}, \dot{\kappa} = \dot{1}, \dot{2}$ and for $SU(4)_{C}$ group $\mu, \nu, \rho, \tau, \lambda, \chi = 1,2,3,4$. For  $SU_{C}(3)_{C} \subset SU(4)_{C}$ group, we use the same symbols for the indices as that of $SU(4)_C$  but with a bar on top,  for example, $ \bar{\mu}, \bar{\nu} = 1,2,3 $. While writing the gauge bosons and the covariant derivatives, we use index $a$ to represent the Lorentz index.

In the PS model, the fermions belong to the  representations ${\bm\Psi_L}_{\mu \alpha}=(2,1,4)_{k}$ and ${\bm\Psi_R}_{\mu \dot{\alpha}}=(1,2,4)_{k}$ that can be written explicitly as follows: 
\vspace{-2pt}
\begin{equation}
\bm\Psi_{L,R}=\begin{pmatrix}
u_r &u_g &u_b &\nu \\
d_r &d_g &d_b &e
\end{pmatrix}_{L,R}.
\end{equation}

\noindent Here $k \;(=1,2,3)$ is the generation index. In group index notation the scalar fields can be written as:
\vspace*{-2pt} 
\begin{eqnarray}
\begin{aligned}\label{Higgs}
& (2,2,1) = \bm\Phi_{\alpha}^{\dot{\alpha}}, \;\;\;\;\;\;\;\;\;\;(2,2,15) = \bm\Sigma^{\nu \;\dot{\alpha}}_{\mu \; \alpha},  \\ 
& (1,3,10) = \bm\Delta_{R\; \mu \nu \; \dot{\alpha}}^{\;\;\;\;\;\;\;\;\;\; \dot{\beta}} , \;\;\;(3,1,10) = \bm\Delta_{L\; \mu \nu \; \alpha}^{\;\;\;\;\;\;\;\;\;\; \beta} .
\end{aligned}
\end{eqnarray}

\noindent The SM decomposition of these fields are given by:
\vspace{-2pt}
\begin{align}
(2,2,1)&=(1,2,\frac{1}{•2})+(1,2,-\frac{1}{•2}),
\\
(2,2,15)&=(1,2,\frac{1}{•2})+(1,2,-\frac{1}{•2})+(3,2,\frac{1}{•6})+(\overline{3},2,-\frac{1}{•6})+(3,2,\frac{7}{•6})+(\overline{3},2,-\frac{7}{•6})
\nonumber \\& 
+(8,2,\frac{1}{•2})+(8,2,-\frac{1}{•2}),
\\
(1,3,10)&= (1,1,0)+(1,1,-1)+(1,1,-2)+(3,1,\frac{2}{3})+(3,1,-\frac{•1}{3•})+(3,1,-\frac{4•}{3•})
\nonumber \\& 
+(6,1,\frac{4•}{3•})+(6,1,\frac{1•}{3•})+(6,1,-\frac{2•}{3•}),
\\
(3,1,10)&=(1,3,-1)+(3,3,-\frac{1}{3•})+(6,3,\frac{1}{•3}).
\end{align}

\subsection{Gauge boson mass spectrum}\label{gauge}
In the PS model, there are in total 21 gauge bosons, $W_{L\; a} \equiv$(3,1,1) of $SU(2)_{L}$, $W_{R\; a} \equiv$(1,3,1) of $SU(2)_{R}$  and $V_{a} \equiv$(1,1,15) of $SU(4)_{C}$. The decomposition of these fields under the SM are:
\vspace{-12pt}
\begin{align}
&(3,1,1)=(1,3,0),
\\
&(1,3,1)=(1,1,1)+(1,1,0)+(1,1,-1),
\\
&(1,1,15)=(1,1,0)+(3,1,\frac{2}{3})+(\overline{3},1,-\frac{2}{3})+(8,1,0).
\end{align}

\noindent The gauge bosons $W_{R}$ are the right-handed analogue of the three SM $SU(2)_{L}$ gauge bosons, $W_{L}$. The decomposition of 15$\subset SU(4)_{C}$ under the group $SU(3)_{C} \times U(1)_{B-L} \subset SU(4)_{C}$ is $15=1(0)+3(+4/3)+\overline{3}(-4/3)+8(0)$, where  8(0) are the massless gluons of $SU(3)_{C}$. The triplets, $X_a\equiv 3(+4/3)$ and $X^{\ast}_a\equiv \overline{3}(-4/3)$ with non-zero $B-L$ quantum numbers are the exotic particles (leptoquark vector bosons). Contrary to the Grand Unified Theories (GUT) based on simple groups, the leptoquark gauge bosons of the PS model do not mediate proton decay as explained below. The transition between quarks and leptons are given by the following interactions that is part of the total Lagrangian: 

\begin{equation}\label{xmu}
\mathcal{L}_{X} \supset \frac{g_{4}}{\sqrt{2}} \{ X_a( \overline{u} \gamma^a \nu + \overline{d} \gamma^a e )  +  X^{\ast}_a( \overline{u}^{c} \gamma^a \nu^{c} + \overline{d}^{c} \gamma^a e^{c} )  \}.
\end{equation}

\noindent
Since $U(1)_{B-L}$ is already a part of the gauge symmetry, $B-L$ is a conserved quantity. In addition to this, the above gauge interactions of the leptoquarks Eq. \eqref{xmu} has the accidental global $B+L$ symmetry, these two conserved quantities ensure the conservation of both $B$ and $L$ separately, this is why the gauge bosons of PS group do not mediate proton decay. On the other hand, minimal $SU(5)$ GUT model is ruled out due to too rapid proton decay mediated by the gauge leptoquarks. Since one can assign specific baryon and lepton numbers to these gauge bosons, in contrast to $SO(10)$ model, proton decay does not take place via these gauge bosons. Unification scale in minimal $SO(10)$ model needs to be really high $> 5\times 10^{15}$ GeV to save the theory from too rapid proton decay.

The spontaneous symmetry breaking $G_{224}\rightarrow G_{213}$ that does not break the $SU(2)_L$ group, the $W_{L\; a}$ gauge bosons remain massless in this stage. Due to this breaking, among the 18 (15 of $SU(4)_{C}$ and 3 of $SU(2)_{R}$) massless gauge bosons, 9 of them become massive after eating up the 9 Goldstone bosons (will be identified at the later part of the text), from the field $\bm\Delta_R$ and the other 9 of them (8 of $SU(3)_{C}$ and 1 of $U(1)_{Y}$) remain massless. Here we compute the mass spectrum of the gauge bosons. Following Ref.\cite{Li:1973mq}   the covariant derivative can be written as
\vspace{-2pt}
\begin{align}
D_a \bm\Delta_R= \partial_a \bm\Delta_{R\mu\nu\;\dot{\alpha}}^{\;\;\;\;\;\;\;\;\;\dot{\beta}} 
&
-i g_R W_{a R\dot{\alpha}}^{\;\;\;\;\dot{\gamma}} \bm\Delta_{R\mu\nu\;\dot{\gamma}}^{\;\;\;\;\;\;\;\;\;\dot{\beta}} 
+i g_R W_{a R\dot{\gamma}}^{\;\;\;\;\dot{\beta}} \bm\Delta_{R\mu\nu\;\dot{\alpha}}^{\;\;\;\;\;\;\;\;\;\dot{\gamma}}
\nonumber \\ &
-i g_C X_{a\;\mu}^{\rho} \bm\Delta_{R\rho\nu\;\dot{\alpha}}^{\;\;\;\;\;\;\;\;\;\dot{\beta}}
-i g_C X_{a\;\nu}^{\rho} \bm\Delta_{R\rho\mu\;\dot{\alpha}}^{\;\;\;\;\;\;\;\;\;\dot{\beta}},
\end{align}

\noindent where $a$ represents the Lorentz index. When the  PS symmetry gets broken spontaneously  by the VEV of the $\bm\Delta_R$ field, using this covariant derivative the gauge boson mass spectrum can be computed to be:
\vspace{-8pt}
\begin{align}
&M_{W^{\pm}_R}=  \sqrt{2} g_R v_R,\\
&M_{V^{(i)}}=  \sqrt{2} g_C v_R .
\end{align}

\noindent Here $i=9-14$ and their electric charge are $\pm 2/3$. The third component, $W_R^{(3)}$ of the (1,3,1) gauge boson  mixes with the $V^{(15)}$ component from (1,1,15),  then in the basis $\{ W_R^{(3)}, V^{(15)} \}$  the mass squared matrix is given by:
\vspace{-5pt}
\begin{align}
M^2=2 \begin{pmatrix}
g^2_R v^2_R&-g_R \overline{g}_C v^2_R \\
-g_R \overline{g}_C v^2_R&\overline{g}^2_C v^2_R
\end{pmatrix},
\end{align}

\noindent where we have defined $\overline{g}_c=\sqrt{3/2} \;g_C$. One can easily calculate the two eigenvalues of this matrix,  one of the eigenvalues is zero and the corresponding eigenstate is given by
\vspace{-2pt}
\begin{align}
A_a=\frac{•1}{•\sqrt{•g^2_R+ \overline{g}^2_C}} \left( \overline{g}_C W_{R\;a}^{(3)} + g_R X^{(15)}_a \right).
\end{align}

\noindent This is the massless gauge boson of $U(1)_{Y}$ group.   Its orthogonal eigenstate acquires mass given by $\sqrt{2} v_R \sqrt{•g^2_R+ \overline{g}^2_C}$. In addition, for the unbroken $SU(3)_C$ group, the massless gauge bosons, the gluons are identified with $V^{(i)}$ ($i=1-8$) fields.

\subsection{Peccei-Quinn symmetry}\label{section:PQ}
On top of the PS gauge symmetry we assume the existence of global Peccei-Quinn (PQ) symmetry, $U(1)_{PQ}$~\cite{Peccei:1977hh,Peccei:1977ur,Weinberg:1977ma,Wilczek:1977pj} (for a relation between leptonic CP violation with strong CP phase in the context of left-right symmetric models see Ref. \cite{Kuchimanchi:2014ota}). The PQ symmetry naturally solves the strong CP problem and simultaneously provides the axion solution to the dark matter problem~\cite{Kim:2009xp,Kim:2008hd}. So the complete symmetry of our theories are either  $G_{224} \times U_{PQ}(1)$ or  $G_{224P} \times U_{PQ}(1)$. The SM singlet present in $\bm \Delta_R$ that breaks the PS symmetry and the singlet $\mathcal{S}$, each can break one $U(1)$ symmetry. As a result, even though $\bm \Delta_R$ multiplet carries PQ charge, it cannot simultaneously break both $U(1)_{B-L}$ and $U(1)_{PQ}$.  
If the VEV of the singlet, $\langle \mathcal{S} \rangle=v_{S} > v_R$, then this VEV breaks the $U(1)_{PQ}$. On the contrary, if $v_R > v_S$ is assumed,  a combination of the $B-L$ and $PQ$ symmetry remains unbroken, which is further broken by the VEV of $\mathcal{S}$.  Hence the presence of an additional SM singlet field ($\mathcal{S}$) carrying non-trivial charge under PQ symmetry  is required. 

Due to the presence of the $U(1)_{PQ}$ symmetry, the complex scalar fields carry PQ charge fixed by the charges of the fermions, which consequently puts additional restrictions on the Higgs potential and also in the Yukawa Lagrangian, this reduces the number of parameters in the Higgs potential as well as in the Yukawa sector significantly. For example, if PQ symmetry is not imposed, each of these $\Phi$ and $\Sigma$ fields can have two independent Yukawa coupling matrices. However, the presence of the PQ symmetry restricts one of such Yukawa coupling terms, hence instead of four, only two Yukawa coupling matrices determine the charged fermion spectrum, makes the theory  predictive.

The VEV of the singlet field, $\langle \mathcal{S} \rangle$ breaks the PQ symmetry at the scale $M_{PQ}$ and phenomenological requirement of this scale is $M_{PQ} \sim 10^{11-13}$ GeV. The multiplets (2,2,1) and (2,2,15) are assumed to be complex and have non-zero charges under the PQ group. We choose the following charge assignment of the fermion and Higgs fields under $U(1)_{PQ}$:
\FloatBarrier
\begin{table}[th]
\label{table:PQ}
\centering
\begin{tabular}{|c|c|c|c|c|c|c|c|}
\hline
fields &$\bm\Phi$(2,2,1) &$\bm\Sigma$(2,2,15)&$\bm\Delta_R$(1,3,10)&$\bm\Delta_L$(3,1,10)&$\bm\Psi_L$(2,1,4)&$\bm\Psi_R$(1,2,4)  &$\mathcal{S}$(1,1,1)  \\ [1ex] \hline
$Q_{PQ}$ &+2&+2&-2&+2&+1&-1&+4 \\ [0.5ex] \hline
\end{tabular}
\caption{  $U(1)_{PQ}$ charge assignment of the scalars.}
\end{table}

\section{Fermion masses and mixings}\label{chapter02}

In this section we discuss the fermion masses and mixings in the PS model. The model under consideration is very predictive in explaining the data in the fermion sector. The Yukawa part of the Lagrangian in our set-up is given by:
\vspace*{-2pt}
\begin{equation}\label{yukawa}
\mathcal{L_{Y}} = {Y_1}_{ij}\;  \overline{\bm\Psi}_{Li} \bm\Phi \bm\Psi_{Rj}  +{Y_{15}}_{ij}\;  \overline{\bm\Psi}_{Li} \bm\Sigma \bm\Psi_{Rj}   +\frac{1}{2} \{ {Y^R_{10}}_{ij} \bm\Psi^T_{Ri} C \bm\Delta^{\ast}_R \bm\Psi_{Rj} + R \leftrightarrow L \} + h.c
\end{equation}

\noindent
where, $Y_1, Y_{15}$ and $Y^{R,L}_{10}$ are the Yukawa coupling matrices resulting due to the interactions of the fermions with the (2,2,1), (2,2,15), (1,3,10) and (3,1,10)  multiplets respectively. Generically $Y_1$ and $Y_{15}$  are  general  complex matrices and due to Majorana nature, $Y^{R,L}_{10}$ are complex symmetric. When parity is imposed (see Eq. \eqref{parity}) the matrices $Y_1$ and $Y_{15}$  become Hermitian and $Y^{R,L}$ become identical, i.e,
\vspace{-2pt}
\begin{equation}\label{restriction}
Y_1=Y^{\dagger}_1,\;\;Y_{15}=Y^{\dagger}_{15},\;\;Y^R_{10}=Y^L_{10}=Y_{10}=Y^T_{10}.
\end{equation}

\noindent For the analysis of the fermion masses and mixings we restrict ourselves to the case when parity summery is realized since this significantly reduces the number of parameters in the fermion sector due to constraints mentioned in Eq. \eqref{restriction}, so our model is highly  predictive.   

  The VEV of the (1,3,10) multiplet $\langle \bm\Delta_{R} \rangle$ breaks the $G_{224}$ group down to the SM group  and, generates the right-handed Majorana neutrino masses given by $v_{R} Y_{10}$. The Higgs fields $\bm\Phi$ and $\bm\Sigma$ each contains two doublets of $SU(2)_L$ that  acquire non-zero VEVs  and are responsible for generating charged fermion masses. From the Lagrangian one can write down the fermion mass matrices as: 
\vspace*{-2pt}
\begin{align}
&M_{u} = k_{u} Y_1 + v_{u} Y_{15},\;\;\;\; M_{d} = k_{d} Y_1 + v_{d} Y_{15} ,\\
&M_{D} = k_{u} Y_1 - 3 v_{u} Y_{15} ,\;\; M_{e} = k_{d} Y_1 - 3 v_{d} Y_{15}, \\
&~~~~~~~~~~~~~~~~~~~~~ M_{R} = v_{R} Y_{10}.
 \end{align}

\noindent
$M_{u}$, $M_{d}$ are the up-type and down-type quark mass matrices, $M_{e}$ is the charged lepton mass matrix,  $M_{D}$ is the neutrino Dirac mass matrix and $M_{R}$ is the right-handed Majorana neutrino mass matrix. $k_{u,d}$, $v_{u,d}$ are the VEVs of the four doublets. $k_{u,d}$ ($v_{u,d}$) are the up-type and down-type VEVs of the multiplet $\Phi (2,2,1)$ ($\Sigma (2,2,15)$).  In general these VEVs are complex and there is one common phase for $k_{u}$ and $k_{d}$ and different phases for each of  $v_{u}$ and $v_{d}$. Only two relative phases will be physical and we bring these phases ($\theta_{1,2}$) with $v_{u}$ and $v_{d}$. The analysis done in Sec. \ref{section:doubletmass} shows that the VEV ratios are complex and can not be made real. One can absorb the VEVs into the coupling matrices and redefine them, leaving two relevant VEV ratios ($r_{1,2}$). Following these arguments, we can rewrite the mass matrices as, 
\vspace*{-2pt}
\begin{align}
&M_{u} =  M_1 + e^{i \theta_1} M_{15},\;\;\;\; M_{d} = r_1 M_1 + r_2 e^{i \theta_2} M_{15} ,\\
&M_{D} =  M_1 -3 e^{i \theta_1} M_{15},\;\; M_{e} = r_1 M_1 -3 r_2 e^{i \theta_2} M_{15} ,\\
&~~~~~~~~~~~~~~~~~~~~~ M_{R} = v_{R} Y_{10} ,
 \end{align}

\noindent
where we have defined $M_1 = k_u Y_1$, $M_{15} = v_u Y_{15}$, $r_1 = k_d/k_u$ and $r_2 = v_d/v_u$. As mentioned  earlier, due to parity symmetry the matrices $M_1$ and $M_{15}$ are Hermitian, so without loss of generality one can take the $M_1$ matrix to be diagonal and real (3 real parameters) and  one can also rotate away the two phases from the $M_{15}$ matrix leaving only one phase in it (5 real and 1 complex parameters). So in total there are 11 magnitudes and 3 phases i.e, 14 free parameters in the charged fermion sector to fit 13 observables for the case of hard CP-violation\footnote{For spontaneous  CP-violation scenario, the Yukawa coupling matrices are real, so there are  11 magnitudes and 2 phases i.e, 13 free parameters to fit 13 observables. In the next section we will perform numerical study to fit the fermion masses and mixings in the charged fermion sector. Our finding is that the spontaneous CP-violation case is unable to reproduce the observables (we found large total $\chi^2 \sim 125$), so from now on we will only consider the hard CP-violation case. }. The fit result in the charged fermion sector is presented in Sec. \ref{fermionfit}.

Let us now discuss the neutrino sector. The right-handed Majorana mass matrix is complex symmetric matrix and the corresponding Yukawa coupling matrix $Y_{10}$ is arbitrary since it decouples from the charged fermion sector which is unlike the case of $SO(10)$ models \footnote{For  fits to fermion masses and mixings within the $SO(10)$ framework see for example Refs.    \cite{Babu:1992ia, Babu:2005ia, Bajc:2005zf, Bertolini:2006pe, Joshipura:2011nn, Dueck:2013gca, Fukuyama:2015kra, Babu:2016cri, Babu:2016bmy, Meloni:2016rnt, Saad:2017wgd, Ohlsson:2018qpt, Babu:2018tfi, Babu:2018qca, Ohlsson:2019sja}.}. In unified theories due to the presence of right-handed neutrinos seesaw  mechanism is a very good candidate to explain the extremely small observed light neutrino masses. One should note that due to the presence of  terms linear in $\bm\Delta_L$ in the Higgs  potential (Eq. \eqref{V}), this field will acquire a small induced VEV, $v_L$ as aforementioned, which would be responsible for generating left-handed Majorana neutrino mass, $M_L=v_L Y_{10}$ (type-II seesaw contribution). In this paper, we assume the dominance of type-I seesaw scenario, then  the light neutrino mass matrix is given by the type-I  seesaw~\cite{Minkowski:1977sc} formula, 

\vspace{-8pt}
\begin{equation}\label{Mnu}
\mathcal{M}_{\nu} = -  M_{D} M_{R}^{-1} M^{T}_{D}.
\end{equation} 
\noindent
Inverting the type-I seesaw formula one can express $M_{R}$ as,
\begin{equation}\label{MR}
M_{R} = -M^{T}_{D} \mathcal{M}_{\nu}^{-1}  M_{D} .
\end{equation}

There is no new parameter in the $M_{D}$ matrix and is completely fixed by the charged fermion sector. The light neutrino mass matrix, $\mathcal{M}_{\nu}$ can be diagonalized as 
\begin{equation}
\mathcal{M}_{\nu}= U_{\nu} \Lambda_{\nu} U^{T}_{\nu},
\end{equation} 
\noindent with  \vspace{-8pt}
\begin{equation}
\Lambda_{\nu} = \rm{diag}(m_{1},m_{2},m_{3}),
\end{equation} 

\noindent
with the eigenvalues being real and in the basis where the charged lepton mass matrix is diagonal,
\vspace{-5pt}
\begin{equation}\label{U}
U_{\nu} = U_{\rm{PMNS}} \; \rm{diag}(e^{-i \alpha},e^{-i \beta},1) 
\end{equation} 
\noindent
where $\alpha$ and $\beta$ are Majorana phases and $U_{\rm{PMNS}}$ is the CKM type mixing matrix with only one Dirac type phase $\delta$ in it.

We assume normal hierarchy \footnote{For inverted ordering we have not found any solution that can generate successful baryon asymmetry, so we only concentrate on normal ordering.} in the light neutrino sector, which leads up to a good approximation, $m_{2} \sim \sqrt{\Delta m^{2}_{sol}}$ and $m_{3} \sim \sqrt{\Delta m^{2}_{atm}}$ for neutrino masses \footnote{As we have assumed normal hierarchy, the lightest left-handed neutrino mass gets restricted in the range  $0 \leq m_{1} \lesssim 70 \% \;m_{2}$.} . The quantities ($\Delta m^{2}_{sol}$, $\Delta m^{2}_{atm}$, $\theta^{\rm{PMNS}}_{ij}$) in the neutrino sector have already been measured experimentally with good accuracy. The quantities $m_{1}$, $\alpha$, $\beta$ and $\delta $ are yet to be determined experimentally. So in Eq. \eqref{MR}, using the experimentally measured quantities in the neutrino sector, the right-handed Majorana mass matrix can be determined as  a function of these  four unknown quantities. In Sec. \ref{BviaLPS},   we will explain the algorithm we follow while searching for the allowed parameter space to reproduce successful leptogenesis in this model and also present our results.


\section{Baryogenesis via Leptogenesis}\label{chapter03}
In unified theories the Baryogenesis via Leptogenesis \cite{Fukugita:1986hr}  is a natural candidate to explain  the observed matter-antimatter asymmetry ~\cite{Sakharov:1967dj}. This simple mechanism can be implemented in theories where light neutrino mass is generated via seesaw mechanism. For  studies on leptogenesis in the framework of $G_{224}/SO(10)$ see for example Refs.  \cite{Pati:2002pe, Fukuyama:2002ch, Majee:2007uv, Buccella:2010jc, Blanchet:2010kw,  Okada:2011ar, Buccella:2012kc, Altarelli:2013aqa, Addazi:2015yna, DiBari:2017uka}. In this mechanism, the baryon asymmetry of the universe is generated by the lepton asymmetry which is initially produced dynamically and later converted into the baryon asymmetry via the $(B+L)$-violating sphaleron process ~\cite{Kuzmin:1985mm} that exists in the SM. Computing the baryon-asymmetric parameter involves solving the coupled Boltzmann equations. The asymmetry is generated when the decay rates of the heavy neutrinos $<H$  ($H$ being the Hubble expansion rate), so leptogenesis is expected to occur at a temperature of order of the mass of the lightest right-handed heavy neutrino, $M_{1}$. For hierarchical spectrum of the right-handed neutrinos, i.e, $M_{1} \ll M_{2} < M_{3}$, the lightest heavy neutrino is responsible for generating the baryon asymmetry and known as $N_{1}$-dominated leptogenesis (for reviews on leptogenesis see for example Refs.  \cite{Buchmuller:2004nz, Fong:2013wr}). In this work we  concentrate on $N_{1}$-dominated leptogenesis.  In the literature it has been pointed out that  flavor can play significant role in the mechanism of  leptogenesis. Flavored leptogenesis has been studied in great details in the literature, see for example Refs.  \cite{Barbieri:1999ma,Abada:2006fw,Abada:2006ea,Blanchet:2006be,Antusch:2006cw,Nardi:2006fx,Dev:2014laa,Dev:2017trv, Moffat:2018wke} for earlier works.

The minimum required reheating temperature of the universe depends on the details of the flavor structure of the lepton asymmetry. Without taking into account the flavor effects, the lower bound to produce successful baryon asymmetry is $M_{1} > 10^{9}$ GeV ~\cite{Davidson:2002qv}. Including the flavor effects relaxes this lower bound a little bit (for details see for example Refs.~\cite{Abada:2006fw, Moffat:2018wke}). Approximate analytical solutions of the Boltzman equations  have been derived that are in good agreement with the exact solutions (see for example Ref.~\cite{Abada:2006ea}). While scanning over  the parameter space in search for successful leptogenesis we apply these analytical solutions to compute the baryon asymmetry. The analytical formula depends on the interaction rate of the charged lepton Yukawa couplings ~\cite{Nardi:2006fx}. We are interested in the two different regions, first, when only the tau Yukawa coupling is in equilibrium which corresponds to the region $10^{9}\rm{GeV}  \lesssim  M_{1} \lesssim 10^{12}$ GeV. In this first case, the flavor effects play vital role. The second region where no charged lepton Yukawa couplings are in equilibrium that corresponds to the case  $M_{1} \gtrsim 10^{12}$ GeV. In this second case all flavors are indistinguishable and is no different than the one flavor scenario.

Here we briefly summarize the required approximate analytical solutions for our analysis   that are derived in the literature as mentioned above. In the regime where flavors are indistinguishable, the CP asymmetry generated by the $N_1$ decay is
\vspace{-2pt}
\begin{align}
\epsilon_1=\frac{1}{8\pi} \sum_{j\neq 1}\frac{Im[(Y_{D}^{\dagger}Y_D)^2_{j1}]}{(Y_{D}^{\dagger}Y_D)_{11}} \;g\left( \frac{M^2_j}{M^2_1}\right),
\end{align}

\noindent where,
\vspace{-2pt}
\begin{align}
g(x)=\sqrt{x}\left[\frac{1}{1-x}+1-(1+x)ln\left(\frac{1+x}{x} \right)\right].
\end{align}

\noindent  Beside the CP parameter $\epsilon_1$, the final asymmetry depends on the wash-out parameter,
\vspace{-2pt}
\begin{align}
K=\frac{\widetilde{m}_1}{\widetilde{m}^{\ast}},
\end{align}

\noindent with $\widetilde{m}^{\ast} \sim 10^{-3}$ eV and
\vspace{-2pt}
\begin{align}
\widetilde{m}_1= \frac{(Y_{D}^{\dagger}Y_D)_{11} v^2}{M_1}.
\end{align}

\noindent In the strong wash-out regime, i.e, for $K>>1$, the lepton asymmetry is given by the following approximate formula
\vspace{-2pt} 
\begin{align}
Y_{\mathcal{L}}\simeq 0.3 \frac{\epsilon_1}{g^{\ast}} \left( \frac{0.55\times 10^{-3} \rm{eV}}{\widetilde{m}_1} \right)^{1.16},
\end{align}

\noindent with $g^{\ast}$ being the effective number of spin-degrees of freedom in thermal equilibrium, which is $\sim 108$ in the SM with a single generation of right-handed neutrinos. With these the baryon asymmetry is given by $Y_{\mathcal{B}}\simeq 12/37 \;Y_{\mathcal{L}}$. Another useful relation is $\eta_B = 7.04\; Y_{\mathcal{B}}$, where $\eta_B$ is the number of baryons and anti-baryons normalized to the number of photons. On the other hand, in the weak wash-out regime, the approximate analytical formula is,
\vspace{2pt}
\begin{align}
Y_{\mathcal{L}}\simeq 0.3 \frac{\epsilon_1}{g^{\ast}} \left( \frac{\widetilde{m}_1}{3.3\times 10^{-3} \rm{eV}} \right).
\end{align}

On the contrary, the regime where the flavor effects are important, the CP asymmetry in the $\alpha$-th flavor is given by
\vspace{-2pt}
\begin{align}
\epsilon_{\alpha \alpha}= \frac{1}{8\pi (Y_{D}^{\dagger}Y_D)_{11}} \sum_{j\neq 1} Im[(Y_{D}^{\dagger})_{1\alpha}(Y_{D}^{\dagger}Y_D)_{1 j}(Y_{D}^{T})_{j\alpha}]\;  g\left( \frac{M^2_j}{M^2_1}\right).
\end{align}

\noindent And the wash-out parameter is 
\vspace{-1pt}
\begin{align}
K_{\alpha \alpha}\simeq \frac{\widetilde{m}_{\alpha \alpha}}{10^{-3} \rm{eV}},\;\; \widetilde{m}_{\alpha 1}=\frac{|(Y_D)_{\alpha 1}|^2 v^2}{M_1}
\end{align}

\noindent that parametrizes the decay rate of $N_1$ to the $\alpha$-th flavor. In the strong wash-out regime for all flavor, i.e, $K_{\alpha \alpha}>>1$, the total asymmetry generated is given by, $Y_{\mathcal{L}}=\sum_{\alpha} Y_{\alpha \alpha}$, where the approximate analytical formula for each flavor, $Y_{\alpha \alpha}$ is
\vspace{1pt} 
\begin{align}
Y_{\alpha \alpha}\simeq 0.3 \frac{\epsilon_{\alpha \alpha}}{g^{\ast}} \left( \frac{0.55\times 10^{-3} \rm{eV}}{\widetilde{m}_{\alpha \alpha}} \right)^{1.16}.
\end{align}

\noindent And in the weak wash-out regime the formula is,
\vspace{1pt} 
\begin{align}
Y_{\alpha \alpha}\simeq 1.5  \frac{\epsilon_{\alpha \alpha}}{g^{\ast}} (\frac{\widetilde{m}_1}{3.3\times 10^{-3} \rm{eV}}) (\frac{\widetilde{m}_{\alpha \alpha}}{3.3\times 10^{-3} \rm{eV}}).
\end{align}

  The Baryon asymmetric  parameter has been measured experimentally which is $\eta_{B}= (5.7 \pm 0.6)\times 10^{-10}$ \footnote{$90 \%$ CL - deuterium only.}   ~\cite{Iocco:2008va, Fong:2011yx}. Since this scenario of generating baryon asymmetry requires the right-handed neutrino mass scale to be high, for this analysis we fix the PS breaking  scale to be $v_R= 10^{14}$ GeV as discussed before in the text.

\section{Fit to fermion masses and mixings and parameter space for successful Leptogenesis}\label{chapter04}

\subsection{Numerical analysis of the charged fermion sector}\label{fermionfit}
In this sub-section we show our fit results of the fermion masses and mixings in the charged fermion sector.  For optimization purpose we do a $\chi^{2}$-analysis. The pull and $\chi^{2}$-function are defined as:  
\vspace{-5pt}
\begin{align}
P_{i} &= \frac{O_{i\; \rm{th}}-E_{i\; \rm{exp}}}{\sigma_{i}}, \\
\chi^{2} &= \sum_{i} P_{i}^{2},
\end{align}

\noindent
where $\sigma_{i}$ represent experimental 1$\sigma$ uncertainty  and $O_{i\ \rm{th}}$, $E_{i\; \rm{exp}}$ and $P_{i}$ represent the theoretical prediction, experimental central value and pull of an observable $i$. We fit the values of the observables at the PS breaking scale, $M_{PS}=10^{14}$ GeV. To get the PS scale values of the observables, we take the central values at the $M_{Z}$ scale from Table-1 of Ref.~\cite{Antusch:2013jca} and run the RGEs \cite{Machacek:1983fi, Arason:1991ic} to get the inputs at the high  scale.  For the associated one sigma uncertainties of the observables at the PS scale, we keep the same percentage uncertainty with respect to the central value of each quantity as that of the $M_{Z}$ scale. For the charged lepton Yukawa couplings, a relative uncertainty of $0.1 \%$ is assumed in order to take into account the theoretical uncertainties, for example threshold  effects at the PS scale. The inputs are shown in the Table \ref{table:1} where the fit results are presented.

\begin{table}[t]
\centering
\footnotesize
\resizebox{0.7\textwidth}{!}{
\begin{tabular}{|c|c|c|c|}
\hline  \hline
\pbox{10cm}{Masses (in GeV) and \\  CKM parameters}  & \pbox{10cm}{~~~Inputs \\(at $\mu= M_{PS}$)} & Best fit values & Pulls \\ [1ex] \hline
$m_{u}/10^{-3}$  & $0.48 \pm 0.16 $ & 0.48  & 0.009 \\ \hline
$m_{c}$ & $0.26 \pm 0.008$ & 0.26 & -0.03 \\ \hline
$m_{t}$  & $80.78 \pm 0.69$ & 80.78 & 0.001  \\ \hline
$m_{d}/10^{-3}$  & $1.24 \pm 0.12$  & 1.26 & 0.020 \\ \hline
$m_{s}/10^{-3}$ & $23.50 \pm 1.23$ & 22.21 & -1.04 \\ \hline
$m_{b}$   & $1.09 \pm 0.009$ & 1.09 & 0.11 \\ \hline
$m_{e}/10^{-3}$  & $0.482669 \pm 0.004826 $  & 0.482645 & -0.05 \\ \hline
$m_{\mu}/10^{-3}$  & $ 101.8943 \pm 1.0189 $  & 101.898 & 0.03 \\ \hline
$m_{\tau}$  & $ 1.732205 \pm 0.017322 $ & 1.73223 & 0.01  \\ \hline
$\theta^{\rm{CKM}}_{12}/10^{-2}$  & $22.543 \pm 0.071$ & 22.541 & -0.02 \\ \hline
$\theta^{\rm{CKM}}_{23}/10^{-2}$  & $4.783 \pm 0.072$ & 4.799 & 0.22 \\ \hline
$\theta^{\rm{CKM}}_{13}/10^{-2}$ & $0.413 \pm 0.014$ & 0.412 & -0.01 \\ \hline
$\delta^{\rm{CKM}}$  & $1.207 \pm 0.054$ & 1.198 & -0.15 \\ [0.5ex] \hline  \hline
\end{tabular}
}
\caption{  $\chi^2$ fit of the observables in the charged fermion sector. This best fit correspond to $\chi^2=1.2$ for 13 observables. For charged leptons, a relative uncertainty of $0.1\%$ is assumed to take into account the  uncertainties, for example  threshold corrections at the PS scale. }\label{table:1}
\end{table}

As noted before, for this case we have 14 parameters: 11 magnitudes and 3 phases.  We perform the $\chi^{2}$ function minimization and the best minimum corresponds to total $\chi^{2} = 1.2$ is obtained for 13 observables which is a good fit \footnote{Note that the total $\chi^2 \neq 0$ even though the number of parameters is 1 more than the number of observables, it is because among the 14 parameters 3 of them are phases that can only be varied between 0 to $2\pi$. So if the theory were CP-conserving, there exits only 11 free parameters to fit 12 observables, 9 charged fermion masses and the three CKM mixing angles, hence a very constrained system.}.  The  result corresponding to the best fit is  shown in Table \ref{table:1}. The values of the  parameters corresponding to the best fit are:
\vspace*{0pt} 
\begin{align}
&\theta_1 = 7.83759\cdot10^{-4},\; \theta_2 =-3.131385,\; r_{1}= 1.29347\cdot10^{-2},\; r_{2}= -9.13047\cdot10^{-3}, \\
&M_1= \left(
\begin{array}{ccc}
 0.2988234 & 0. & 0. \\
 0. & 5.066234 & 0. \\
 0. & 0. & 94.801891
\end{array}
\right) \rm{GeV}, \\
&M_{15}= \left(
\begin{array}{ccc}
 -0.212786 & 0.367673 & -2.85309 \\
 0.367673 & -3.53464 & -11.8404-0.699369 i \\
 -2.85309 & -11.8404+0.699369 i & -15.8963 \\
\end{array}
\right) \rm{GeV} \label{Y15}.
\end{align}

\subsection{Parameter space for successful Leptogenesis}\label{BviaLPS}
Using the seesaw formula Eq. \eqref{Mnu}, one can in principle fit all the neutrino observables since the matrix $M_{R}$ which is in general a complex symmetric matrix contains 6 complex parameters. Instead, we will follow an alternative procedure. The right-handed neutrino mass matrix is given by inverting the seesaw formula Eq. \eqref{MR}. After the fitting of the fermion masses and mixings has been done, the Dirac neutrino mass matrix gets fixed unambiguously. For our fit, this Dirac neutrino mass matrix is 

\vspace*{-10pt}
\begin{equation}\label{MD}
M_{D} =\left(
\begin{array}{ccc}
 0.937182 +0.00050032 i & -1.10302-0.000864501 i &
   8.55928 +0.00670841 i \\
 -1.10302-0.000864501 i & 15.6702 +0.00831092 i &
   35.5195 +2.12595 i \\
 8.55928 +0.00670841 i & 35.5228 -2.07027 i &
   142.491 +0.0373765 i
\end{array}
\right) \rm{GeV}.
\end{equation}

\noindent
Then for observed known values of $\Delta m^{2}_{sol,atm}$ and  $\sin^{2}\theta^{\rm{PMNS}}_{ij}$  we are left with 4 unknown parameters $m_{1}$, $\alpha$, $\beta$ and $\delta$ so one can express the right-handed Majorana mass matrix as a function of these four free parameters, $M_R=M_R(m_1,\alpha,\beta,\delta)$, this is why the baryon asymmetric parameter in leptogenesis mechanism is also become a function of these parameters only: $\eta_B=\eta_B(m_1, \alpha, \beta, \delta)$. We search for the parameter space $\{ m_1, \alpha, \beta, \delta \}$ that corresponds to successful leptogenesis. While  hunting for the parameter space, the algorithm we follow is: we vary the experimentally measured quantities ($\Delta m^{2}_{sol,atm},  \sin^{2}\theta^{\rm{PMNS}}_{ij}$) in the neutrino sector within the 2$\sigma$ allowed range. In Eq.  \eqref{U} the Dirac phase $\delta$ is varied in  the range [0, 2$\pi$] whereas the Majorana phases $\alpha, \beta$  are varied within [0, $\pi$], these are the physical ranges for these phases (for details see Ref. \cite{deGouvea:2008nm}).  Baryon asymmetric parameter is  computed in a basis where both the charged lepton and the right-handed neutrino mass matrices are real and diagonal. We diagonalize these mass matrices as,

\vspace*{-10pt}
\begin{equation}
M_{e}= U_{e_{L}} \Lambda_{e} U^{\dagger}_{e_{R}} \;,\;\; M_{R}= U_{\nu_{R}} \Lambda_{R} U^{T}_{\nu_{R}},
\end{equation}
\noindent
with $\Lambda_{e} = \rm{diag}(m_{e},m_{\mu},m_{\tau})$ and $\Lambda_{R} = \rm{diag}(M_{1},M_{2},M_{3})$. In this basis, the Dirac neutrino mass matrix is given by $U^{\dagger}_{e_{L}} M_{D} U^{T}_{\nu_{R}}$ where 
\vspace*{-0pt}
\begin{equation}
U_{e_{L}} =\left(
\begin{array}{ccc}
 0.964706 & -0.259692+0.00589944 i & 0.0432075
   +0.00025877 i \\
 0.246127 +0.00525722 i & 0.947897 & 0.201767
   +0.0132524 i \\
 -0.0934313+0.00250479 i & -0.184011+0.0125101 i &
   0.97839
\end{array}
\right),
\end{equation}

\noindent
which is fixed from the fit parameters in the charged fermions and $U_{\nu_{R}}$ can be computed as a function of the free parameters $m_1, \alpha, \beta, \delta$. The inputs in the neutrino sector are taken from \cite{Fogli:2012ua} and shown in Table \ref{table:2}.

\begin{table}[th!]
\centering
\begin{tabular}{|c|c|c|}
\hline \hline
Quantity & $1\sigma$ range & $2\sigma$ range \\ \hline 
$\Delta m^{2}_{sol}/10^{-5} eV^{2}$ &7.32-7.80 &7.15-8.00  \\ \hline
$\Delta m^{2}_{atm}/10^{-3} eV^{2}$ &2.33-2.49 &2.27-2.55 \\ \hline
$\sin^{2}\theta^{\rm{PMNS}}_{12}/10^{-1}$  &2.91-3.25 &2.75-3.42 \\ \hline
$\sin^{2}\theta^{\rm{PMNS}}_{23}/10^{-1}$ &3.65-4.10 &3.48-4.48 \\ \hline 
$\sin^{2}\theta^{\rm{PMNS}}_{13}/10^{-2}$ &2.16-2.66 &1.93-2.90   \\ [1ex] \hline \hline
\end{tabular}
\caption{  Observables in the neutrino sector taken from \cite{Fogli:2012ua}. }\label{table:2}
\end{table}
 
While scanning over the parameter space, if $10^{9} \rm{GeV} \lesssim  M_{1} \lesssim 10^{12}$ GeV, we compute the baryon asymmetric parameter by taking into account the flavor effects and for the regime $M_{1} \gtrsim 10^{12}$ GeV, calculating $\eta_B$ involving the case where flavors are indistinguishable. We remind  the readers that for this high scale leptogenesis study, we have fixed the PS breaking scale to be $10^{14}$ GeV. Since the Majorana mass for the right-handed neutrinos are given by $v_R\;Y_R$, for perturbitivity reason, we put a cut-off of $M_{3} \lesssim 2\cdot10^{14}$ GeV. For both the scenarios, unflavored or flavored, we use the formula for the  strong wash-out regime when the wash-out parameter $>1$ ($K$ and $K_{\alpha \alpha}$) and the formula for weak wash-out regime when it is $< 1$ (instead of $\gg 1$ and $\ll 1$ respectively). It is to be mentioned that our investigation shows that the parameter space only permits solutions in the strong wash-out regime, so all the results presented below are solutions in the strong wash-out regime.

\begin{figure}
\begin{subfigure}{.3\textwidth}
  \centering
    \caption*{$m_1=1$ meV (Flavored)}
  \includegraphics[scale=0.3]{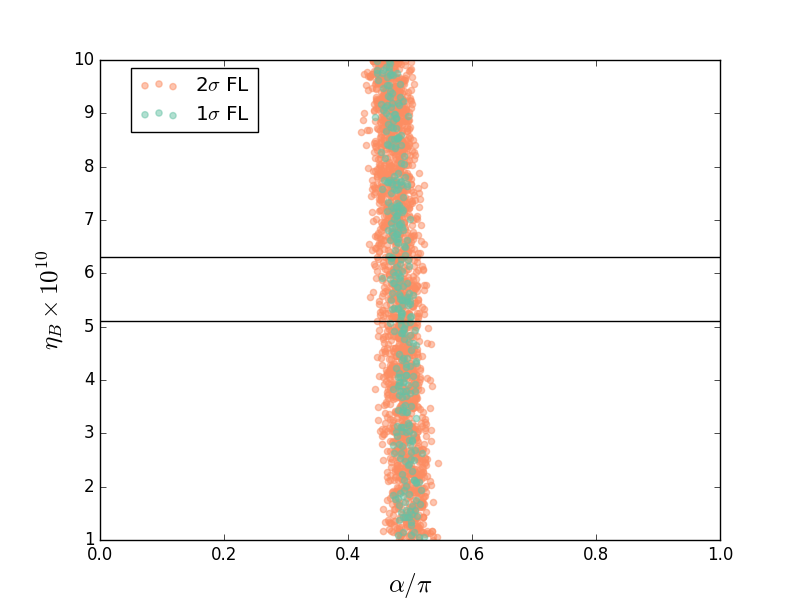}
\end{subfigure}
 \hspace{20pt}
\begin{subfigure}{.3\textwidth}
  \centering
  \caption*{$m_1=2$ meV (Flavored)}
  \includegraphics[scale=0.3]{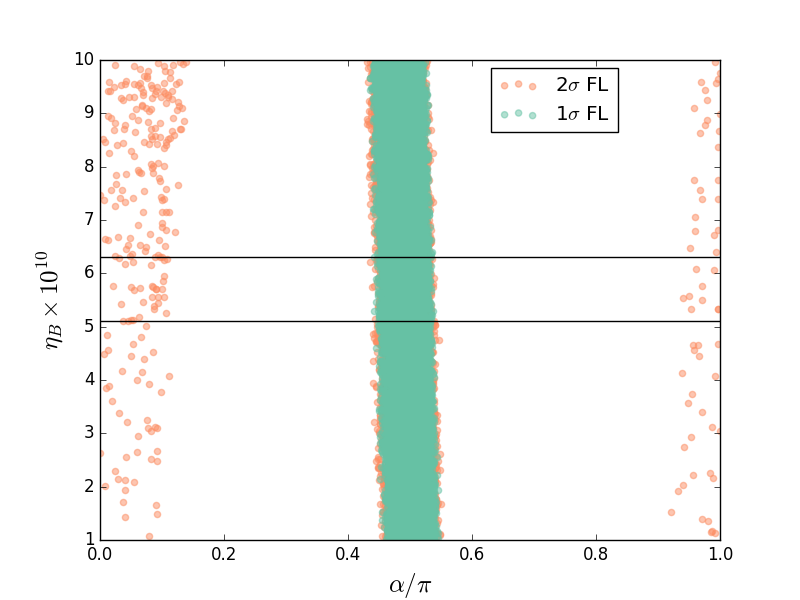}
\end{subfigure}
 \hspace{20pt}
\begin{subfigure}{.3\textwidth}
  \centering
   \caption*{$m_1=2$ meV (Unflavored)}
  \includegraphics[scale=0.3]{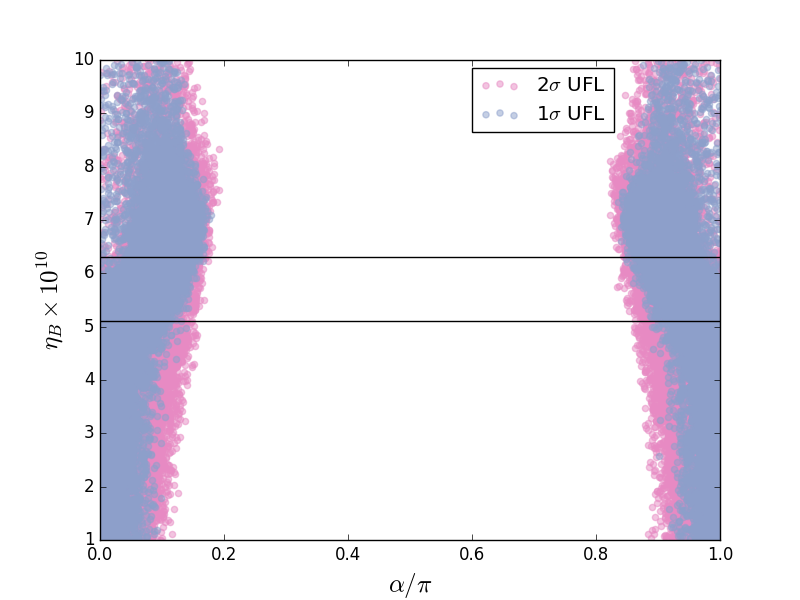}
\end{subfigure} 
\begin{subfigure}{.3\textwidth}
  \centering
  \includegraphics[scale=0.3]{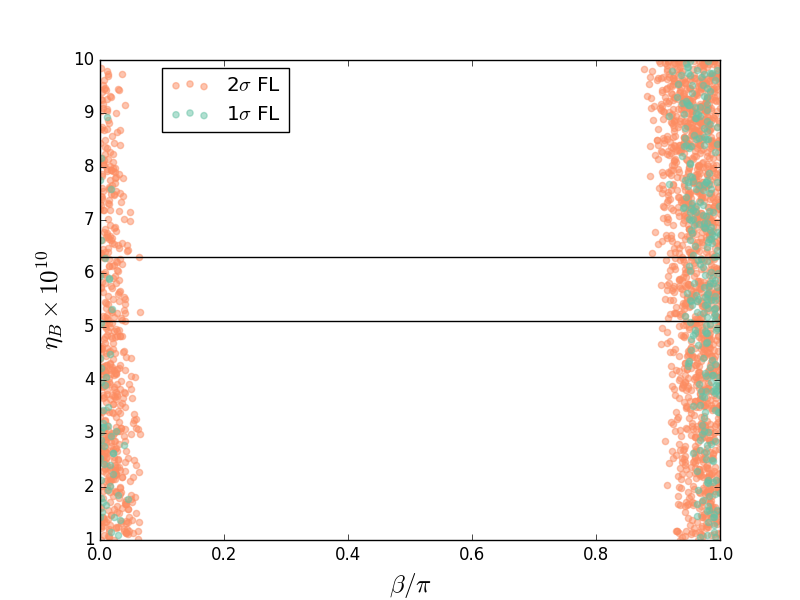}
\end{subfigure}
 \hspace{20pt}
\begin{subfigure}{.3\textwidth}
  \centering
  \includegraphics[scale=0.3]{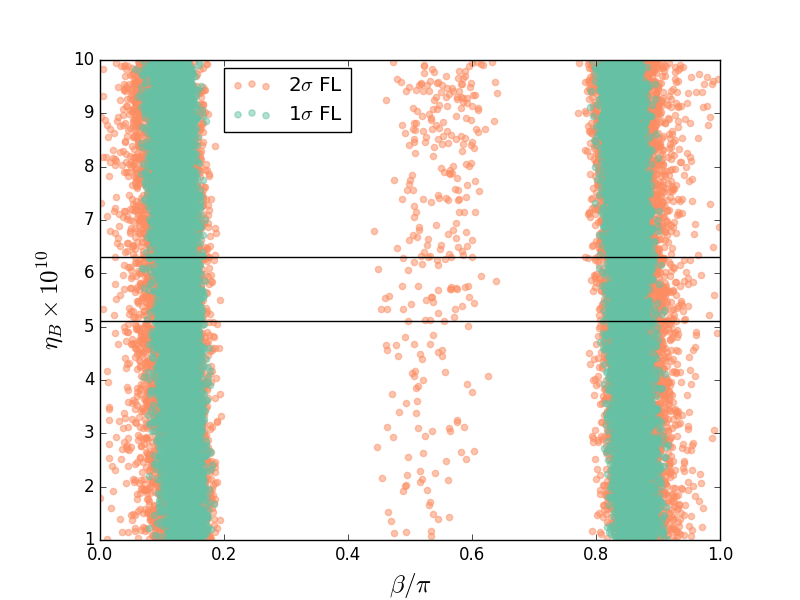}
\end{subfigure}
 \hspace{20pt}
\begin{subfigure}{.3\textwidth}
  \centering
  \includegraphics[scale=0.3]{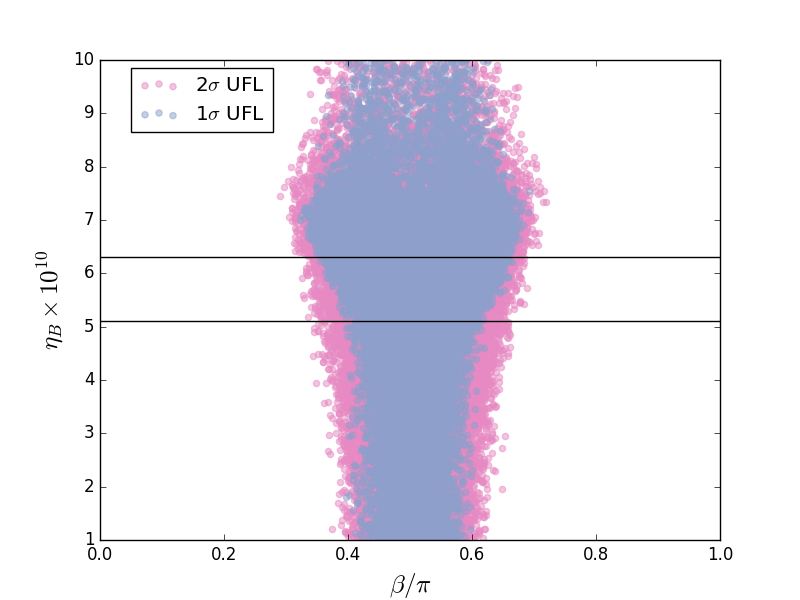}
\end{subfigure} 
\begin{subfigure}{.3\textwidth}
  \centering
  \includegraphics[scale=0.3]{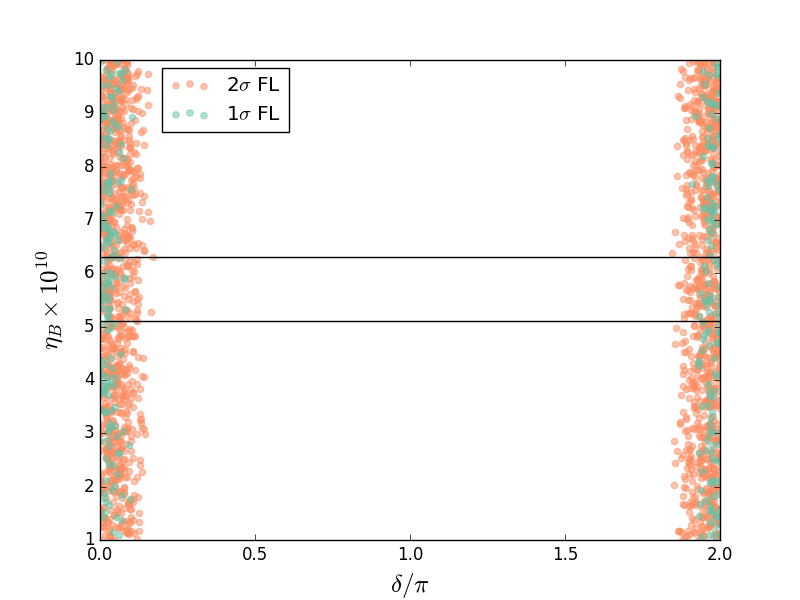}
\end{subfigure}
 \hspace{20pt}
\begin{subfigure}{.3\textwidth}
  \centering
  \includegraphics[scale=0.3]{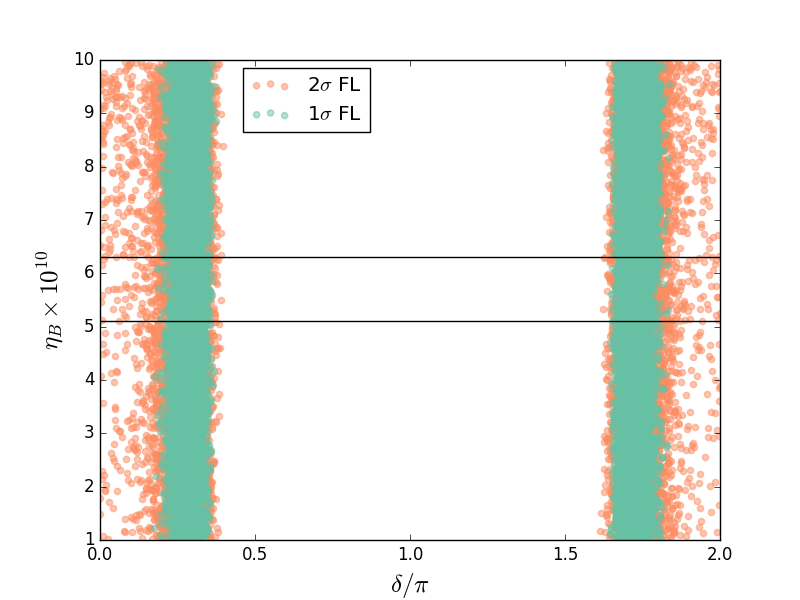}
\end{subfigure}
 \hspace{20pt}
\begin{subfigure}{.3\textwidth}
  \centering
  \includegraphics[scale=0.3]{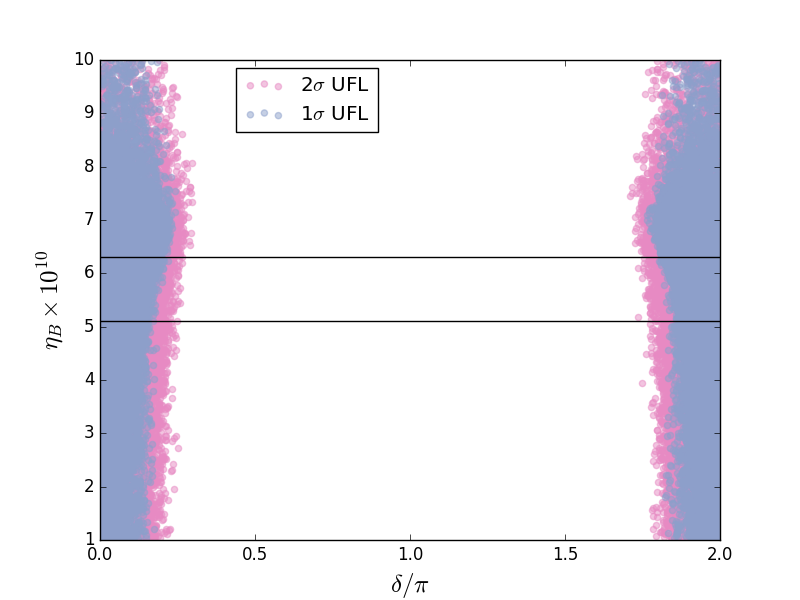}
\end{subfigure} 
\caption{As mentioned in the text, the baryon asymmetric parameter is a function of the four unknown quantities,  $\eta_B=\eta_B(m_1, \alpha, \beta, \delta)$. Allowed parameter space for these unknown quantities $\alpha, \beta, \delta$  permitted by the successful generation of baryon asymmetric parameter $\eta_B$ are presented here for   two different values of  $m_{1}=1, 2$ meV . While searching for the parameter space, the other quantities in the neutrino sector, $\Delta m^{2}_{sol,atm},  \sin^{2}\theta^{\rm{PMNS}}_{ij}$ that have been measured experimentally, are varied within their 2$\sigma$ experimental allowed range. The horizontal black  lines represent the experimental 1$\sigma$ range of $\eta_B$.   The green and orange set correspond to leptogenesis scenario where flavor effects are important, whereas, the blue and pink set is the flavor blind solutions. For these two different scenarios, green and blue represent solutions where  $\Delta m^{2}_{sol,atm},  \sin^{2}\theta^{\rm{PMNS}}_{ij}$ are varied within experimental 1$\sigma$ range and orange and pink within 2$\sigma$ range.}
\label{fig:001}
\end{figure}

We now discuss the results of leptogenesis in our framework. In  Fig. \ref{fig:001}, $\eta_B$ is plotted against $\alpha$, $\beta$ and $\delta$ phases respectively for the two different values of $m_1=1,2$ meV. While keeping $m_1$ fixed,  the other three parameters are varied over the whole range as mentioned before. Similar plots for another two fixed values of  $m_1=0.8$ and 4 meV are presented in  Fig. \ref{fig:84} in  Appendix \ref{A}.  From these plots, it is clear that whether or not flavor effects are involved,  depending on that, the allowed region in the parameter space is pretty  much different. The general behaviour is as follows, for larger values of $m_{1}$, the parameter space gets more populated for both the flavored and unflavored cases. The reason for this is, for larger values of  $m_1$ the heaviest right-handed neutrino mass $M_3$ becomes smaller. Note that, for this high scale leptogenesis study we kept the $v_R$ scale to be fixed at $10^{14}$ GeV.  For perturbatively of  the right-handed Yukawa couplings in the Majorana mass matrix $M_R= v_R\;Y_R$,  we restricted ourselves to the case of $M_3 \leq 2\times 10^{14}$ GeV. To reproduce the SM light neutrino mass in type-I seesaw scenario, $M_3$ tends to have values $\gtrsim 10^{14}$ GeV. This is why, larger the $m_1$, $M_3$ lies in the lower values and hence, valid solutions  in our frameworks are mostly realized in this region of the parameter space and we demonstrated this behaviour in  Fig. \ref{fig:84} in  Appendix \ref{B}, where correspondence between baryon asymmetry $\eta_B$ and right-handed mass spectrum is presented.

 From these plots, we find that successful leptogenesis cannot be realized in this framework for $m_1 < 0.8$ meV.  Comparing the flavored and unflavored solutions,  for smaller values of $m_1$, the parameter space is mostly preferred by flavored leptogenesis scenario. For example, setting $m_1=0.8$ meV, even though no solution can be found when all the neutrino observables are within their $1\sigma$ range,  a very small portion of the parameter space still permits baryon asymmetry in the right range provided that not all the varied quantities are  restricted within $1\sigma$ range.  If $m_1$ is set to a higher value,  for example $m_1=1$ meV, again only solutions exits for flavored leptogenesis scenario but in this case solutions are permitted even if all the varied quantities of the neutrino observables are within $1\sigma$ range. For even higher values of the lightest left-handed neutrino mass, parameter space allows solutions for both   flavored and unflavored  leptogenesis scenarios. We demonstrate such case by setting $m_1=2$ and 4 meV. Our investigation shows that, when $m_1$ is set to higher and higher values, the parameter space gets even more and more crowded. It is interesting to note that the  regions in the parameter space corresponding to these two  different scenarios of leptogenesis  are distinct and higher the value of $m_1$, more the overlapping is realized in the parameter space.  The relation of the baryon asymmetry with the CP-violating phases $\alpha, \beta, \delta$ are also due to the same reason. Since $M_i$ are expressed as a function of the set $\{m_1, \alpha, \beta, \delta\}$, for all values of such a parameter set, the condition $M_3 \leq 2\times 10^{14}$  GeV is not satisfied. The specific regions of the parameter space  that satisfy the demanded perturbatively condition returns solutions as demonstrated in Figs. \ref{fig:001} and \ref{fig:84}.

 In Appendices \ref{C} and \ref{D}, we present additional plots Figs. \ref{fig:002} and \ref{fig:003}  to show the correlation between  some of the physical quantities to the baryon asymmetric parameter for these two cases  with $m_1=1$ and 2 meV.  In Fig. \ref{fig:002},  the permitted region  for $m_{\beta}$ and $m_{\beta \beta}$ to have successful leptogenesis is shown, where $m_{\beta}=\sum_{i} |U_{\nu \;\;e i}|^{2} m_{i}$ is the effective mass parameter for the beta-decay and $m_{\beta \beta}= | \sum_{i} U_{\nu \;\;e i}^{2} m_{i} |$ is the effective mass parameter for neutrinoless double beta decay. The correlations between the Dirac phase $\delta$ and the  angle $\theta_{13}$ is presented in Fig. \ref{fig:003}. All the plots presented here are the result of  $10^8$ iterations.

\begin{figure}[t!]
\centering
\includegraphics[scale=0.8]{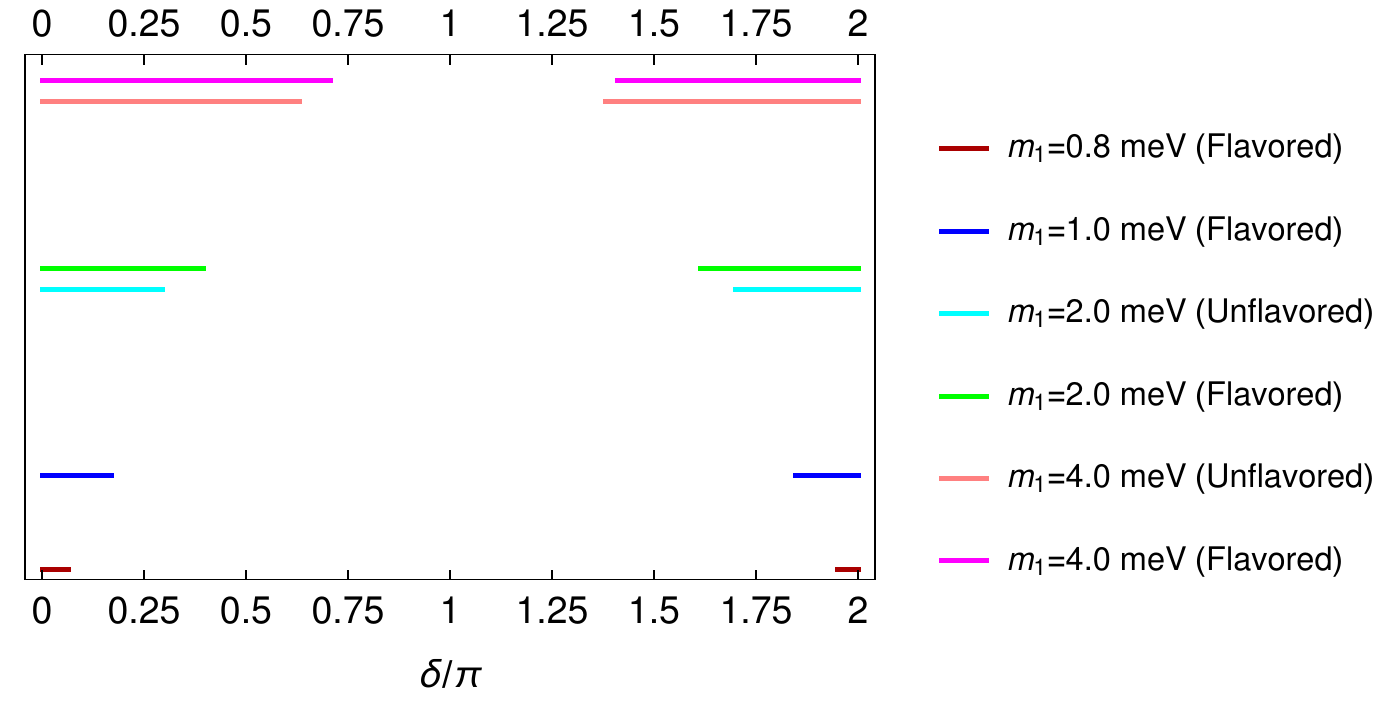}
\caption{Allowed range of the Dirac type CP violating phase $\delta$ for successful leptogenesis for different values of $m_1$.  }\label{DEL}
\end{figure}

\begin{table}[th!]
\centering
\resizebox{0.7\textwidth}{!}{
\begin{tabular}{|c|c|c|c|}
\hline  \hline
parameters & \multicolumn{2}{c|}{$10^{9} \rm{GeV} \lesssim  M_{1} \lesssim 10^{12}$ GeV} & \multicolumn{1}{c|}{$M_{1} \gtrsim 10^{12}$ GeV} \\\cline{2-4}  & $m_1=1$ meV & $m_1=2$ meV  & $m_1=2$ meV  \\ [1ex] \hline
$\alpha$ &1.52000&1.58856&0.17877 \\ \hline
$\beta$ &3.05225&0.41436&1.89040 \\ \hline
$\delta$ &-0.03128&0.96204&0.45498 \\ \hline
$\Delta m^{2}_{sol}/10^{-5} eV^{2}$ &7.60680&7.62805&7.54618 \\ \hline
$\Delta m^{2}_{atm}/10^{-3} eV^{2}$ &2.37437&2.33256&2.42017 \\ \hline
$\sin^{2}\theta^{\rm{PMNS}}_{12}$ &0.29188&0.29219&0.30002 \\ \hline
$\sin^{2}\theta^{\rm{PMNS}}_{23}$ &0.36578&0.39725&0.37940 \\ \hline
$\sin^{2}\theta^{\rm{PMNS}}_{13}$ &0.02581&0.02213& 0.02478 \\ \hline
$\eta_{B}/10^{-10}$  &5.65&5.74&6.29 \\ [0.5ex] \hline  \hline
\end{tabular}}
\caption{  Benchmark points for computing  baryon asymmetric parameter is presented. $\eta_{B}$ is computed by taking into account the flavor effects if $10^{9} \rm{GeV} \lesssim  M_{1} \lesssim 10^{12}$ GeV or in the flavor indistinguishable regime if  $M_{1} \gtrsim 10^{12}$ GeV. Two different values of the lightest left-handed neutrino masses are considered, $m_{1}=1$ and 2 meV, where for the second case, solutions exists for both flavored and unflavored scenarios.   }\label{BM}
\end{table}

In the neutrino sector, among the four different experimentally unmeasured quantities, particularly the Dirac type phase  $\delta$ is the most important one, since it has the potential to be measured in the upcoming neutrino experiments. In Fig. \ref{DEL},  the allowed range for this CP violating phase to have successful leptogenesis is presented for different values of the lightest  neutrino mass $m_1$.  Benchmark points corresponding to few different cases are  presented in Table \ref{BM}.

\section{The Higgs potential and scalar mass spectrum}\label{chapter05}

\subsection{The Higgs potential}\label{section:potential}

In this sub-section we construct the complete scalar potential with $G_{224} \times U(1)_{PQ}$ symmetry. As mentioned earlier, the field $\bm\Delta_{L}$ which is present if the  group is $G_{224P}$ but need not be present if the gauge group is $G_{224}$ instead. But for generality, we construct the scalar potential containing $(2,2,1), (2,2,15), (1,3,10)$ and $(3,1,10)$ fields that respects $G_{224} \times U(1)_{PQ}$ symmetry and then discuss the additional constraints introduced by imposing the parity symmetry. For $G_{224}$ with the absence of $(3,1,10)$ one can set $\bm\Delta_L =0$ to obtain the relevant terms in the potential. The most general Higgs potential respecting $G_{224} \times U(1)_{PQ}$ symmetry with the scalars given in Eq. \eqref{Higgs} is: 
\vspace*{-5pt}
\begin{equation}\label{V}
V = V_{\bm\Phi} + V_{\bm\Sigma} + V_{\bm\Delta} + V_{\bm\Phi \bm\Sigma} + V_{\bm\Phi \bm\Delta} + V_{\bm\Sigma \bm\Delta} + V_{\bm\Phi \bm\Sigma \bm\Delta} + V_{\mathcal{S}} ,
\end{equation}

\noindent  with,
\vspace{-5pt} 
\begin{align}
V_{\bm\Phi} &= -\mu^{2}_{\Phi}\; \bm\Phi_{\alpha}^{ \dot{\alpha}} \bm\Phi^{\ast \alpha}_{\dot{\alpha}} + \lambda_{1 \Phi}\; \bm\Phi_{\alpha}^{ \dot{\alpha}} \bm\Phi^{\ast \alpha}_{ \dot{\alpha}} \bm\Phi_{\beta}^{ \dot{\beta}} \bm\Phi^{\ast \beta}_{ \dot{\beta}} + \lambda_{2 \Phi}\; \bm\Phi_{\alpha}^{ \dot{\alpha}} \bm\Phi^{\ast \alpha}_{ \dot{\beta}} \bm\Phi_{\beta}^{ \dot{\beta}} \bm\Phi^{\ast \beta}_{ \dot{\alpha}},
\\ 
V_{\bm\Sigma} &= -\mu^{2}_{\Sigma}\; \bm\Sigma^{\nu\;\dot{\alpha}}_{\mu \;\alpha} \bm\Sigma^{\ast \;\mu\;\alpha }_{\nu\;\dot{\alpha}} 
+
   \lambda_{1 \Sigma}\; \bm\Sigma^{\nu\;\dot{\alpha}}_{\mu \;\alpha } \bm\Sigma^{\ast \;\mu\;\alpha }_{\nu\;\dot{\alpha}} \bm\Sigma^{\tau\;\dot{\beta}}_{\rho \;\beta } \bm\Sigma^{\ast \;\rho\;\beta }_{\tau\;\dot{\beta}} 
+
  \lambda_{2 \Sigma}\; \bm\Sigma^{\nu\;\dot{\alpha}}_{\mu \;\alpha } \bm\Sigma^{\ast \;\rho\;\alpha }_{\tau\;\dot{\alpha}} \bm\Sigma^{\mu\;\dot{\beta}}_{\nu \;\beta } \bm\Sigma^{\ast \;\tau\;\beta }_{\rho\;\dot{\beta}}
\nonumber \\&
+
  \lambda_{3 \Sigma}\; \bm\Sigma^{\nu\;\dot{\alpha}}_{\mu \;\alpha } \bm\Sigma^{\ast \;\tau\;\alpha }_{\rho\;\dot{\alpha}} \bm\Sigma^{\rho\;\dot{\beta}}_{\tau \;\beta } \bm\Sigma^{\ast \;\mu\;\beta }_{\nu\;\dot{\beta}} 
+
\lambda_{4 \Sigma}\; \bm\Sigma^{\nu\;\dot{\alpha}}_{\mu \;\alpha } \bm\Sigma^{\ast \;\rho\;\alpha }_{\nu\;\dot{\alpha}} \bm\Sigma^{\tau\;\dot{\beta}}_{\rho \;\beta } \bm\Sigma^{\ast \;\mu\;\beta }_{\tau\;\dot{\beta}} 
+
 \lambda_{5 \Sigma}\; \bm\Sigma^{\nu\;\dot{\alpha}}_{\mu \;\alpha } \bm\Sigma^{\ast \;\mu\;\alpha }_{\tau\;\dot{\alpha}} \bm\Sigma^{\rho\;\dot{\beta}}_{\nu \;\beta } \bm\Sigma^{\ast \;\tau\;\beta }_{\rho\;\dot{\beta}}
\nonumber \\ &
+
 \lambda_{6 \Sigma}\; \bm\Sigma^{\nu\;\dot{\alpha}}_{\mu \;\alpha } \bm\Sigma^{\ast \;\tau\;\alpha }_{\rho\;\dot{\alpha}} \bm\Sigma^{\mu\;\dot{\beta}}_{\tau \;\beta } \bm\Sigma^{\ast \;\rho\;\beta }_{\nu\;\dot{\beta}}
+
\lambda_{7 \Sigma}\; \bm\Sigma^{\nu\;\dot{\alpha}}_{\mu \;\alpha } \bm\Sigma^{\ast \;\mu\;\alpha }_{\nu\;\dot{\beta}} \bm\Sigma^{\tau\;\dot{\beta}}_{\rho \;\beta } \bm\Sigma^{\ast \;\rho\;\beta }_{\tau\;\dot{\alpha}} 
+  \lambda_{8 \Sigma}\; \bm\Sigma^{\nu\;\dot{\alpha}}_{\mu \;\alpha } \bm\Sigma^{\ast \;\rho\;\alpha }_{\tau\;\dot{\beta}} \bm\Sigma^{\mu\;\dot{\beta}}_{\nu \;\beta } \bm\Sigma^{\ast \;\tau\;\beta }_{\rho\;\dot{\alpha}}
 \nonumber  \\ &
 +
\lambda_{9 \Sigma}\; \bm\Sigma^{\nu\;\dot{\alpha}}_{\mu \;\alpha } \bm\Sigma^{\ast \;\rho\;\alpha }_{\nu\;\dot{\beta}} \bm\Sigma^{\tau\;\dot{\beta}}_{\rho \;\beta } \bm\Sigma^{\ast \;\mu\;\beta }_{\tau\;\dot{\alpha}} 
+
 \lambda_{10 \Sigma}\; \bm\Sigma^{\nu\;\dot{\alpha}}_{\mu \;\alpha } \bm\Sigma^{\ast \;\mu\;\alpha }_{\tau\;\dot{\beta}} \bm\Sigma^{\rho\;\dot{\beta}}_{\nu \;\beta } \bm\Sigma^{\ast \;\tau\;\beta }_{\rho\;\dot{\alpha}}
 \nonumber  \\ & 
 +
\lambda_{11 \Sigma}\; \bm\Sigma^{\nu\;\dot{\alpha}}_{\mu \;\alpha } \bm\Sigma^{\mu\;\dot{\gamma}}_{\nu \;\gamma } \epsilon^{\;\alpha \gamma} \epsilon_{\;\dot{\alpha} \dot{\gamma}} \bm\Sigma^{\ast \;\tau\;}_{\rho\;\beta \dot{\beta}} \bm\Sigma^{\ast \;\rho\;\kappa }_{\tau\;\dot{\kappa}}  \epsilon_{\;\beta \kappa} \epsilon^{\;\dot{\beta} \dot{\kappa}} 
+
 \lambda_{12 \Sigma}\; \bm\Sigma^{\nu\;\dot{\alpha}}_{\mu \;\alpha } \bm\Sigma^{\rho\;\dot{\gamma}}_{\tau\;\gamma }  \epsilon^{\;\alpha \gamma} \epsilon_{\;\dot{\alpha} \dot{\gamma}} \bm\Sigma^{\ast \;\mu\;\beta }_{\nu\;\dot{\beta}} \bm\Sigma^{\ast \;\tau \;\kappa }_{\rho\;\dot{\kappa}} \epsilon_{\;\beta \kappa} \epsilon^{\;\dot{\beta} \dot{\kappa}} 
\nonumber  \\&
+
\lambda_{13 \Sigma}\; \bm\Sigma^{\nu\;\dot{\alpha}}_{\mu \;\alpha } \bm\Sigma^{\rho\;\dot{\gamma}}_{\nu\;\gamma } \epsilon^{\;\alpha \gamma} \epsilon_{\;\dot{\alpha} \dot{\gamma}} \bm\Sigma^{\ast \;\tau\;\beta }_{\rho\;\dot{\beta}} \bm\Sigma^{\ast \;\mu \;\kappa }_{\tau\;\dot{\kappa}} \epsilon_{\;\beta \kappa} \epsilon^{\;\dot{\beta} \dot{\kappa}} 
+
 \lambda_{14 \Sigma}\; \bm\Sigma^{\nu\;\dot{\alpha}}_{\mu \;\alpha } \bm\Sigma^{\tau\;\dot{\gamma}}_{\rho\;\gamma } \epsilon^{\;\alpha \gamma} \epsilon_{\;\dot{\alpha} \dot{\gamma}} \bm\Sigma^{\ast \;\mu\;\beta }_{\tau\;\dot{\beta} } \bm\Sigma^{\ast \;\rho \;\kappa }_{\nu\;\dot{\kappa}} \epsilon_{\;\beta \kappa} \epsilon^{\;\dot{\beta} \dot{\kappa}},
\\ 
V_{\bm\Delta} &= \{-\mu^{2}_{\Delta_R}\; \bm\Delta_{R\mu \nu \; \dot{\alpha}}^{\;\;\;\;\;\;\;\; \dot{\beta}} \bm\Delta_{R\; \dot{\beta}}^{\ast\mu \nu \; \dot{\alpha}} 
+ 
\lambda_{1R}\; \bm\Delta_{R\mu \nu \; \dot{\alpha}}^{\;\;\;\;\;\;\;\; \dot{\beta}} \bm\Delta_{R\; \dot{\beta}}^{\ast\mu \nu \; \dot{\alpha}} \bm\Delta_{R\rho \tau \; \dot{\gamma}}^{\;\;\;\;\;\;\;\; \dot{\kappa}} \bm\Delta_{R\; \dot{\kappa}}^{\ast\rho \tau \; \dot{\gamma}}
+
 \lambda_{2R}\; \bm\Delta_{R\mu \nu \; \dot{\alpha}}^{\;\;\;\;\;\;\;\; \dot{\beta}} \bm\Delta_{R\; \dot{\kappa}}^{\ast\mu \nu \; \dot{\gamma}} \bm\Delta_{R\rho \tau \; \dot{\beta}}^{\;\;\;\;\;\;\;\; \dot{\alpha}} \bm\Delta_{R\; \dot{\gamma}}^{\ast\rho \tau \; \dot{\kappa}}
\nonumber \\ &+ \lambda_{3R}\; \bm\Delta_{R\mu \nu \; \dot{\alpha}}^{\;\;\;\;\;\;\;\; \dot{\beta}} \bm\Delta_{R\; \dot{\gamma}}^{\ast\mu \nu \; \dot{\kappa}} \bm\Delta_{R\rho \tau \; \dot{\kappa}}^{\;\;\;\;\;\;\;\; \dot{\gamma}} \bm\Delta_{R\; \dot{\beta}}^{\ast\rho \tau \; \dot{\alpha}} +
\lambda_{4R}\; \bm\Delta_{R\mu \nu \; \dot{\alpha}}^{\;\;\;\;\;\;\;\; \dot{\beta}} \bm\Delta_{R\; \dot{\beta}}^{\ast\nu \rho \; \dot{\alpha}} \bm\Delta_{R\rho \tau \; \dot{\gamma}}^{\;\;\;\;\;\;\;\; \dot{\kappa}} \bm\Delta_{R\; \dot{\kappa}}^{\ast\tau \mu \; \dot{\gamma}}
\nonumber \\ &+ \lambda_{5R}\; \bm\Delta_{R\mu \nu \; \dot{\alpha}}^{\;\;\;\;\;\;\;\; \dot{\beta}} \bm\Delta_{R\; \dot{\kappa}}^{\ast\nu \rho \; \dot{\gamma}} \bm\Delta_{R\rho \tau \; \dot{\beta}}^{\;\;\;\;\;\;\;\; \dot{\alpha}} \bm\Delta_{R\; \dot{\gamma}}^{\ast\tau \mu \; \dot{\kappa}}+ \;R\leftrightarrow L \}  
+
  \lambda_{6}\; \bm\Delta_{R\mu \nu \; \dot{\alpha}}^{\;\;\;\;\;\;\;\; \dot{\beta}} \bm\Delta_{R\; \dot{\beta}}^{\ast\mu \nu \; \dot{\alpha}} \bm\Delta_{L\rho \tau \; \alpha}^{\;\;\;\;\;\;\;\; \beta} \bm\Delta_{L\; \beta}^{\ast\rho \tau \; \alpha} 
\nonumber \\&
+
 \lambda_{7}\; \bm\Delta_{R\mu \nu \; \dot{\alpha}}^{\;\;\;\;\;\;\;\; \dot{\beta}} \bm\Delta_{R\; \dot{\beta}}^{\ast\nu \rho \; \dot{\alpha}} \bm\Delta_{L\rho \tau \; \alpha}^{\;\;\;\;\;\;\;\; \beta} \bm\Delta_{L\;\beta}^{\ast\tau \mu \;\alpha}
 + \lambda_{8}\; \bm\Delta_{R\mu \nu \; \dot{\alpha}}^{\;\;\;\;\;\;\;\; \dot{\beta}} \bm\Delta_{R\; \dot{\beta}}^{\ast\rho \tau \; \dot{\alpha}} \bm\Delta_{L\rho \tau \; \alpha}^{\;\;\;\;\;\;\;\; \beta} \bm\Delta_{L\; \beta}^{\ast\mu \nu \; \alpha}
\nonumber \\&+  (\widetilde{\lambda_{9}}\; \bm\Delta_{R\mu \nu \;\dot{\alpha}}^{\;\;\;\;\;\;\;\dot{\beta}} \bm\Delta_{R\rho \tau \;\dot{\beta}}^{\;\;\;\;\;\;\;\dot{\alpha}} \bm\Delta_{L\lambda \chi \;\alpha}^{\;\;\;\;\;\;\;\beta} \bm\Delta_{L\zeta \omega \;\beta}^{\;\;\;\;\;\;\;\alpha} \epsilon^{\mu \rho \lambda \zeta} \epsilon^{\nu \tau \chi \omega} + \widetilde{\lambda_{9}}^{\ast}\; \bm\Delta_{R\dot{\alpha}}^{\ast\mu \nu \;\dot{\beta}} \bm\Delta_{R\dot{\beta}}^{\ast\rho \tau \;\dot{\alpha}} \bm\Delta_{L\alpha}^{\ast\lambda \chi \;\beta} \bm\Delta_{L\beta}^{\ast\zeta \omega \;\alpha} \epsilon_{\mu \rho \lambda \zeta} \epsilon_{\nu \tau \chi \omega}) , \label{Vdelta}
\\ 
V_{\bm\Phi \bm\Sigma} &= \alpha_{1}\; \bm\Phi_{\alpha}^{ \dot{\alpha}} \bm\Phi^{\ast \alpha}_{ \dot{\alpha}} \bm\Sigma_{\mu \;\beta }^{\nu\;\dot{\beta}} \bm\Sigma^{\ast \mu \;\beta }_{\nu\;\dot{\beta}}
+
  \alpha_{2}\; \bm\Phi_{\alpha}^{ \dot{\alpha}} \bm\Phi^{\ast \alpha}_{ \dot{\beta}} \bm\Sigma_{\mu \;\beta }^{\nu\;\dot{\beta}} {\bm\Sigma^{\ast}}_{\nu\;\dot{\alpha} }^{\mu \;\beta } 
+
 \alpha_{3}\; \bm\Phi_{\alpha}^{ \dot{\alpha}} \bm\Phi^{\ast \beta}_{\dot{\alpha}} \bm\Sigma_{\mu \;\beta }^{\nu\;\dot{\beta}} {\bm\Sigma^{\ast}}_{\nu\;\dot{\beta} }^{\mu \;\alpha }
+
\alpha_{4}\; \bm\Phi_{\alpha}^{ \dot{\alpha}} \bm\Phi^{\ast \beta}_{ \dot{\beta}} \bm\Sigma_{\mu \;\beta }^{\nu\;\dot{\beta}} {\bm\Sigma^{\ast}}_{\nu\;\dot{\alpha} }^{\mu \;\alpha } 
\nonumber \\ &
+  
( \widetilde{\alpha}_{5}\; \bm\Phi_{\alpha}^{ \dot{\alpha}} \bm\Phi_{\beta}^{ \dot{\beta}} {\bm\Sigma^{\ast}}_{\mu\;\dot{\alpha}}^{\nu \;\alpha } {\bm\Sigma^{\ast}}_{\nu\;\dot{\beta}}^{\mu \;\beta }
+
 {\widetilde{\alpha}}^{\ast}_{5}\; \bm\Phi^{\ast \alpha}_{ \dot{\alpha}} \bm\Phi^{\ast \beta}_{ \dot{\beta}} \bm\Sigma_{\mu \;\alpha }^{\nu\;\dot{\alpha}} \bm\Sigma_{\nu \; \beta }^{\mu\;\dot{\beta}} )
+
( \widetilde{\alpha}_{6}\; \bm\Phi_{\alpha}^{ \dot{\alpha}} \bm\Phi_{\beta}^{ \dot{\beta}} {\bm\Sigma^{\ast}}_{\mu\;\dot{\beta}}^{\nu \; \alpha } {\bm\Sigma^{\ast}}_{\nu\;\dot{\alpha}}^{\mu \; \beta }
+
 {\widetilde{\alpha}}^{\ast}_{6}\; \bm\Phi^{\ast \alpha}_{ \dot{\alpha}} \bm\Phi^{\ast \beta}_{ \dot{\beta}} \bm\Sigma_{\mu \;\alpha }^{ \nu\;\dot{\beta}} \bm\Sigma_{\nu \; \beta }^{\mu\;\dot{\alpha}} )   ,
\\  
V_{\bm\Phi \bm\Delta} &= \{ \beta_{1R}\; \bm\Phi_{\alpha}^{ \dot{\alpha}} \bm\Phi^{\ast \alpha}_{ \dot{\alpha}} \bm\Delta_{R\mu \nu \dot{\beta}}^{\;\;\;\;\;\; \dot{\gamma}} \bm\Delta_{R\;\dot{\gamma}}^{\ast\mu \nu \dot{\beta}} 
+
 \beta_{2R}\; \bm\Phi_{\alpha}^{ \dot{\alpha}} \bm\Phi^{\ast \alpha }_{\dot{\beta}} \bm\Delta_{R\mu \nu \dot{\alpha}}^{\;\;\;\;\;\; \dot{\gamma}} \bm\Delta_{R \;\dot{\gamma}}^{\ast\mu \nu \dot{\beta}}  + \;R\leftrightarrow L \}
\nonumber \\ &
+
  (\widetilde{\beta}_{3} \bm\Phi_{\alpha}^{ \dot{\alpha}} \bm\Phi_{\beta}^{ \dot{\beta}} \epsilon_{\alpha \kappa} \epsilon^{\dot{\alpha} \dot{\kappa}} \bm\Delta^{\ast\mu \nu \; \dot{\beta}}_{R\dot{\kappa}} \bm\Delta_{L\mu \nu \;\kappa}^{\;\;\;\;\;\;\;\; \beta}   
+
\widetilde{\beta}^{\ast}_{3} \bm\Phi^{\ast \alpha}_{ \dot{\alpha}} \bm\Phi^{\ast \beta}_{ \dot{\beta}} \epsilon^{\beta \kappa}  \epsilon_{\dot{\beta} \dot{\kappa}}\bm\Delta_{R\mu \nu \; \dot{\beta}}^{\;\;\;\;\;\;\;\; \dot{\kappa}} \bm\Delta_{L\kappa}^{\ast\mu \nu \; \beta}  ) ,
\\  
V_{\bm\Sigma \bm\Delta} &= \{
\gamma_{1R}\; \bm\Sigma_{\rho \;\alpha }^{\tau\;\dot{\alpha}} {\bm\Sigma^{\ast}}_{\tau\;\dot{\alpha} }^{\rho \;\alpha } \bm\Delta_{R\mu \nu \dot{\beta}}^{\;\;\;\;\;\; \dot{\gamma}} \bm\Delta_{R\dot{\gamma}}^{\ast\mu \nu \dot{\beta}}
+
 \gamma_{2R}\; \bm\Sigma_{\rho \;\alpha }^{\tau\;\dot{\alpha}} {\bm\Sigma^{\ast}}_{\tau\;\dot{\alpha} }^{\mu \;\alpha } \bm\Delta_{R\mu \nu \dot{\beta}}^{\;\;\;\;\;\; \dot{\gamma}} \bm\Delta_{R\dot{\gamma}}^{\ast\nu \rho \dot{\beta}}
+
\gamma_{3R}\; \bm\Sigma_{\rho \; \alpha }^{\tau\;\dot{\alpha}} {\bm\Sigma^{\ast}}_{\mu \;\dot{\alpha}}^{\rho \;\alpha } \bm\Delta_{R\tau \nu \dot{\beta}}^{\;\;\;\;\;\; \dot{\gamma}} \bm\Delta_{R\dot{\gamma}}^{\ast\nu \mu \dot{\beta}}
\nonumber \\ &
+
\gamma_{4R}\; \bm\Sigma_{\rho \;\alpha }^{\tau\;\dot{\alpha}} {\bm\Sigma^{\ast}}_{\mu\;\dot{\alpha} }^{\nu \; \alpha } \bm\Delta_{R\tau \nu \dot{\beta}}^{\;\;\;\;\;\; \dot{\gamma}} \bm\Delta_{R\dot{\gamma}}^{\ast\rho \mu \dot{\beta}}
+
\gamma_{5R}\; \bm\Sigma_{\rho \; \alpha }^{\tau\;\dot{\alpha}} {\bm\Sigma^{\ast}}_{\tau\;\dot{\beta} }^{\rho \;\alpha } \bm\Delta_{R\mu \nu \dot{\alpha}}^{\;\;\;\;\;\; \dot{\gamma}} \bm\Delta_{R\dot{\gamma}}^{\ast\mu \nu \dot{\beta}}
+
 \gamma_{6R}\; \bm\Sigma_{\rho \; \alpha }^{\tau\;\dot{\alpha}} {\bm\Sigma^{\ast}}_{\tau\;\dot{\beta} }^{\mu \; \alpha } \bm\Delta_{R\mu \nu \dot{\alpha}}^{\;\;\;\;\;\; \dot{\gamma}} \bm\Delta_{R\dot{\gamma}}^{\ast\nu \rho \dot{\beta}}
\nonumber \\ &
+
\gamma_{7R}\; \bm\Sigma_{\rho \;\alpha }^{\tau\;\dot{\alpha}} {\bm\Sigma^{\ast}}_{\mu\;\dot{\beta} }^{\rho \;\alpha } \bm\Delta_{R\tau \nu \dot{\alpha}}^{\;\;\;\;\;\; \dot{\gamma}} \bm\Delta_{R\dot{\gamma}}^{\ast\nu \mu \dot{\beta}}
+
\gamma_{8R}\; \bm\Sigma_{\rho \;\alpha }^{\tau\;\dot{\alpha}} {\bm\Sigma^{\ast}}_{\mu \;\dot{\beta}}^{\nu \;\alpha } \bm\Delta_{R\tau \nu \dot{\alpha}}^{\;\;\;\;\;\; \dot{\gamma}} \bm\Delta_{R\dot{\gamma}}^{\ast\rho \mu \dot{\beta}} 
+ \;R\leftrightarrow L \} 
\nonumber \\&
+
(\widetilde{\gamma}_{9R}\;  \bm\Sigma_{\mu \;\alpha }^{\nu\;\dot{\alpha}} \bm\Sigma_{\rho \;\beta }^{\tau\;\dot{\beta}} \epsilon^{\alpha \beta} \epsilon_{\dot{\alpha} \dot{\beta}}  \bm\Delta_{R\nu \lambda\;\dot{\kappa}}^{\;\;\;\;\;\;\;\; \dot{\gamma}} \bm\Delta_{R\tau \chi\;\dot{\gamma}}^{\;\;\;\;\;\;\;\; \dot{\kappa}} \epsilon^{\mu \rho \lambda \chi}+\widetilde{\gamma}^{\ast}_{9R}\;  {\bm\Sigma^{\ast}}_{\nu\;\dot{\alpha}}^{\mu \;\alpha } {\bm\Sigma^{\ast}}_{\tau\;\dot{\beta}}^{\rho \;\beta } \epsilon_{\alpha \beta} \epsilon^{\dot{\alpha} \dot{\beta}}  \bm\Delta^{\ast\nu \lambda \dot{\gamma}}_{R\; \dot{\kappa}} \bm\Delta_{R\; \dot{\gamma}}^{\ast\tau \chi\; \dot{\kappa}} \epsilon_{\mu \rho \lambda \chi} )
\nonumber \\ &
+
(\widetilde{\gamma}_{10R}\;  \bm\Sigma_{\mu \;\alpha }^{\nu\;\dot{\alpha}} \bm\Sigma_{\rho \; \beta }^{\tau\;\dot{\beta}}  \epsilon^{\alpha \beta} \epsilon_{\dot{\alpha} \dot{\kappa} } {\bm\Delta}_{R\nu \lambda\;\dot{\beta}}^{\;\;\;\;\;\;\;\; \dot{\gamma}} {\bm\Delta}_{R\tau \chi\;\dot{\gamma}}^{\;\;\;\;\;\;\;\;\; \dot{\kappa}} \epsilon^{\mu \rho \lambda \chi} + \widetilde{\gamma}^{\ast}_{10R}\; {\bm\Sigma^{\ast}}_{\nu\;\dot{\alpha}}^{\mu \;\alpha } {\bm\Sigma^{\ast}}^{\rho\;\beta}_{\tau\;  \dot{\beta}}   \epsilon_{\alpha \beta} \epsilon^{\dot{\alpha} \dot{\kappa}} \bm\Delta_{R\; \dot{\gamma}}^{\ast\nu \lambda\;\dot{\beta}} \bm\Delta_{R\; \dot{\kappa}}^{\ast\tau \chi\;\dot{\gamma}} \epsilon_{\mu \rho \lambda \chi})
\nonumber \\&
+
(\widetilde{\gamma}_{9L}\; 
  {\bm\Sigma^{\ast}}_{\nu\;\dot{\alpha}}^{\mu \;\alpha } {\bm\Sigma^{\ast}}_{\tau\;\dot{\beta}}^{\rho \;\beta } \epsilon_{\alpha \beta} \epsilon^{\dot{\alpha} \dot{\beta}}  \bm\Delta_{L\mu \lambda\; \kappa}^{\;\;\;\;\;\;\;\;\gamma} \bm\Delta_{L \rho \chi \gamma}^{\;\;\;\;\;\;\;\;\kappa} \epsilon^{\nu \tau \lambda \chi}
+
\widetilde{\gamma}^{\ast}_{9L}\;
 \bm\Sigma_{\mu \;\alpha }^{\nu\;\dot{\alpha}} \bm\Sigma_{\rho \;\beta }^{\tau\;\dot{\beta}} \epsilon^{\alpha \beta} \epsilon_{\dot{\alpha} \dot{\beta}}  \bm\Delta_{L\;\kappa}^{\ast \mu \lambda\;\gamma} \bm\Delta_{L\;\gamma}^{\ast \rho \chi\;\kappa} \epsilon_{\nu \tau \lambda \chi} )  
 \nonumber \\&
+
(\widetilde{\gamma}_{10L}\; 
  {\bm\Sigma^{\ast}}_{\nu\;\dot{\alpha}}^{\mu \;\alpha } {\bm\Sigma^{\ast}}_{\tau\;\dot{\beta}}^{\rho \;\beta } \epsilon_{\alpha \kappa} \epsilon^{\dot{\alpha} \dot{\beta}}  \bm\Delta_{L\mu \lambda\; \gamma}^{\;\;\;\;\;\;\;\;\kappa} \bm\Delta_{L \rho \chi \beta}^{\;\;\;\;\;\;\;\;\gamma} \epsilon^{\nu \tau \lambda \chi}
+
\widetilde{\gamma}^{\ast}_{10L}\;
 \bm\Sigma_{\mu \;\alpha }^{\nu\;\dot{\alpha}} \bm\Sigma_{\rho \;\beta }^{\tau\;\dot{\beta}} \epsilon^{\alpha \kappa} \epsilon_{\dot{\alpha} \dot{\beta}}  \bm\Delta_{L\;\kappa}^{\ast \mu \lambda\;\gamma} \bm\Delta_{L\;\gamma}^{\ast \rho \chi\;\beta} \epsilon_{\nu \tau \lambda \chi} )  
\nonumber \\ &
+ 
  (\widetilde{\eta}_{1}\; \bm\Sigma^{\nu\;\dot{\alpha}}_{\mu \;\alpha } \bm\Sigma_{\rho\;\dot{\beta}}^{\ast \tau \;\beta }  \bm\Delta_{R\nu \lambda  \;\dot{\alpha}}^{\;\;\;\;\;\;\;\;\; \dot{\beta}} \bm\Delta_{L\tau \chi \;\beta}^{\;\;\;\;\;\;\;\;\; \alpha}\epsilon^{\mu \rho \lambda \chi}  + \widetilde{\eta}^{\ast}_{1}\; \bm\Sigma_{\mu\;\dot{\alpha}}^{\ast \nu\;\alpha } \bm\Sigma_{\rho \;\beta }^{\tau\;\dot{\beta}} \bm\Delta_{R  \;\dot{\beta}}^{\ast\mu \lambda \dot{\alpha}} \bm\Delta_{L\;\alpha}^{\ast\rho \chi \beta} \epsilon_{\nu \tau \lambda \chi})
\nonumber \\ &
+ 
  (\widetilde{\eta}_{2}\; \bm\Sigma^{\nu\;\dot{\alpha}}_{\mu \;\alpha } \bm\Sigma^{\mu\;\dot{\beta}}_{\nu \;\beta } \epsilon^{\alpha \kappa} \epsilon_{\dot{\alpha} \dot{\kappa}} \bm\Delta_{R\lambda \chi\; \dot{\beta}}^{\;\;\;\;\;\;\; \;\dot{\kappa}} \bm\Delta_{L \;\kappa}^{\ast\lambda \chi\;\beta}   + \widetilde{\eta}^{\ast}_{2}\;  \bm\Sigma_{\nu\;\dot{\alpha}}^{\ast \mu \; \alpha } \bm\Sigma_{\mu\;\dot{\beta}}^{\ast \nu \;\beta }  \epsilon_{\alpha \kappa} \epsilon^{\dot{\alpha} \dot{\kappa}} \bm\Delta_{R\;\dot{\kappa}}^{\ast\lambda \chi \;\dot{\beta}} {\bm\Delta}_{L\lambda \chi\; \beta}^{\;\;\;\;\;\;\;\; \;\kappa}    )  
\nonumber \\ &
+  
 (\widetilde{\eta}_{3}\; \bm\Sigma^{\nu\;\dot{\alpha}}_{\mu \;\alpha } \bm\Sigma^{\mu\;\dot{\beta}}_{\rho \;\beta } \epsilon^{\alpha \kappa} \epsilon_{\dot{\alpha} \dot{\kappa}} {\bm\Delta}_{R\nu \tau\; \dot{\beta}}^{ \dot{\kappa}} \bm\Delta_{L \;\kappa}^{\ast\tau \rho\;\beta} + \widetilde{\eta}^{\ast}_{3}\;  \bm\Sigma_{\nu\;\dot{\alpha}}^{\ast \mu \; \alpha } \bm\Sigma_{\mu\;\dot{\beta}}^{\ast \rho \; \beta }  \epsilon_{\alpha \kappa} \epsilon_{\dot{\alpha} \dot{\kappa}} \bm\Delta_{R \;\dot{\kappa}}^{\ast\nu \tau\;\dot{\beta}} {\bm\Delta}_{L\tau \rho\; \beta}^{ \;\;\;\;\;\;\;\;\kappa}     )
\nonumber \\ &+   (\widetilde{\eta}_{4}\; \bm\Sigma^{\nu\;\dot{\alpha}}_{\mu \;\alpha } \bm\Sigma^{\tau\;\dot{\beta}}_{\rho \;\beta } \epsilon^{\alpha \kappa} \epsilon_{\dot{\alpha} \dot{\kappa}} {\bm\Delta}_{R\;\nu \tau\; \dot{\beta}}^{\;\;\;\;\;\;\;\dot{\kappa}} \bm\Delta_{L \;\kappa}^{\ast\mu \rho\beta}    + \widetilde{\eta}^{\ast}_{4}\;   \bm\Sigma_{\nu\;\dot{\alpha}}^{\ast \mu \; \alpha } \bm\Sigma_{\tau \;\dot{\beta}}^{\ast \rho \;\beta}  \epsilon_{\alpha \kappa} \epsilon^{\dot{\alpha} \dot{\kappa}} \bm\Delta_{R \;\dot{\kappa}}^{\ast\nu\tau\;\dot{\beta}} {\bm\Delta}_{L\mu\rho\; \beta}^{\;\;\;\;\;\;\;\;\kappa} ) , \label{Vsigmadelta}
\\  
V_{\bm\Phi \bm\Sigma \bm\Delta} &=  \{
( \widetilde{\chi}_{1R}\; \bm\Phi^{\ast \alpha}_{ \dot{\alpha}} \bm\Sigma_{\mu \;\alpha }^{\nu\;\dot{\alpha}} \bm\Delta_{R\nu \rho \dot{\beta}}^{\;\;\;\;\; \dot{\gamma}} \bm\Delta_{R\dot{\gamma}}^{\ast\rho \mu \dot{\beta}} 
+ 
\widetilde{\chi}^{\ast}_{1R}\; \bm\Phi_{\alpha}^{ \dot{\alpha}} \bm\Sigma^{\ast \nu \; \alpha }_{\mu\;\dot{\alpha}} \bm\Delta_{R\nu \rho \dot{\beta}}^{\;\;\;\;\;\; \dot{\gamma}} \bm\Delta_{R\dot{\gamma}}^{\ast\rho \mu \dot{\beta}} )
\nonumber \\ &
+
( \widetilde{\chi}_{2R}\; \bm\Phi^{\ast \alpha}_{ \dot{\alpha}} \bm\Sigma_{\mu \;\alpha }^{\nu\;\dot{\beta}} \bm\Delta_{R\nu \rho \dot{\beta}}^{\;\;\;\;\; \dot{\gamma}} \bm\Delta_{R\dot{\gamma}}^{\ast\rho \mu \dot{\alpha}} 
+
 \widetilde{\chi}^{\ast}_{2R}\; \bm\Phi_{\alpha}^{ \dot{\alpha}} \bm\Sigma^{\ast \nu \;\alpha }_{\mu\;\dot{\beta}} \bm\Delta_{R\nu \rho \dot{\alpha}}^{\;\;\;\;\;\; \dot{\gamma}} \bm\Delta_{R\dot{\gamma}}^{\ast\rho \mu \dot{\beta}} )  + \;R\leftrightarrow L \}
\nonumber \\ &
+
 (\widetilde{\chi}_{3}\; \bm\Phi_{\alpha}^{ \dot{\alpha}} \bm\Sigma_{\mu \;\beta }^{\nu\;\dot{\beta}} \epsilon^{\alpha \kappa} \epsilon_{\dot{\alpha} \dot{\kappa}} {\bm\Delta}_{R\nu\tau\;\dot{\beta}}^{\;\;\;\;\;\;\;\; \dot{\kappa}} \bm\Delta_{L\;\kappa}^{\ast\tau \mu\; \beta} 
+ 
 \widetilde{\chi}^{\ast}_{3}\; \bm\Phi^{\ast \alpha}_{ \dot{\alpha}} \bm\Sigma_{\nu\;\dot{\beta}}^{\ast \mu \;\beta } \epsilon_{\alpha \kappa}\epsilon^{\dot{\alpha} \dot{\kappa}} \bm\Delta_{R\nu \tau \;\dot{\kappa}}^{\ast\;\;\;\;\;\;\; \dot{\beta}} {\bm\Delta}_{L\tau\mu\;\beta}^{\;\;\;\;\;\;\;\; \;\kappa}  ) ,
\\ 
V_{\mathcal{S}} &= -\mu^{2}_{\mathcal{S}}\; \mathcal{S} \mathcal{S}^{\ast} + \lambda_{\mathcal{S}}\; \mathcal{S} \mathcal{S}^{\ast} \mathcal{S} \mathcal{S}^{\ast}
+ 
( \xi_{1}\; \bm\Phi_{\alpha}^{\dot{\alpha}} \bm\Phi^{\ast \alpha}_{\dot{\alpha}}+ \xi_{2}\; \bm\Sigma_{\mu \;\alpha}^{\nu\;\dot{\alpha}} {\bm\Sigma^{\ast}}_{\nu\;\alpha}^{\mu \; \dot{\alpha}} + \{ \xi_{3R}\; \bm\Delta_{R\mu \nu \dot{\alpha}}^{\;\;\;\;\; \dot{\beta}} \bm\Delta^{\ast\mu \nu \dot{\alpha}}_{R\; \dot{\beta}}  + \;R\leftrightarrow L \}  ) \mathcal{S} \mathcal{S}^{\ast}
\nonumber \\ &
+ 
(\widetilde{\zeta} \bm\Phi_{\alpha}^{ \dot{\alpha}} \bm\Phi_{\beta}^{ \dot{\beta}} \epsilon^{\alpha \beta} \epsilon_{\dot{\alpha} \dot{\beta}} \mathcal{S}^{\ast} +
 \widetilde{\zeta}^{\ast} \bm\Phi^{\ast \alpha}_{ \dot{\alpha}} \bm\Phi^{\ast \beta}_{ \dot{\beta}} \epsilon_{\alpha \beta} \epsilon^{\dot{\alpha} \dot{\beta}} \mathcal{S}) 
+
 (\widetilde{\omega} \bm\Sigma_{\mu \;\alpha }^{\nu\;\dot{\alpha}} \bm\Sigma^{\mu\;\dot{\beta}}_{\nu \;\beta } \epsilon^{\alpha \beta} \epsilon_{\dot{\alpha} \dot{\beta}} \mathcal{S}^{\ast} 
 +
  \widetilde{\omega}^{\ast} {\bm\Sigma^{\ast}}_{\mu\;\alpha}^{\nu \; \dot{\alpha}} {\bm\Sigma^{\ast}}^{\mu \;\beta }_{\nu\;\dot{\beta}}  \epsilon_{\alpha \beta} \epsilon^{\dot{\alpha} \dot{\beta}} \mathcal{S}).
\end{align}

To differentiate the complex couplings from the real ones in the potential we put tilde on the top of the complex ones. All the index contractions are shown explicitly. The parameters with dimension of mass are $\mu_{\phi}, \mu_{\Sigma}, \mu_{\Delta}, \mu_{\mathcal{S}}, \widetilde{\zeta}, \widetilde{\omega}$. To find the maximum possible number of invariants of each kind one needs to use the group theoretical rules of tensor product decomposition (for details see Ref. ~\cite{Slansky:1981yr}). Note that in general there can be more gauge invariant terms in the Higgs potential however are  absent in our theory due to the presence of the global $U(1)_{PQ}$ symmetry. Below we discuss the constraints on the cubic and quartic couplings in the potential due to additional left-right parity symmetry.   \\

\noindent
\textbf{ Scalar potential in the left-right parity symmetric limit}\label{section:LR}

\noindent If the parity symmetry is assumed to be a good symmetry then there are further restrictions on the potential Eq. \eqref{V}. Under left-right parity, the fermions and the scalar fields transform as 
\vspace{-0pt}
\begin{align}\label{parity}
\bm\Psi_L \longleftrightarrow \bm\Psi_R,  \;\; \bm\Phi \longleftrightarrow \bm\Phi^{\ast},\;\; \bm\Sigma \longleftrightarrow \bm\Sigma^{\ast},\;\; \bm\Delta_{R} \longleftrightarrow \bm\Delta_{L},\;\; \mathcal{S} \longleftrightarrow \mathcal{S}^{\ast}.
\end{align}

\noindent The terms that are achieved by  $R\leftrightarrow L$ in Eq. \eqref{V} have exactly the same coupling constants, for example, $\mu^2_{\Delta_L}=\mu^2_{\Delta_R}$,  $\lambda_{iL}=\lambda_{iR}$ ($i=1-5$) and so on.  Also due to the invariance under parity, some of the complex couplings in the potential will become real, they are:
\begin{align}
\widetilde{\alpha}_{5,6},\; \widetilde{\beta}_3,\; \widetilde{\eta}_{4,5,6},\; \widetilde{\chi}_{3},\; \widetilde{\zeta},\; \widetilde{\omega}\; \in \mathbb{R}.
\end{align}

\noindent
The only six couplings in the potential that remain complex are 
\begin{align}
\widetilde{\lambda}_9,\; \widetilde{\gamma}_{9,10},\; \widetilde{\eta}_{1},\; \widetilde{\chi}_{1,2} \in  \mathbb{C}.
\end{align}

\noindent Note that, under parity, if the singlet  field is odd, i.e, instead of $\mathcal{S} \longleftrightarrow \mathcal{S}^{\ast}$, if the transformation property is $\mathcal{S} \longleftrightarrow -\mathcal{S}^{\ast}$, then the cubic couplings $\widetilde{\zeta}$ and $\widetilde{\omega}$ become purely imaginary.  If the VEV of the  parity odd singlet is $v_S > v_R$, then the parity breaking scale and the $SU(2)_R$ breaking scale can be decoupled and in this scenario the PS breaking scale can be as low as $10^6$ GeV as mentioned earlier.

\subsection{The scalar mass spectrum}

In this sub-section, we compute the Higgs mass spectrum after the PS symmetry is broken.  

\vspace{5pt}
\noindent
\textbf{ Mass spectrum of $\Delta_{R}$ scalar fields}\label{section:deltamass}

\vspace*{5pt}
\noindent
The Yukawa  Lagrangian of the theory is given in Eq. \eqref{yukawa}, where the first two terms are the Dirac type Yukawa couplings. The third term generates the right-handed neutrino Majorana masses when the PS symmetry is broken by the VEV $\langle(1,3,10)\rangle$. Expanding this term of the Yukawa coupling one gets (here $\bm\Delta$ represents $\bm\Delta_R$):
\vspace{-2pt}
\begin{align}
\mathcal{L}_{Majorana} &= \frac{1}{2} {Y^R_{10}}_{ij} \{ \nu^T_{Ri}C \nu_{Rj} \bm\Delta^{\ast}_{\nu\nu} - e^T_{Ri}C e_{Rj} \bm\Delta^{\ast}_{ee} - \frac{(e^T_{Ri} \nu_{Rj} + \nu^T_{Ri}C e_{Rj})}{\sqrt{2}} \bm\Delta^{\ast}_{e\nu}
+ u^T_{Ri}C u_{Rj} \bm\Delta^{\ast}_{uu} 
\nonumber \\ &- d^T_{Ri}C d_{Rj} \bm\Delta^{\ast}_{dd} -\frac{( u^T_{Ri}C d_{Rj} + d^T_{Ri}C u_{Rj} )}{\sqrt{2}} \bm\Delta^{\ast}_{ud}  
+ \frac{( u^T_{Ri}C \nu_{Rj} + \nu^T_{Ri}C u_{Rj} )}{\sqrt{2}} \bm\Delta^{\ast}_{u\nu}
\nonumber \\ & -
 \frac{( e^T_{Ri}C d_{Rj} + d^T_{Ri}C e_{Rj} )}{\sqrt{2}} \bm\Delta^{\ast}_{de} - \frac{( d^T_{Ri}C \nu_{Rj} + \nu^T_{Ri}C d_{Rj} + e^T_{Ri}C u_{Rj} + u^T_{Ri}C e_{Rj} )}{2} \bm\Delta^{\ast}_{ue} \} + h.c
\end{align}

\noindent with the following identification: 
\vspace{-2pt}
\begin{align}\label{identify}
&\bm\Delta^{\ast}_{\nu\nu}(1,1,0) = \bm\Delta_{\dot{2}}^{\ast44\; \dot{1}} ;\;\;\;\;\;\;\; \bm\Delta^{\ast}_{ee}(1,1,2) = \bm\Delta_{\dot{1}}^{\ast44\;  \dot{2}} ;\;\;\;\; \bm\Delta^{\ast}_{e\nu}(1,1,1) = \sqrt{2}\; \bm\Delta_{\dot{1}}^{\ast44\; \dot{1}}\; ; \\ 
&\bm\Delta^{\ast}_{uu}(\overline{6},1,-\frac{4}{3}) = \bm\Delta_{ \dot{2}}^{\ast\bar{\mu} \bar{\nu}\; \dot{1}} ;\;\;\; \bm\Delta^{\ast}_{dd}(\overline{6},1,\frac{2}{3}) = \bm\Delta_{ \dot{1}}^{\ast\bar{\mu} \bar{\nu}\;\dot{2}} ;\;\;\; \bm\Delta^{\ast}_{ud}(\overline{6},1,-\frac{1}{3}) = \sqrt{2}\; \bm\Delta_{ \dot{1}}^{\ast\bar{\mu} \bar{\nu}\; \dot{1}}\; ; \\
&\bm\Delta^{\ast}_{u\nu}(\overline{3},1,-\frac{2}{3}) = \sqrt{2}\; \bm\Delta_{ \dot{2}}^{\ast\bar{\mu}4\; \dot{1}} ;\;\;\; \bm\Delta^{\ast}_{de}(\overline{3},1,\frac{4}{3}) = \sqrt{2}\; \bm\Delta_{\dot{1}}^{\ast\bar{\mu}4\; \dot{2}} ;\;\;\; \bm\Delta^{\ast}_{ue}(\overline{3},1,\frac{1}{3}) = 2\; \bm\Delta_{ \dot{1}}^{\ast\bar{\mu}4\; \dot{1}} .
\end{align}

\noindent 
Only the neutral component of $\Delta_R$ gets  VEV, $v_{R}=\langle \bm\Delta_{\nu\nu} \rangle$.  With this identification and by minimizing the potential Eq. \eqref{V}, one can compute the mass spectrum of $\bm\Delta_{R}$. The PS breaking  minimization conditions is found to be:
\vspace{-2pt}
\begin{align}
\frac{\partial V_{\Delta}}{\partial v_{R}} &=v_R [2 v_{R}^{2} (\lambda_{1R} + \lambda_{3R} + \lambda_{4R})-\mu^{2}_{\Delta}] = 0 .
\end{align}

\noindent Choosing the non-trivial solution with $v_R \neq 0$, this equation is used to eliminate $\mu^{2}_{\Delta}$ from the potential. Imposing this extremum condition back to the potential we find the following mass spectrum for $\bm\Delta_{R}$:
\vspace{-2pt}
\begin{align}
& m^{2}_{\Delta_{\nu\nu}} = 2 \; v^{2}_{R}\; (\lambda_{1R}+\lambda_{3R}+\lambda_{4R}) ,\\ 
&m^{2}_{\Delta_{ee}} = 4 \; v^{2}_{R}\; (\lambda_{2R}+\lambda_{5R}) ,\\ 
&m^{2}_{\Delta_{e\nu}} = 0, \\
& m^{2}_{\Delta_{uu}} = -2 \; v^{2}_{R}\; \lambda_{4R} ,\\ 
&m^{2}_{\Delta_{dd}} = 2 \; v^{2}_{R}\; (\lambda_{2R}-\lambda_{3R}-\lambda_{4R}) , \\
&m^{2}_{\Delta_{ud}} = -2 \; v^{2}_{R}\; (\lambda_{3R}+\lambda_{4R}), \\
& m^{2}_{\Delta_{u\nu}} = 0 ,\\
&m^{2}_{\Delta_{de}} = 2 \; v^{2}_{R}\; (\lambda_{2R}- \lambda_{3R}- \frac{\lambda_{4R}}{2} + \lambda_{5R}) ,\\ 
 &m^{2}_{\Delta_{ue}} = -2 \; v^{2}_{R}\; (2\; \lambda_{3R}+\lambda_{4R}). 
\end{align}

\noindent
There is a  mass relation which is given by: 
\vspace{-5pt}
\begin{equation}
m^{2}_{\Delta_{ee}} = m^{2}_{\Delta_{de}} - m^{2}_{\Delta_{ud}} + m^{2}_{\Delta_{uu}}.
\end{equation}

\noindent There exist seven physical Higgs states $\bm\Delta_{\nu\nu}, \bm\Delta_{ee},\bm\Delta_{uu}, \bm\Delta_{dd}, \bm\Delta_{ud}, \bm\Delta_{de}, \bm\Delta_{ue}$ and three Nambu-Goldstone boson states  $\bm\Delta_{e\nu}, \bm\Delta_{u\nu}$ and $i (\bm\Delta_{44 \;\dot{1}}^{\;\;\;\;\;\; \dot{2}}-\bm\Delta_{\dot{2}}^{\ast44 \; \dot{1}})/2 \equiv \bm\Delta_G$. As mentioned in Sec. \ref{gauge}, due to  the $G_{224}\rightarrow G_{213}$ breaking,  9 of the gauge bosons become massive after eating up the 9 Goldstone bosons. These Goldstone bosons correspond to  $\bm\Delta_{e\nu}$ , $\bm\Delta_{u\nu}$ and $\bm\Delta_G$ (real field) fields. We note that these sextets can have rich phenomenology if their masses are relatively low, for example, these sextets can be responsible for generating baryon asymmetry after the sphaleron decoupling, see Ref. \cite{Babu:2006xc,Babu:2006wz,Babu:2008rq,Babu:2013yca}. By considering the sextet masses at the TeV scale, flavor physics constraints are also computed in Ref. \cite{Fortes:2013dba}.

If both the PS and  PQ symmetry breaking are taken into  account together, where the PQ symmetry is broken by the complex singlet VEV, $\langle \mathcal{S} \rangle=v_{S}$ the minimization conditions are 
\vspace{-8pt}
\begin{align}
\frac{\partial V}{\partial v_{R}} &= v_R [2 v_{R}^{2} (\lambda_{1R} + \lambda_{3R} + \lambda_{4R}+v^2_S \xi_{3R})-\mu^{2}_{\Delta}]= 0 \;\; \text{\rm and}
 \\ 
\frac{\partial V}{\partial v_{S}} &= v_S [2 v^2_S \lambda_{\mathcal{S}}+v^2_R \xi_{3R} -\mu^2_{\mathcal{S}}] = 0 .
\end{align}

\noindent Assuming the general symmetry breaking solutions $v_S \neq 0$ and $v_R \neq 0$, these equations can be used to solve for $\mu^{2}_{\Delta}$ and $\mu^2_{\mathcal{S}}$.   Using these stationary  conditions like before one can easily derive the mass spectrum for the $\bm\Delta_R$ and $\mathcal{S}$ fields. The mass spectrum essentially remains unchanged except $\bm\Delta_{\nu\nu}$ mixes with the real part of the singlet field.  The two by two mass squared matrix of this mixing in the basis $\{ \bm\Delta_{\nu\nu}, Re[\mathcal{S}]\}$ is computed to be: 
\vspace{-0pt}
\begin{align}
\begin{pmatrix}
2 \; v^{2}_{R}\; (\lambda_{1R}+\lambda_{3R}+\lambda_{4R}) & 2v_S v_R \xi_{3R} \\
2v_S v_R \xi_{3R} & 4 v^2_S \lambda_{\mathcal{S}}
\end{pmatrix}.
\end{align}

\noindent The imaginary part of $\mathcal{S}$ remains massless after the PQ symmetry breaking. After EW  symmetry breaking, this field will eventually mix with the components from the four doublets coming from  $\bm\Phi$ and $\bm\Sigma$ and receive a mass of the order of $v_{ew}/v_S$. Since $v_{ew} \ll v_S$, this field will remain essentially massless and can be identified as the axion field, which is the dark matter candidate in our model.

\noindent \\
\textbf{ The doublet $(1,2,\pm 1/2)$ mass square matrix}\label{section:doubletmass}

\vspace*{5pt}
\noindent
In the model, there are two complex bi-doublets  (2,2,1) and (2,2,15) that contain four $SU_{L}(2)$ doublets. Among them, two of them are   $\bm\Phi_{\alpha}^{ \dot{1}}$ and $\bm\Sigma_{\alpha}^{ \dot{1}}\equiv -\frac{2}{\sqrt{3}} \bm\Sigma_{4\;\alpha}^{4\; \dot{1}}$ that have the quantum number  $(1,2,-1/2)$ under the SM group and the other  two are  $\bm\Phi_{\alpha}^{ \dot{2}}$ and $\bm\Sigma_{\alpha}^{ \dot{2}}\equiv -\frac{2}{\sqrt{3}}\bm\Sigma_{4\;\alpha}^{4\; \dot{2}}$ which have quantum number of  $(1,2,+1/2)$. Writing as, 
\vspace*{-2pt}
\begin{equation}
h_{\alpha}^{(i)} = \{ \bm\Phi_{\alpha}^{ \dot{1}},\bm\Sigma_{\alpha}^{ \dot{1}}, \bm\Phi^{\ast \beta}_{ \dot{2}} \epsilon_{\beta \alpha}, \bm\Sigma^{\ast\beta}_{ \dot{2}} \epsilon_{\beta \alpha} \}
\end{equation}
\noindent and similarly
\begin{equation}
\bar{h}^{(i) \alpha} = \{ \bm\Phi^{\ast \alpha}_{ \dot{1}}, \bm\Sigma^{\ast \alpha}_{ \dot{1}}, \bm\Phi_{\beta}^{ \dot{2}} \epsilon^{\beta \alpha}, \bm\Sigma_{\beta}^{ \dot{2}} \epsilon^{\beta \alpha}  \}
\end{equation}

\noindent
the doublet mass squared matrix, $\mathcal{D}$ in the flavor basis can be found from the Higgs potential as
\vspace*{-2pt}
\begin{equation}
\bar{h}^{\alpha (j)} \mathcal{D}_{ij}  h^{(i)}_{\alpha} .
\end{equation}

\noindent It is straightforward to compute this doublet mass square matrix,
\vspace{2pt}
\begin{align}
\mathcal{D}= \begin{pmatrix}
 -\mu^{2}_{\phi} + v^{2}_{R}\;(\beta_{1} + \beta_{2})+v^{2}_{S}\; \xi_{1} & -\frac{\sqrt{3}}{2} v^{2}_{R}\;(\tilde{\chi}^{\ast}_{1} + \tilde{\chi}^{\ast}_{2})& 2\; v_{S}\; \tilde{\zeta}  & 0 \\
-\frac{\sqrt{3}}{2} v^{2}_{R}\;(\tilde{\chi}_{1} + \tilde{\chi}_{2}) &  -\mu^{2}_{\Sigma} + v^{2}_{R}\; A_{2} + v^{2}_{S}\; \xi_{2} & 0 & 2\; v_{S}\; \tilde{\omega}  \\
2\; v_{S}\; \tilde{\zeta}^{\ast}  & 0 & -\mu^{2}_{\phi} + v_{R}^{2}\; \beta_{1}+v^{2}_{S} \xi_{1} &  -\frac{\sqrt{3}}{2} v^{2}_{R}\; \tilde{\chi}_{1} \\
0 &2\; v_{S}\; \tilde{\omega}^{\ast} &-\frac{\sqrt{3}}{2} v^{2}_{R}\; \tilde{\chi}^{\ast}_{1} & -\mu^{2}_{\Sigma} + v^{2}_{R}\;A_{1}+v^{2}_{S}\; \xi_{2}
\end{pmatrix}
\end{align}

\noindent where we have defined
\begin{equation}
A_{1} = \gamma_{1}+ \frac{3}{4} (\gamma_{2}+\gamma_{3}+\gamma_{4}), \; A_{2} = A_{1}+\gamma_{5}+\frac{3}{4} (\gamma_{6}+\gamma_{7}+\gamma_{8}).
\end{equation}

\noindent Recall that if parity symmetry is imposed, $\widetilde{\zeta}$ and  $\widetilde{\omega}$ will be real but $\widetilde{\chi}_{1,2}$ entering in this matrix will remain complex, so in general $\mathcal{D}$ will have two independent phases entering in this matrix.

The Hermitian matrix, $\mathcal{D}$ can be diagonalized as $\mathcal{D}= U \Lambda U^{\dagger}$, where $U$ is an unitary matrix ($\Lambda$ is the diagonal matrix containing real eigenvalues) that relates the flavor basis, $h_{\alpha}^{(i)}$ and mass basis, $h_{\alpha}^{\prime (i)}$ states,
\vspace{-5pt}
\begin{eqnarray}
\bar{h}^{\alpha (i)} \mathcal{D}_{ij} h_{\alpha}^{(j)} = \bar{h}^{\alpha (i)} U_{il} \Lambda_{lk} U^{\ast}_{jk} h_{\alpha}^{(j)} = \bar{h}^{\alpha \prime (i)} \Lambda_{ij} h_{\alpha}^{\prime (j)}.
\end{eqnarray}
\vspace{-3pt}
\noindent That is,  
\begin{equation}
h_{\alpha}^{\prime (k)} = U^{\ast}_{jk}  h_{\alpha}^{(j)}.
\end{equation}

\noindent The doublet mass matrix  written here is before the EW phase transition, so the SM Higgs doublet will correspond to the zero eigenvalue solution, which can be found by imposing the fine tuning condition 
$det(\mathcal{D})=0$. One can write the SM Higgs doublet that is a linear combination of the four doublets as,
\begin{align}
H \equiv h^{\prime (1)}_{\alpha} = U^{\ast}_{j1} h^{(j)}_{\alpha},\;\;\; \textrm{that\; gives},\;\;\; h^{(i)}_{\alpha} = U_{ji} h^{\prime (j)}_{\alpha}.
\end{align}

\noindent When the SM doublet acquires VEV, $\langle H\rangle = v_{\rm{EW}}$, the EW phase transition takes place and one gets, 
\vspace{-5pt}
\begin{align}\label{vevs}
\langle h^{(1)}_{\alpha}\rangle = U_{11} v_{\rm{EW}} \equiv \alpha \; v_{\rm{EW}},\;\;\; 
\langle h^{(2)}_{\alpha} \rangle = U_{12} v_{\rm{EW}} \equiv \beta \; v_{\rm{EW}}, \\
\langle h^{(3)}_{\alpha} \rangle = U_{13} v_{\rm{EW}} \equiv \gamma \; v_{\rm{EW}}, \;\;\;
\langle  h^{(4)}_{\alpha} \rangle = U_{14} v_{\rm{EW}} \equiv \delta \; v_{\rm{EW}}.
\end{align}

\noindent By finding the matrix elements $U_{ij}$ it can be shown that the combinations $\alpha
 \gamma^{\ast} $ and $\beta  \delta^{\ast} $ will remain complex and so all the VEVs in Eq. \eqref{vevs} cannot be taken to be real.  This is why the VEV ratios of the doublets that appear in the fermion mass matrices are in general complex. This conclusion is also applicable for the case with parity symmetry imposed, since $\widetilde{\chi}_{1,2}$ that are complex couplings will introduce two independent phases in $\mathcal{D}$.

\noindent \\
\textbf{ The color triplet $(3,2,\pm \frac{1}{6})$ mass square matrix}

\vspace*{5pt}
\noindent
The color triplets are $\bm\Sigma^{4\;\dot{1}}_{\bar{\mu}\;\alpha}$ and $\bm\Sigma^{\bar{\mu}\;\dot{2}}_{4\;\alpha }$ that are $(3,2,+1/6)$ and $(\overline{3},2,-1/6)$ under the SM group respectively. The mass square matrix is given as follows

\begin{align}\label{tripA}
\begin{pmatrix}
\bm\Sigma^{4\;\dot{1}}_{\bar{\mu}\;\alpha } & \bm\Sigma^{\ast 4 \;\beta }_{\bar{\mu}\;\dot{2}} \epsilon_{\beta \alpha} 
\end{pmatrix} 
\begin{pmatrix}
-\mu^2_{\Sigma}+ v^2_R(\gamma_1+\gamma_3+\gamma_5+\gamma_7)+v^2_{\mathcal{S}}\xi_2 & 2\; v_{\mathcal{S}}\; \widetilde{\omega}\\
2\; v_{\mathcal{S}}\; \widetilde{\omega}^{\ast} & -\mu^2_{\Sigma}+ v^2_R(\gamma_1+\gamma_2)+v^2_{\mathcal{S}}\; \xi_2
\end{pmatrix}
\begin{pmatrix}
\bm\Sigma^{\ast\bar{\mu}\;\alpha }_{4\;\dot{1}}  \\
\bm\Sigma^{\bar{\mu}\;\dot{2}}_{4\;\sigma } \epsilon^{\sigma \alpha}
\end{pmatrix}.
\end{align} 

\noindent Note that if the parity symmetry is imposed, all the matrix elements in this mass squared matrix will become real.

\noindent \\
\textbf{ The color triplet $(3,2,\pm \frac{7}{6})$ mass square matrix}

\vspace*{5pt}
\noindent
The color triplets are $\bm\Sigma^{4\;\dot{2}}_{\bar{\mu}\;\alpha }$ and $\bm\Sigma^{\bar{\mu}\;\dot{1}}_{4\;\alpha }$ that are $(3,2,+7/6)$ and $(\overline{3},2,-7/6)$ under the SM group respectively. The mass square matrix is given as follows

\begin{align}\label{tripB}
\begin{pmatrix}
\bm\Sigma^{4\;\dot{2}}_{\bar{\mu}\;\alpha } & \bm\Sigma^{\ast 4 \;\beta }_{\bar{\mu}\;\dot{1}} \epsilon_{\beta \alpha} 
\end{pmatrix} 
\begin{pmatrix}
-\mu^2_{\Sigma}+ v^2_R(\gamma_1+\gamma_3)+v^2_{\mathcal{S}}\xi_2 & -2\; v_{\mathcal{S}}\; \widetilde{\omega}\\
-2\; v_{\mathcal{S}}\; \widetilde{\omega}^{\ast} & -\mu^2_{\Sigma}+ v^2_R(\gamma_1+\gamma_2+\gamma_5+\gamma_6)+v^2_{\mathcal{S}}\; \xi_2
\end{pmatrix}
\begin{pmatrix}
\bm\Sigma^{\ast\bar{\mu}\;\alpha }_{4\;\dot{2}}  \\
\bm\Sigma^{\bar{\mu}\;\dot{1}}_{4\;\sigma } \epsilon^{\sigma \alpha}
\end{pmatrix}.
\end{align}

\noindent Again if the parity symmetry is imposed, all the matrix elements in this mass squared matrix will become real.

\noindent \\
\textbf{ The color octet $(8,2,\pm \frac{1}{2})$ mass square matrix}

\vspace*{5pt}
\noindent
The color octets are $\bm\Sigma^{\bar{\nu}\;\dot{1}}_{\bar{\mu}\;\alpha }$ and $\bm\Sigma^{\bar{\nu}\;\dot{2}}_{\bar{\mu}\;\alpha }$ that are (8,2,-1/2) and (8,2,+1/2) under the SM group respectively. The mass square matrix is given as follows

\begin{align}\label{oct}
\begin{pmatrix}
\bm\Sigma^{\bar{\nu}\;\dot{1}}_{\bar{\mu}\;\alpha } & \bm\Sigma^{\ast \bar{\nu} \;\beta }_{\bar{\mu}\;\dot{2}} \epsilon_{\beta \alpha} 
\end{pmatrix} 
\begin{pmatrix}
-\mu^2_{\Sigma}+ v^2_R(\gamma_1+\gamma_5)+v^2_{\mathcal{S}}\xi_2 & 2\; v_{\mathcal{S}}\; \widetilde{\omega}\\
2\; v_{\mathcal{S}}\; \widetilde{\omega}^{\ast} & -\mu^2_{\Sigma}+ v^2_R \; \gamma_1+v^2_{\mathcal{S}}\; \xi_2
\end{pmatrix}
\begin{pmatrix}
\bm\Sigma^{\ast\bar{\mu}\;\alpha }_{\bar{\nu}\;\dot{1}}  \\
\bm\Sigma^{\bar{\mu}\;\dot{2}}_{\bar{\nu}\;\sigma } \epsilon^{\sigma \alpha}
\end{pmatrix}.
\end{align}

\noindent Like the color triplet cases, if parity is a good  symmetry, this mass squared matrix will become real.

\noindent \\
\textbf{ The  mass spectrum of $\bm\Delta_L$ field}

\vspace*{5pt}
\noindent The identification of the multiplets  of the $(3,1,10^{\ast})$ field under the SM group is (here $\bm\Delta$ represents $\bm\Delta_L$):
\vspace{-5pt}
\begin{align}
\bm\Delta^{\ast}_{qq}(\overline{6},3,-\frac{1}{3})=\bm\Delta^{\ast \overline{\mu}\overline{\nu}\;\beta}_{\alpha},\;\;\;
\bm\Delta^{\ast}_{ql}(\overline{3},3,\frac{1}{3})=\bm\Delta^{\ast \overline{\mu}4\;\beta}_{\alpha},\;\;\;
\bm\Delta^{\ast}_{ll}(1,3,-1)=\bm\Delta^{\ast 44\;\beta}_{\alpha}.
\end{align}

\noindent The  mass spectrum of these fields are given as follows:
\vspace{-5pt}
\begin{align}
&m^2_{\Delta_{ll}}= -\mu^2_{\Delta_L} + v^2_R\; (\lambda_{6L} +\lambda_{7L}+\lambda_{8L}) + v^2_{\mathcal{S}} \;\xi_{L3} \\
&m^2_{\Delta_{qq}}= -\mu^2_{\Delta_L} + v^2_R\; \lambda_{6L} + v^2_{\mathcal{S}} \;\xi_{L3} \\
&m^2_{\Delta_{ql}}= - \mu^2_{\Delta_L} + v^2_R\; ( \lambda_{6L} +\frac{\lambda_{7L}}{2}) + v^2_{\mathcal{S}} \;\xi_{L3}. 
\end{align}

\section{Baryon number violation} \label{chapter06}
\subsection{Nucleon decay} 

\begin{figure}[th!]
\centering
\includegraphics[scale=0.45]{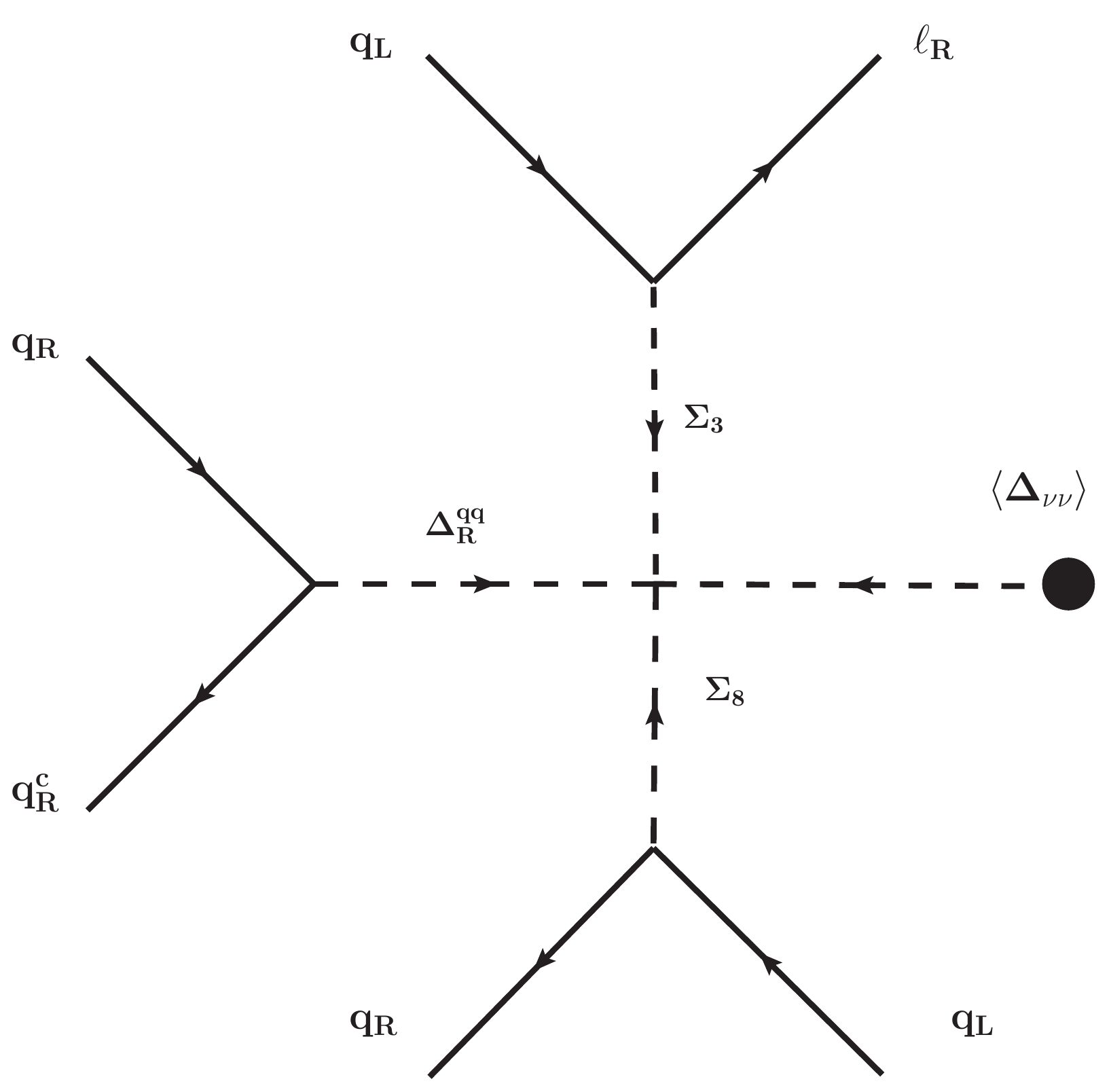}\hspace{10pt}
\includegraphics[scale=0.45]{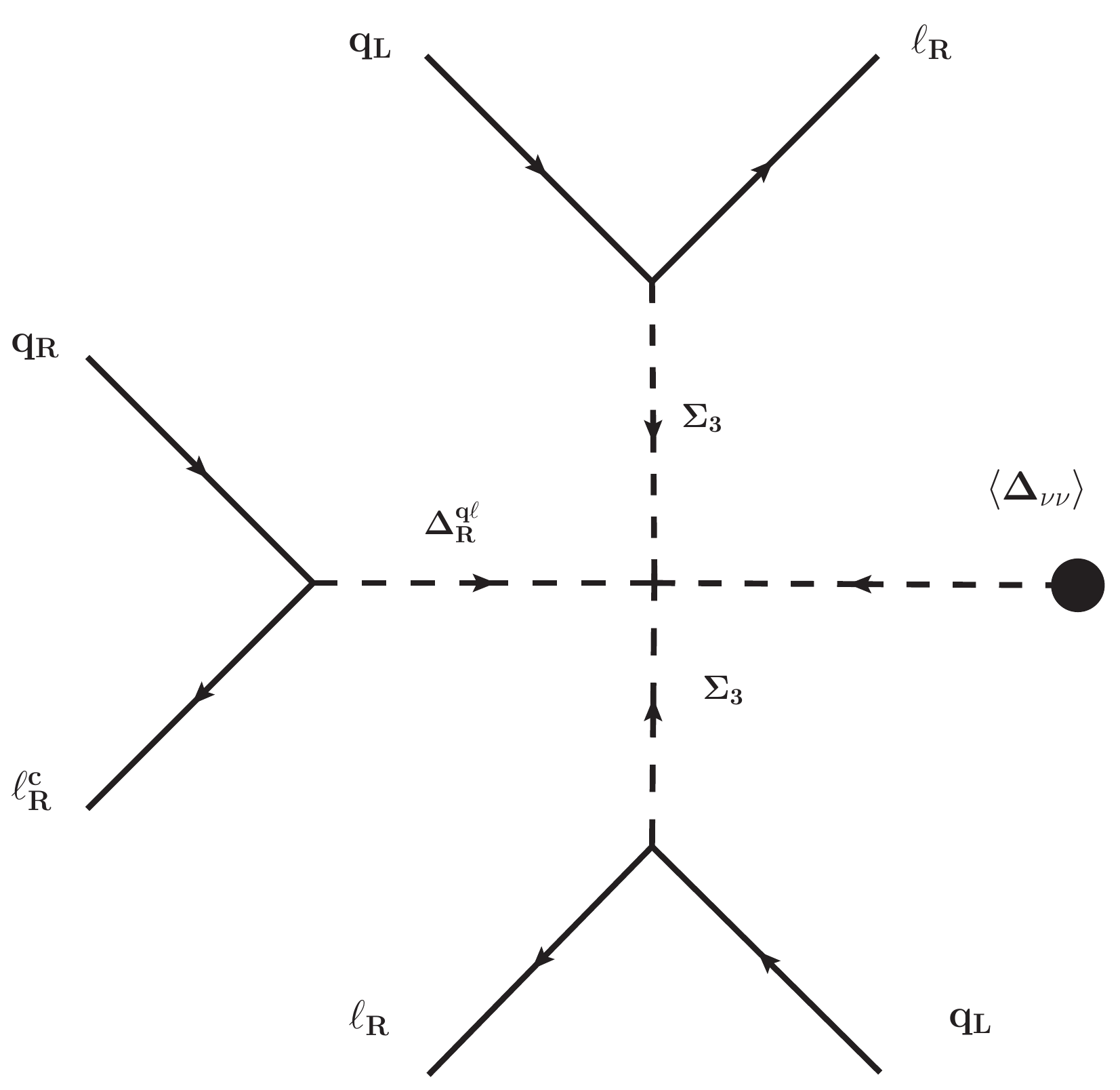}
\caption{Feynman diagrams for nucleon decay with the $v_R=\langle \bm\Delta_R \rangle$ VEV insertions. The left diagram induces nucleon decay processes like, nucleon $\rightarrow$ lepton + mesons and the right digram, nucleon $\rightarrow$ lepton + lepton + antilepton processes. }\label{d9}
\end{figure}

Though nucleon decay is not mediated by the gauge bosons of the PS group, depending on the details of the scalar sector, nucleon may decay. A PS model with scalars (2,2,1), (1,3,10) and (3,1,10),  nucleon is absolutely stable. The reason for the stability is due to the existence of a hidden discrete symmetry~\cite{Mohapatra:1980qe} in the model $q_{\mu}\rightarrow e^{i \pi/3} q_{\mu}$, $\bm\Delta_{\mu \nu} \rightarrow e^{-2 i \pi/3} \bm\Delta_{\mu \nu}$, $\bm\Delta_{\mu 4}\rightarrow e^{i \pi/3} \bm\Delta_{\mu 4}$. The Lagrangian is invariant under this discrete symmetry even after SSB. But
the scalar sector Eq. \eqref{eq:Higgs} that we is considered  in this work, which also contains (2,2,15) multiplet,  in principle can lead to baryon(B) and lepton(L) violating processes by nucleon decay~\cite{Pati:1983zp,Pati:1983jk}. This happens due to the presence of some specific quartic terms in the scalar potential Eq. \eqref{V}. In our model, the part of the potential $V_{\Sigma \Delta}$ in Eq. \eqref{Vsigmadelta} contains terms that can cause the nucleon to decay. The  terms with coupling coefficients $\widetilde{\gamma}_{9}, \widetilde{\gamma}_{10}, \widetilde{\eta}_{1}$  in Eq. \eqref{Vsigmadelta}, in combination with the  Yukawa interactions Eq. \eqref{yukawa} are responsible for $|\Delta (B-L)|=2$ processes when the symmetry gets broken spontaneously by $\langle \bm\Delta_R \rangle$. These $(B+L)$ conserving processes cause the proton to decay into leptons and mesons.   The Feynman diagrams associated with  such quartic terms involving processes like $3q\rightarrow q q^c \ell$ ($p,n \rightarrow \ell+$ mesons, with $\ell=e^{-}, \mu^{-}, \nu_{e}, \nu_{\mu}$; meson$=\pi, K$, etc.) contain $SU(3)_{C}$ triplets, $\bm\Sigma_3$ and octets, $\bm\Sigma_8$ originating from the multiplet (2,2,15). The Feynman diagrams corresponding to these processes are  as shown in Fig. \ref{d9} (left diagram).

For PS model with this minimal set of scalars, another Feynman diagram that contributes to the nucleon decay can be constructed by replacing the color octet $\Sigma_8$ by a color triplet $\Sigma_3$ and the sextet $\Delta_6$  by color triplet $\Delta_3$ as shown in Fig. \ref{d9} (right diagram). This kind of diagrams will lead to nucleon decay, $3q \rightarrow \ell \ell \ell^c$. These  processes shown in Fig. \ref{d9} are generated by the dimension nine ($d=9$) operators. Shortly we will show that in our set-up, $d=9$ operators only give rise to the decay processes of the type nucleon$\to$ lepton+ meson(s)  but not nucleon $\rightarrow$ lepton + lepton + antilepton processes since these three lepton decays always involve $\nu_R$ in the final state   and hence are extremely suppressed.
 
\begin{figure}[th!]
\centering
\includegraphics[scale=0.45]{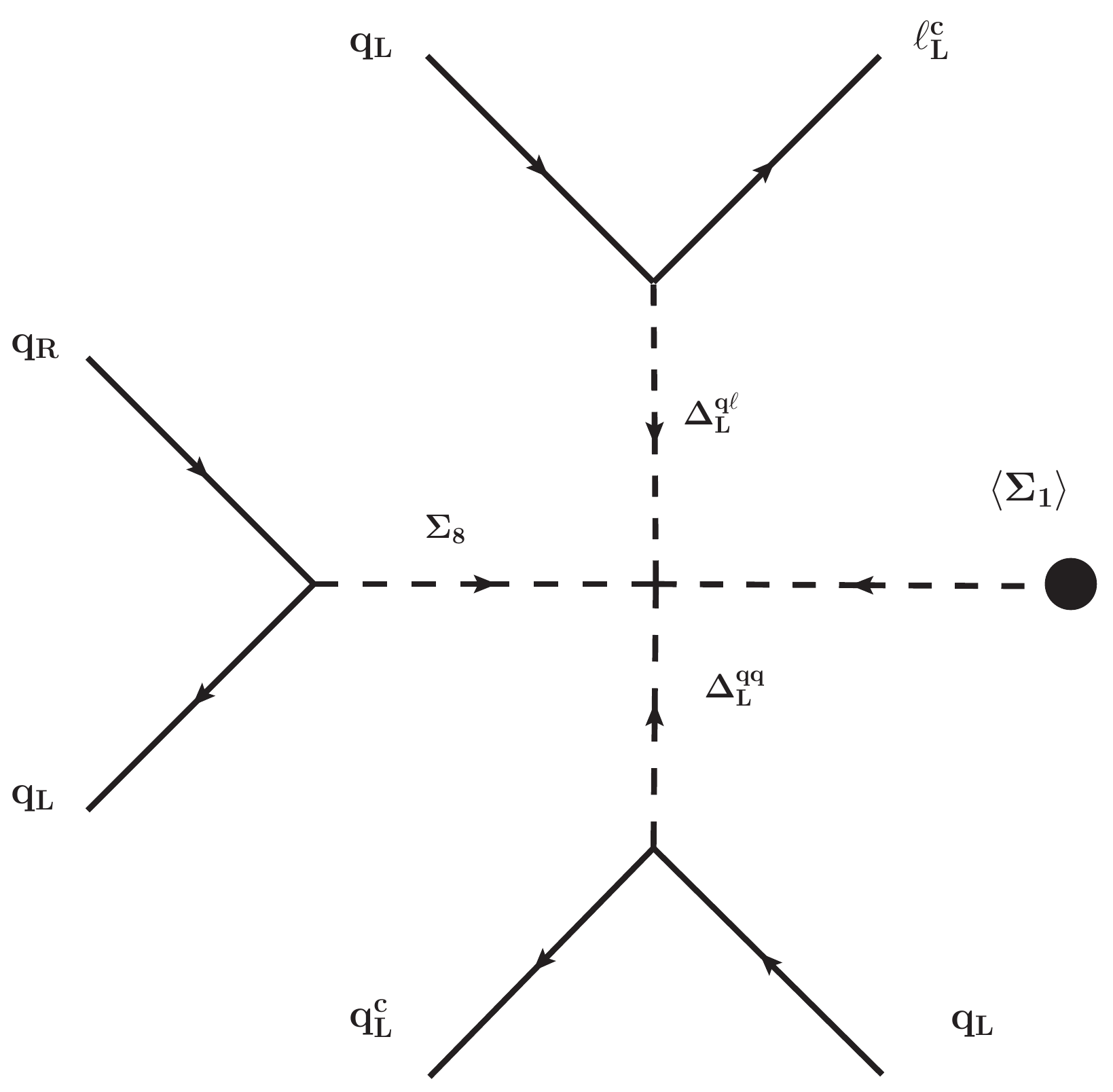}\hspace{10pt}
\includegraphics[scale=0.45]{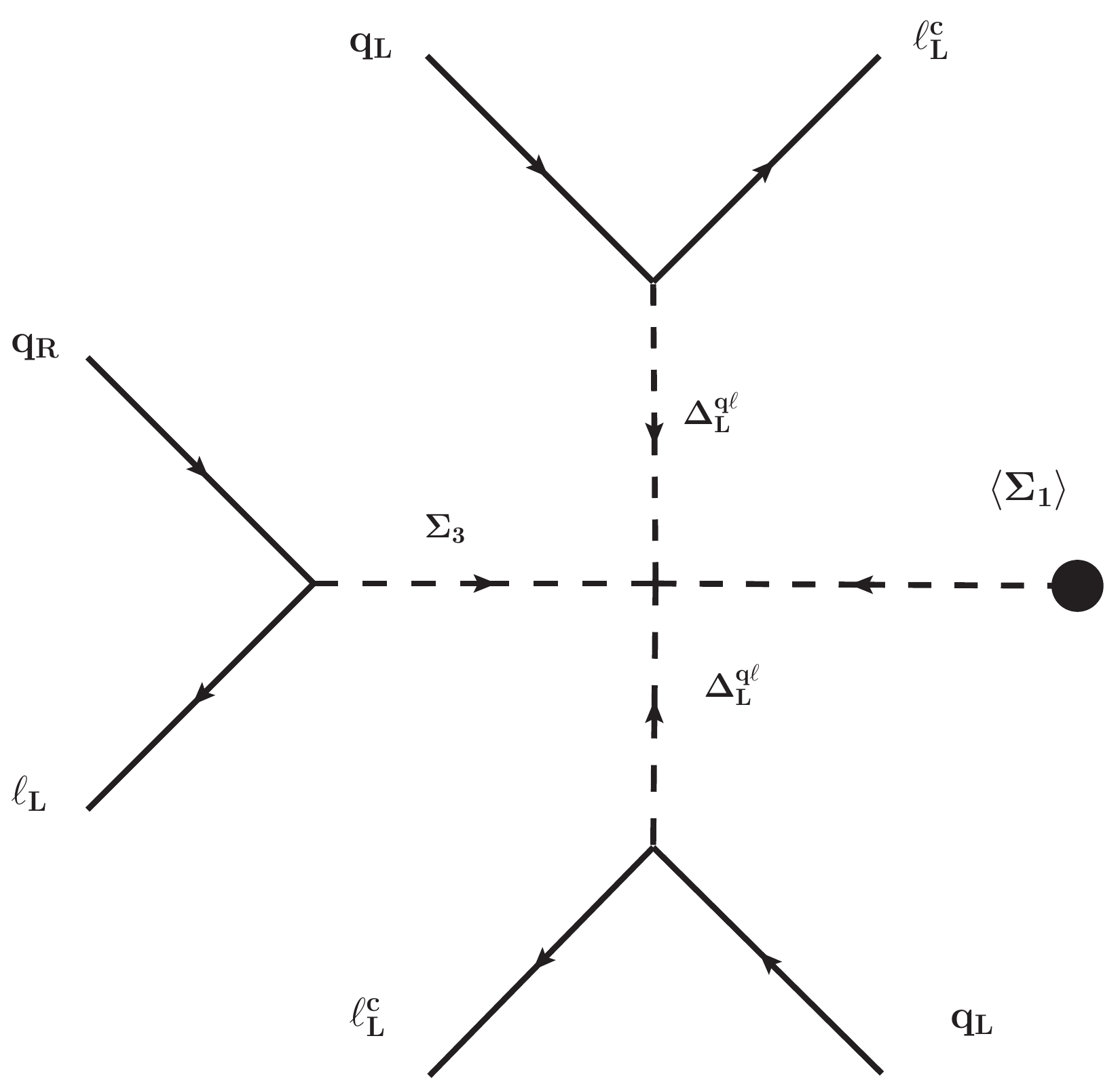}
\caption{Feynman diagrams for nucleon decay with the SM doublet VEV insertions. The left diagram induces nucleon decay processes like, nucleon $\rightarrow$ antilepton + mesons and the right digram, nucleon $\rightarrow$ lepton + antilepton + antilepton processes. }\label{d10}
\end{figure}

However,  three lepton decay processes of nucleon can take place in our model via the $d=10$ operators \cite{Rudaz:1992fi,ODonnell:1993kdg,Brahmachari:1994dj}. The Feynman diagrams corresponding to nucleon decay processes mediated by $d=10$ operators are shown in Fig. \ref{d10}.  These decay modes give rise to: nucleon$\to$ antilepton + meson and nucleon$\to$ lepton + antilepton+ antilepton.     Below we present the effective Lagrangians corresponding to  $d=9$ and $d=10$ and discuss the different nucleon decay modes and compute the branching fractions in certain approximations. For operator analysis regarding baryon and lepton number violation see Ref. \cite{Weinberg:1979sa,Weinberg:1980bf,Wilczek:1979hc,Weldon:1980gi}.
 
\vspace*{5pt}
\noindent
\textbf{$d=9$ proton decay} \\
\noindent
To write down terms  responsible for $d=9$ proton decay,  we expand the part of the scalar potential  that contains terms with quartic couplings:  $\tilde \gamma_{9R}$, $\tilde \gamma_{10R}$, $\tilde \eta_1$  in  Eq. \eqref{Vsigmadelta} in terms of the SM multiplets,

\begin{align}
&2 \tilde \gamma_{9R}^{\ast} v_R \epsilon_2 \epsilon_3\; [ \Delta^{\ast}_{R}(\overline 6,1,\frac{2}{3})\; \{  \Sigma^{\ast}(8,2,\frac{1}{2})\; \Sigma^{\ast}(\overline 3,2,-\frac{7}{6}) - \Sigma^{\ast}(8,2,-1/2) \;\Sigma^{\ast}(\overline 3,2,-\frac{1}{6})  \} 
\nonumber \\
&+ \frac{\Delta^{\ast}_{R}(\overline 3,1,4/3)}{\sqrt{2}}\; \{  \Sigma^{\ast}(\overline 3,2,-\frac{1}{6})\; \Sigma^{\ast}(\overline 3,2,-\frac{7}{6}) - \Sigma^{\ast}(\overline 3,2,-\frac{7}{6}) \;\Sigma^{\ast}(\overline 3,2,-\frac{1}{6})  \} ]  + h.c. \;,
\end{align}

\begin{align}
&-2 \tilde \gamma_{10R}^{\ast} v_R \epsilon_2 \epsilon_3\; [ \frac{\Delta^{\ast}_{R}(\overline 6,1,-\frac{1}{3})}{\sqrt 2}\;   \Sigma^{\ast}(\overline 3,2,-\frac{1}{6})\; \Sigma^{\ast}(8,2,\frac{1}{2}) 
+
 \Delta^{\ast}_{R}(\overline 6,1,\frac{2}{3})\;   \Sigma^{\ast}(\overline 3,2,-\frac{7}{6})\; \Sigma^{\ast}(8,2,\frac{1}{2})
\nonumber \\
&+
\frac{\Delta^{\ast}_{R}(\overline 3,1,\frac{1}{3})}{ 2}\;   \Sigma^{\ast}(\overline 3,2,-\frac{1}{6})\; \Sigma^{\ast}(\overline 3,2,-\frac{1}{6})
+
\frac{\Delta^{\ast}_{R}(\overline 3,1,\frac{4}{3})}{\sqrt 2}\;   \Sigma^{\ast}(\overline 3,2,-\frac{7}{6})\; \Sigma^{\ast}(\overline 3,2,-\frac{1}{6})
 ] 
 + h.c. \;,
\end{align}

\begin{align}
& \tilde \eta_{1}^{\ast} v_R  \epsilon_3\; [ \Sigma^{\ast}(\overline 3,2,-\frac{1}{6})\; \Sigma(8,2,\frac{1}{2}) \Delta^{\ast}_L(\overline 6,3,-\frac{1}{3}) + \Sigma^{\ast}(\overline 3,2,-\frac{1}{6})\; \Sigma(\overline 3,2,-\frac{1}{6}) \Delta^{\ast}_L(\overline 3,3,\frac{1}{3})  ] + h.c. \;,
\end{align}

\noindent   From these, the effective Lagrangian describing the $d=9$ six-fermion vertex that corresponds to nucleon decay can be written down,
 \vspace{-2pt}
 \begin{align}
 \mathcal{L}^{d=9}_{eff}= \mathcal{L}^{(a)}_{eff}+  \mathcal{L}^{(b)}_{eff}+  \mathcal{L}^{(c)}_{eff},
 \end{align}

\noindent with,  
\begin{align} 
\mathcal{L}^{(a)}_{eff}&= 
-(2 \widetilde{\gamma}_{9R} v_R) \epsilon^{\overline{•\mu}\overline{•\rho}\overline{•\lambda}} 
{Y^{\ast}_{15}}_{pq}{Y^{\ast}_{15}}_{kl}{Y^R_{10}}_{mn}
\left(
 \frac{•d^T_{R m \overline{\chi}} C d_{R n \overline{\lambda}}}{•m^2_{\Delta_{R (\overline{6},1,\frac{2}{3})}}} 
 \right)
\Biggl\{ 
\left(
 \frac{•\overline{•u}^{\overline{\chi}}_{R p} u_{L q \overline{\mu}}}{•m^2_{\Sigma_{(8,2,\frac{•1}{•2})}}} 
 \right)
 \left(
 \frac{•\overline{•e}_{R k} d_{L l \overline{\rho}}}{•m^2_{\Sigma_{{(\overline{3},2,-\frac{•7}{•6})}}}} 
 \right)
 \nonumber \\ &-
 \left(
 \frac{•\overline{•d}^{\overline{\chi}}_{R p} u_{L q \overline{\mu}}}{•m^2_{\Sigma_{{(8,2,-\frac{•1}{•2})}}}} 
 \right)
 \left(
 \frac{•\overline{•\nu}_{R k} d_{L l \overline{\rho}}}{•m^2_{\Sigma_{{(\overline{3},2,-\frac{•1}{•6})}}}} 
 \right)
 -
 \left(
 \frac{•\overline{•u}^{\overline{\chi}}_{R p} d_{L q \overline{\mu}}}{•m^2_{\Sigma_{{(8,2,\frac{•1}{•2})}}}} 
 \right)
 \left(
 \frac{•\overline{•e}_{R k} u_{L l \overline{\rho}}}{•m^2_{\Sigma_{{(\overline{3},2,-\frac{•7}{•6})}}}} 
 \right)
 +
 \left(
 \frac{•\overline{•d}^{\overline{\chi}}_{R p} d_{L q \overline{\mu}}}{•m^2_{\Sigma_{{(8,2,-\frac{•1}{•2})}}}} 
 \right)
 \left(
 \frac{•\overline{•\nu}_{R k} u_{L l \overline{\rho}}}{•m^2_{\Sigma_{{(\overline{3},2,-\frac{•1}{•6})}}}} 
 \right)
\Biggr\}
\nonumber \\  &+h.c\;, 
\end{align}

\begin{align}
\mathcal{L}^{(b)}_{eff}&=
-( \widetilde{\gamma}_{10R} v_R) \epsilon^{\overline{•\mu}\overline{•\rho}\overline{•\lambda}} 
{Y^{\ast}_{15}}_{pq}{Y^{\ast}_{15}}_{kl}{Y^R_{10}}_{mn}
\; \times
\nonumber \\& 
\Biggl\{ 
\left(
 \frac{•\overline{•u}^{\overline{\chi}}_{R p} u_{L q \overline{\mu}}}{•m^2_{\Sigma_{{(8,2,\frac{•1}{•2})}}}} 
 \right)
\left [
 \left(
 \frac{•\overline{•\nu}_{R k} d_{L l \overline{\rho}}}{•m^2_{\Sigma_{{(\overline{3},2,-\frac{•1}{•6})}}}} 
 \right)
\left(
 \frac{•d^T_{R m \overline{\chi}} C u_{R n \overline{\lambda}}+u^T_{R m \overline{\chi}} C d_{R n \overline{\lambda}}}{•m^2_{\Delta_{R (\overline{6},1,-\frac{•1}{•3})}}} 
 \right) 
 + 2
 \left(
 \frac{•\overline{•e}_{R k} d_{L l \overline{\rho}}}{•m^2_{\Sigma_{{(\overline{3},2,-\frac{•7}{•6})}}}} 
 \right)
\left(
 \frac{•d^T_{R m \overline{\chi}} C d_{R n \overline{\lambda}}}{•m^2_{\Delta_{R(\overline{6},1,\frac{•2}{•3})}}} 
 \right)
 \right]
\nonumber \\&  
 -
 \left(
 \frac{•\overline{•u}^{\overline{\chi}}_{R p} d_{L q \overline{\mu}}}{•m^2_{\Sigma_{{(8,2,\frac{•1}{•2})}}}} 
 \right)
 \left [
 \left(
 \frac{•\overline{•\nu}_{R k} u_{L l \overline{\rho}}}{•m^2_{\Sigma_{{(\overline{3},2,-\frac{•1}{•6})}}}} 
 \right)
\left(
 \frac{•d^T_{R m \overline{\chi}} C u_{R n \overline{\lambda}}+u^T_{R m \overline{\chi}} C d_{R n \overline{\lambda}}}{•m^2_{\Delta_{R(\overline{6},1,-\frac{•1}{•3})}}} 
 \right)  
 + 2
 \left(
 \frac{•\overline{•e}_{R k} u_{L l \overline{\rho}}}{•m^2_{\Sigma_{{(\overline{3},2,-\frac{•7}{•6})}}}} 
 \right)
\left(
 \frac{•d^T_{R m \overline{\chi}} C d_{R n \overline{\lambda}}}{•m^2_{\Delta_{R(\overline{6},1,\frac{•2}{•3})}}} 
 \right)
 \right]
 \nonumber \\&  
 +
 \left(
 \frac{•\overline{•\nu}_{R p} u_{L q \overline{\mu}}}{•m^2_{\Sigma_{{(\overline{3},2,-\frac{•1}{•6})}}}} 
 \right)
 \left [ \frac{1}{2}
 \left(
 \frac{•\overline{•\nu}_{R k} d_{L l \overline{\rho}}}{•m^2_{\Sigma_{{(\overline{3},2,-\frac{•1}{•6})}}}} 
 \right)
\left(
 \frac{•e^T_{R m} C u_{R n \overline{\lambda}}+\nu^T_{R m} C d_{R n \overline{\lambda}}}{•m^2_{\Delta_{R(\overline{3},1,\frac{•1}{•3})}}} 
 \right)  
 +
 \left(
 \frac{•\overline{•e}_{R k} d_{L l \overline{\rho}}}{•m^2_{\Sigma_{{(\overline{3},2,-\frac{•7}{•6})}}}} 
 \right)
\left(
 \frac{•e^T_{R m} C d_{R n \overline{\lambda}}}{•m^2_{\Delta_{R(\overline{3},1,\frac{•4}{•3})}}} 
 \right)
 \right]
 \nonumber \\&  
 -
 \left(
 \frac{•\overline{•\nu}_{R p} d_{L q \overline{\mu}}}{•m^2_{\Sigma_{{(\overline{3},2,-\frac{•1}{•6})}}}} 
 \right)
 \left [ \frac{1}{2}
 \left(
 \frac{•\overline{•\nu}_{R k} u_{L l \overline{\rho}}}{•m^2_{\Sigma_{{(\overline{3},2,-\frac{•1}{•6})}}}} 
 \right)
\left(
 \frac{•e^T_{R m} C u_{R n \overline{\lambda}}+\nu^T_{R m} C d_{R n \overline{\lambda}}}{•m^2_{\Delta_{R(\overline{3},1,\frac{•1}{•3})}}} 
 \right)  
 +
 \left(
 \frac{•\overline{•e}_{R k} u_{L l \overline{\rho}}}{•m^2_{\Sigma_{{(\overline{3},2,-\frac{•7}{•6})}}}} 
 \right)
\left(
 \frac{•e^T_{R m} C d_{R n \overline{\lambda}}}{•m^2_{\Delta_{R(\overline{3},1,\frac{•4}{•3})}}} 
 \right)
 \right]
\Biggr\} \nonumber \\ & +h.c\;, 
 \end{align}

\begin{align}
\mathcal{L}^{(c)}_{eff}&=
-(\widetilde{\eta}_1 v_R) \epsilon^{\overline{•\zeta}\overline{•\tau}\overline{•\chi}} 
{Y^{\ast}_{15}}_{pq}{Y_{15}}_{kl}{Y^L_{10}}_{mn}
\frac{1}{m^2_{\Sigma_{{(\overline{3},2,-\frac{•1}{•6})}}}}
\Biggl\{ 
\left(
 \overline{•\nu}_{R p} u_{L q \overline{\zeta}}
 \right)
\left(
 \frac{•\overline{•u}^{\overline{\rho}}_{L k} d_{R l \overline{\tau}}}{•m^2_{\Sigma_{{(8,2,\frac{•1}{•2})}}}} 
 \right)
\left(
 \frac{•d^T_{L m \overline{\rho}} C u_{L n \overline{\chi}}+u^T_{L m \overline{\rho}} C d_{L n \overline{\chi}}}{•m^2_{\Delta_{L(\overline{6},1,-\frac{•1}{•3})}}} 
 \right) 
 \nonumber \\&
 +
 \left(
 \overline{•\nu}_{R p} u_{L q \overline{\zeta}}
 \right)
\left(
 \frac{•\overline{•d}^{\overline{\rho}}_{L k} d_{R l \overline{\tau}}}{•m^2_{\Sigma_{{(8,2,\frac{•1}{•2})}}}} 
 \right)
\left(
 \frac{•d^T_{L m \overline{\rho}} C d_{L n \overline{\chi}}}{•m^2_{\Delta_{L(\overline{6},1,-\frac{•1}{•3})}}} 
 \right)
 -
 \left(
\overline{•\nu}_{R p} d_{L q \overline{\zeta}}
 \right)
\left(
 \frac{•\overline{•u}^{\overline{\rho}}_{L k} d_{R l \overline{\tau}}}{•m^2_{\Sigma_{{(8,2,\frac{•1}{•2})}}}} 
 \right)\;
\left(
 \frac{•u^T_{L m \overline{\rho}} C u_{L n \overline{\chi}}}{•m^2_{\Delta_{L(\overline{6},1,-\frac{•1}{•3})}}} 
 \right)
 \nonumber \\&
 +
 \left(
\overline{•\nu}_{R p} u_{L q \overline{\zeta}}
 \right)
\left(
 \frac{•\overline{•\nu}_{L k} d_{R l \overline{\tau}}}{•m^2_{\Sigma_{{(\overline{3},2,-\frac{•1}{•6})}}}} 
 \right)
\left(
 \frac{•e^T_{L m \overline{\rho}} C u_{L n \overline{\chi}}+\nu^T_{L m \overline{\rho}} C d_{L n \overline{\chi}}}{•m^2_{\Delta_{L(\overline{3},1,\frac{•1}{•3})}}} 
 \right)
 +
 \left(
 \overline{•\nu}_{R p} u_{L q \overline{\zeta}}
 \right)
\left(
 \frac{•\overline{•e}_{L k} d_{R l \overline{\tau}}}{•m^2_{\Sigma_{{(\overline{3},2,-\frac{•1}{•6})}}}} 
 \right)
\left(
 \frac{•e^T_{L m \overline{\rho}} C d_{L n \overline{\chi}}}{•m^2_{\Delta_{L(\overline{3},1,\frac{•1}{•3})}}} 
 \right)
 \nonumber \\&
 -
 \left(
 \overline{•\nu}_{R p} d_{L q \overline{\zeta}} 
 \right)
\left(
 \frac{•\overline{•\nu}_{L k} d_{R l \overline{\tau}}}{•m^2_{\Sigma_{{(\overline{3},2,-\frac{•1}{•6})}}}} 
 \right)
\left(
 \frac{•\nu^T_{L m \overline{\rho}} C u_{L n \overline{\chi}}}{•m^2_{\Delta_{L(\overline{3},1,\frac{•1}{•3})}}} 
 \right)
\Biggr\} +h.c\;, 
 \end{align}

\noindent here $k, l, m, n, p, q$ are the generation indices. The terms involving color octets mediate neutron decay via the channels $n \rightarrow \pi^+ e_R^-, K^+ e_R^-, \pi^+ \mu_R^-, K^+ \mu_R^-, \pi^0 \nu_R, K^0 \nu_R$ and proton decay via $p\rightarrow \pi^+ \nu_R, K^+ \nu_R$.  And the terms where the color triplets replacing the color octets, the decay modes are, $n \rightarrow {\nu_L}^c \nu_L \nu_R$, $e_R^+ e_R^- \nu_R$, $\mu_R^+ e_R^- \nu_R$, $e_R^+ \mu_R^- \nu_R$, $\mu_R^+ \mu_R^- \nu_R$, $e_L^+ e_L^- \nu_R$, $e_L^+ \mu_L^- \nu_R$, $\mu_L^+ e_L^- \nu_R$, $\mu_L^+ \mu_L^- \nu_R$ and $p \rightarrow e_R^+ \nu_R \nu_R$, $\mu_R^+ \nu_R \nu_R$, $e_L^+ \nu_L \nu_R$, $\mu_L^+ \nu_L \nu_R$.  There exist also terms that lead to three quark decay as aforementioned.

Note that for all the three lepton decay modes of the nucleon as well as, some of the two body decay modes  with the lepton being the neutrino, these decays can not be observed due to the additional suppressions of large right-handed neutrino mass. For the three lepton decay channels, always one of the leptons is a right-handed neutrino and for the two body decay channels with neutrino as the lepton, it is always the right-handed neutrino. This is not true in general within the PS framework. But in our model due to the additional $U(1)_{PQ}$ symmetry, $\bm\Sigma$ has coupling with $\overline{\bm \psi}_L \bm \psi_R$ and $\bm\Sigma^{\ast}$ has coupling with $\overline{\bm \psi}_R \bm \psi_L$, see Eq. \eqref{yukawa}. Also  PQ charge conservation does not allowed quartic terms of the form $\bm \Sigma^2 {\bm \Delta^{\ast}_R}^2$, rather allows term is of the form $\bm \Sigma^2 {\bm \Delta_R}^2$. The combined effect of these two facts restricts  nucleon decay modes containing only left-handed neutrinos in the final state in our set-up.   However, neutron decay into a lepton and a meson ($n \rightarrow  e_R^- \pi^+,  e_R^- K^+,  \mu_R^- \pi^+, \mu_R^- K^+$) can be within the observable range with specific choice of the parameter space. There will be similar modes of proton decay  ($p \rightarrow  e_R^- \pi^+ \pi^+,  e_R^- K^+ \pi^+,  \mu_R^- \pi^+ \pi^+,  \mu_R^- K^+ \pi^+$) with an additional pion in the final state and hence will be suppressed compared to neutron decay.

On the dimensional ground the decay rate of these $n \rightarrow$ lepton + meson processes is given by: 
\vspace*{-5pt}
\begin{align}
\Gamma^{d=9}_{n\to \ell+\rm{meson}} \sim \frac{1}{8 \pi} \left|\frac{v_R\; \Lambda^5_{QCD}}{M^6}\right|^2 m_p .
\end{align}

\noindent  Here $m_p$ is the mass of the proton and the mass of the Higgs bosons involved are taken to be of the same order and is denoted by $M$. While computing this decay rate,  the amplitude of such processes get multiplied by the factor $\Lambda_{QCD}^5$, here a factor of $\Lambda_{QCD}^3$ enters due to the hadronization of 3 quarks into a nucleon and a factor of $\Lambda_{QCD}^2$ comes into play due to the hadronization of $qq^c$ to a meson (for numerical computations, we take $\Lambda_{QCD}=170$ MeV). Assuming the Higgs bosons masses equal to the PS breaking scale, i.e, $M=v_R$, the decay rate ($\tau = \Gamma^{-1}$) of such processes to be within the observables range ($\tau \sim 10^{34}$ yrs) requires the PS breaking scale to be as low as  $v_R \sim 3.5\times 10^5$ GeV.

For high scale breaking of PS group, the nucleon decay is completely unobservable. On the other hand, low scale PS symmetry breaking can lead to observable nucleon decay modes. As mentioned above, the PS breaking scale can be as low as $10^3$ TeV. From the naive computation of decay rate performed above, it is clear that, for low scale PS breaking, if the    scalar masses that mediate the nucleon decay are somewhat smaller than the breaking scale, which can be made by some choice of the parameters of the theory,  these processes can be within the observable range.

On the other hand, decay rate of the $p \rightarrow$ lepton + mesons processes is given by: 
\vspace*{-5pt}
\begin{align}
\Gamma^{d=9}_{p\to \ell+\rm{mesons}} \sim \frac{1}{8 \pi} \left|\frac{v_R\; \Lambda^7_{QCD}}{M^6}\right|^2 m_p^{-3} .
\end{align}

\noindent The additional factor of $\Lambda_{QCD}^2$ is due to the presence of an extra pion in the final state. By a similar computation one finds that $v_R \sim 9.5\times 10^4$ GeV is required for such processes to be within the observable range. Again this required $v_R$ is computed naively, but in general there is no reason for the masses of the scalar fields that mediate nucleon decay to be degenerate with $v_R$. Even though additional suppression factor is present due to an extra pion in the final state, with some  choice of the model parameters can make these proton decay modes observable. \\

\vspace*{5pt}
\noindent
\textbf{$d=10$ proton decay} \\
\noindent
To write  down the $d=10$ proton decay operators, we also expand the relevant terms in the potential that have quartic couplings $\tilde \gamma_{9L}$ and $\tilde \gamma_{10L}$,

\begin{align}
&2 \tilde \gamma_{9L}^{\ast} \epsilon_3\;\; [ v_1\;
\{
\Sigma_{\alpha=2}(8,2,\frac{1}{2})\; \Delta^{\ast\;\;\gamma}_{L\;\kappa}(\overline 3,3,\frac{1}{3})\; \Delta^{\ast\;\;\kappa}_{L\;\gamma}(\overline 6,3,-\frac{1}{3}) 
+ 
\Sigma_{\alpha=2}(\overline 3,2,-\frac{1}{6})\; \Delta^{\ast\;\;\gamma}_{L\;\kappa}(\overline 3,3,\frac{1}{3})\; \Delta^{\ast\;\;\kappa}_{L\;\gamma}(\overline 3,3,\frac{1}{3}) 
\}
\nonumber \\
&+
v_2\;
\{
\Sigma_{\alpha=1}(8,2,-\frac{1}{2})\; \Delta^{\ast\;\;\gamma}_{L\;\kappa}(\overline 3,3,\frac{1}{3})\; \Delta^{\ast\;\;\kappa}_{L\;\gamma}(\overline 6,3,-\frac{1}{3}) 
+ 
\Sigma_{\alpha=1}(\overline 3,2,-\frac{7}{6})\; \Delta^{\ast\;\;\gamma}_{L\;\kappa}(\overline 3,3,\frac{1}{3})\; \Delta^{\ast\;\;\kappa}_{L\;\gamma}(\overline 3,3,\frac{1}{3}) 
\}
 ] + h.c.,
\end{align}

\begin{align}
& \tilde \gamma_{10L}^{\ast} \epsilon_3\;\; [
 v_1\;
\{
\Sigma_{\beta}(8,2,\frac{1}{2})\; \Delta^{\ast\;\;\gamma}_{L\;\alpha=2}(\overline 3,3,\frac{1}{3})\; \Delta^{\ast\;\; \beta}_{L\;\gamma}(\overline 6,3,-\frac{1}{3}) 
+ 
\Sigma_{\beta}(\overline 3,2,-\frac{1}{6})\; \Delta^{\ast\;\;\gamma}_{L\;\alpha=2}(\overline 3,3,\frac{1}{3})\; \Delta^{\ast\;\; \beta}_{L\;\gamma}(\overline 3,3,\frac{1}{3}) 
\}
\nonumber \\
&+ 
 v_1\;
\{
\Sigma_{\alpha}(8,2,\frac{1}{2})\; \Delta^{\ast\;\;\gamma}_{L\;\kappa}(\overline 6,3,-\frac{1}{3})\; \Delta^{\ast\;\; \beta=1}_{L\;\gamma}(\overline 3,3,\frac{1}{3}) 
+ 
\Sigma_{\alpha}(\overline 3,2,-\frac{1}{6})\; \Delta^{\ast\;\;\gamma}_{L\;\kappa}(\overline 3,3,\frac{1}{3})\; \Delta^{\ast\;\; \beta=1}_{L\;\gamma}(\overline 3,3,\frac{1}{3}) 
\} \epsilon^{\alpha \kappa}
\nonumber \\
&+
 v_2\;
\{
\Sigma_{\beta}(8,2,-\frac{1}{2})\; \Delta^{\ast\;\;\gamma}_{L\;\alpha=1}(\overline 3,3,\frac{1}{3})\; \Delta^{\ast\;\; \beta}_{L\;\gamma}(\overline 6,3,-\frac{1}{3}) 
+ 
\Sigma_{\beta}(\overline 3,2,-\frac{7}{6})\; \Delta^{\ast\;\;\gamma}_{L\;\alpha=1}(\overline 3,3,\frac{1}{3})\; \Delta^{\ast\;\; \beta}_{L\;\gamma}(\overline 3,3,\frac{1}{3}) 
\}
\nonumber \\
&+ 
 v_2\;
\{
\Sigma_{\alpha}(8,2,\frac{1}{2})\; \Delta^{\ast\;\;\gamma}_{L\;\kappa}(\overline 6,3,-\frac{1}{3})\; \Delta^{\ast\;\; \beta=1}_{L\;\gamma}(\overline 3,3,\frac{1}{3}) 
+ 
\Sigma_{\alpha}(\overline 3,2,-\frac{7}{6})\; \Delta^{\ast\;\;\gamma}_{L\;\kappa}(\overline 3,3,\frac{1}{3})\; \Delta^{\ast\;\; \beta=1}_{L\;\gamma}(\overline 3,3,\frac{1}{3}) 
\} \epsilon^{\alpha \kappa}
] \nonumber \\
& + h.c.,
\end{align}

\noindent
Here $v_1\equiv \langle \Sigma^{4\;\dot 1}_{4\;1}\rangle$ and $v_2\equiv \langle \Sigma^{4\;\dot 2}_{4\;2}\rangle$ are  the electroweak scale VEVs of the bi-doublet coming from (2,2,15) multiplet. Note, since these terms are generated due to the electroweak symmetry breaking,  $SU(2)$ indices  are shown explicitly.  The effective Lagrangian describing the $d=10$ six-fermion vertex that corresponds to nucleon decay can be written down,
 \vspace{-2pt}
 \begin{align}
 \mathcal{L}^{d=10}_{eff}= \mathcal{L}^{(d)}_{eff}+  \mathcal{L}^{(e)}_{eff},
 \end{align}

\noindent with, 
\begin{align} 
\mathcal{L}^{(d)}_{eff}&= 
(\widetilde{\gamma}_{9L} ) 
\epsilon^{\overline{•\tau}\overline{•\lambda}
\overline{•\chi}} 
{Y_{15}}_{pq}{Y^{L}_{10}}_{kl}{Y^L_{10}}_{mn}
\Biggl\{ 
\frac{1}{•m^2_{\Delta_{L(\overline{3},3,\frac{1}{3})}} m^2_{\Delta_{L(\overline{6},3,-\frac{1}{3})}}} 
\left[
v_2
\left(
 \frac{•\overline{•u}_{L p}^{\overline{\rho}} u_{R q \overline{\tau}}}{•m^2_{\Sigma_{{(8,2,-\frac{•1}{•2})}}}} 
 \right)
 +
 v_1
 \left(
 \frac{•\overline{•d}_{L p}^{\overline{\rho}} d_{R q \overline{\tau}}}{•m^2_{\Sigma_{{(8,2,\frac{•1}{•2})}}}} 
 \right)
\right]
\nonumber \\&
\left[2
\left( 
 e^T_{L k} C u_{L l \overline{\lambda}}
 \right)
\left(
 d^T_{L m \overline{\rho}} C u_{L n \overline{\chi}}+u^T_{L m \overline{\rho}} C d_{L n \overline{\chi}} 
 \right)
 -
 \left(
 e^T_{L k} C d_{L l \overline{\lambda}}
 \right)
\left(
u^T_{L m \overline{\rho}} C u_{L n \overline{\chi}}
 \right)
 + 2
 \left(
 \nu^T_{L k} C d_{L l \overline{\lambda}}
 \right)
\left(
 d^T_{L m \overline{\rho}} C d_{L n \overline{\chi}}
 \right)
  \right. 
\nonumber  \\&+
 \left.  2
 \left(
 \nu^T_{L k} C d_{L l \overline{\lambda}}
 \right)
\left(
 u^T_{L m \overline{\rho}} C d_{L n \overline{\chi}} + d^T_{L m \overline{\rho}} C u_{L n \overline{\chi}} 
 \right)
\right]
+
\frac{1}{•m^4_{\Delta_{L(\overline{3},3,\frac{1}{3})}}}
\left[
v_2
\left(
 \frac{•\overline{•\nu}_{L p} u_{R q \overline{\tau}}}{•m^2_{\Sigma_{{(\overline{3},2,-\frac{•7}{•6})}}}} 
 \right)
 + v_1
 \left(
 \frac{•\overline{•e}_{L p} d_{R q \overline{\tau}}}{•m^2_{\Sigma_{{(\overline{3},2,-\frac{•1}{•6})}}}} 
 \right)
\right]
\nonumber \\&
2 \left[
\left(
 e^T_{L k} C u_{L l \overline{\lambda}}
 \right)
\left(
 e^T_{L m} C u_{L n \overline{\chi}}
 \right) 
 +
 \left(
 e^T_{L k} C d_{L l \overline{\lambda}}
 \right)
\left(
 \nu^T_{L m} C u_{L n \overline{\chi}}
 \right)
 +
 \left(
 \nu^T_{L k} C d_{L l \overline{\lambda}}
 \right)
\left(
\nu^T_{L m \overline{\rho}} C d_{L n \overline{\chi}} 
 \right)
\right]
\Biggr\} +h.c\;, 
\end{align}

\noindent and  $\mathcal{L}^{(e)}=\sum_i \mathcal{L}^{(e_i)}$ with $i=1-4$, where,

\begin{align}
\mathcal{L}^{(e_1)}_{eff}&= 
(\widetilde{\gamma}_{10L} v_{1}) 
\epsilon^{\overline{•\tau}\overline{•\lambda}
\overline{•\chi}} 
{Y_{15}}_{pq}{Y^{L}_{10}}_{kl}{Y^L_{10}}_{mn}
\; \times
\nonumber \\&
\Biggl\{ 
\frac{1}{m^2_{\Sigma_{{(8,2,\frac{•1}{•2})}}}
m^2_{\Delta_{L(\overline{3},3,\frac{1}{3})}}
m^2_{\Delta_{L(\overline{6},3,-\frac{1}{3})}}}
\{
\left(
\overline{•u}_{L p}^{\overline{\rho}} d_{R q \overline{\tau}} 
 \right)
 \left[
\left(
 \nu^T_{L k} C u_{L l \overline{\lambda}}
 \right)
\left(
d^T_{L m \overline{\rho}} C u_{L n \overline{\chi}}
 \right) 
 +
 \left(
 \nu^T_{L k} C d_{L l \overline{\lambda}} 
 \right)
\left(
u^T_{L m \overline{\rho}} C u_{L n \overline{\chi}}
 \right)
 \right]
\nonumber \\&
+
\left(
 \overline{•d}_{L p}^{\overline{\rho}} d_{R q \overline{\tau}} 
 \right)
 \left[
\left(
 \nu^T_{L k} C u_{L l \overline{\lambda}} 
 \right)
\left(
 d^T_{L m \overline{\rho}} C d_{L n \overline{\chi}}
 \right) 
 +
 \left(
 \nu^T_{L k} C d_{L l \overline{\lambda}}
 \right)
\left(
u^T_{L m \overline{\rho}} C d_{L n \overline{\chi}}
 \right)
 \right]
 \}
\nonumber \\&
+
\frac{1}{m^2_{\Sigma_{{(\overline{3},2,-\frac{•1}{•6})}}}
m^4_{\Delta_{L(\overline{3},3,\frac{1}{3})}}} 
\{
 \left(
\overline{•\nu}_{L p} d_{R q \overline{\tau}}
 \right)
 \left[
\left(
\nu^T_{L k} C u_{L l \overline{\lambda}}
 \right)
 \left(
 e^T_{L k} C u_{L l \overline{\lambda}} 
 \right)
 +
 \left(
 \nu^T_{L k} C d_{L l \overline{\lambda}} 
 \right)
 \left(
\nu^T_{L k} C u_{L l \overline{\lambda}} 
 \right)
 \right]
\nonumber \\&
 +
 \left(
\overline{•e}_{L p} d_{R q \overline{\tau}} 
 \right)
 \left[
\left(
 \nu^T_{L k} C u_{L l \overline{\lambda}} 
 \right)
 \left(
 e^T_{L k} C d_{L l \overline{\lambda}} 
 \right)
 +
 \left(
 \nu^T_{L k} C d_{L l \overline{\lambda}} 
 \right)
 \left(
\nu^T_{L k} C d_{L l \overline{\lambda}}
 \right)
 \right]
 \}
\Biggr\} +h.c\;, 
\end{align}

\begin{align}
\mathcal{L}^{(e_2)}_{eff}&= 
(\widetilde{\gamma}_{10L} v_{2})  
\epsilon^{\overline{•\tau}\overline{•\lambda}
\overline{•\chi}} 
{Y_{15}}_{pq}{Y^{L}_{10}}_{kl}{Y^L_{10}}_{mn}
\; \times
\nonumber \\&
\Biggl\{ 
\frac{1}{m^2_{\Sigma_{{(8,2,-\frac{•1}{•2})}}}
m^2_{\Delta_{L(\overline{3},3,\frac{1}{3})}}
m^2_{\Delta_{L(\overline{6},3,-\frac{1}{3})}}
}
\{
\left(
 \overline{•u}_{L p}^{\overline{\rho}} u_{R q \overline{\tau}}
 \right)
 \left[
\left(
 e^T_{L k} C u_{L l \overline{\lambda}} 
 \right)
\left(
 d^T_{L m \overline{\rho}} C u_{L n \overline{\chi}}
 \right) 
 +
 \left(
 e^T_{L k} C d_{L l \overline{\lambda}}
 \right)
\left(
 u^T_{L m \overline{\rho}} C u_{L n \overline{\chi}}
 \right)
 \right]
\nonumber \\&
+
\left(
 \overline{•d}_{L p}^{\overline{\rho}} u_{R q \overline{\tau}}
 \right)
 \left[
\left(
 e^T_{L k} C u_{L l \overline{\lambda}}
 \right)
\left(
 d^T_{L m \overline{\rho}} C d_{L n \overline{\chi}}
 \right) 
 +
 \left(
 e^T_{L k} C d_{L l \overline{\lambda}}
 \right)
\left(
 u^T_{L m \overline{\rho}} C d_{L n \overline{\chi}}
 \right)
 \right]
 \}
\nonumber \\&
 +
 \frac{1}{m^2_{\Sigma_{{(\overline{3},2,-\frac{•7}{•6})}}}
m^4_{\Delta_{L(\overline{3},3,\frac{1}{3})}} 
 }
 \{
 \left(
 \overline{•\nu}_{L p} u_{R q \overline{\tau}}
 \right)
 \left[
\left(
 e^T_{L k} C u_{L l \overline{\lambda}}
 \right)
 \left(
 e^T_{L k} C u_{L l \overline{\lambda}} 
 \right)
 +
 \left(
 e^T_{L k} C d_{L l \overline{\lambda}} 
 \right)
 \left(
 \nu^T_{L k} C u_{L l \overline{\lambda}}
 \right)
 \right]
\nonumber \\&
 +
 \left(
 \overline{•e}_{L p} u_{R q \overline{\tau}} 
 \right)
 \left[
\left(
 e^T_{L k} C u_{L l \overline{\lambda}}
 \right)
 \left(
 e^T_{L k} C d_{L l \overline{\lambda}} 
 \right)
 +
 \left(
 e^T_{L k} C d_{L l \overline{\lambda}}
 \right)
 \left(
 \nu^T_{L k} C d_{L l \overline{\lambda}} 
 \right)
 \right]
 \}
\Biggr\} +h.c\;, 
\end{align}

\begin{align}
\mathcal{L}^{(e_3)}_{eff}&= 
(\widetilde{\gamma}_{10L} v_{1}) 
\epsilon^{\overline{•\tau}\overline{•\lambda}
\overline{•\chi}} 
{Y_{15}}_{pq}{Y^{L}_{10}}_{kl}{Y^L_{10}}_{mn}
\; \times
\nonumber \\&
\Biggl\{ 
\frac{1}{m^2_{\Sigma_{{(8,2,\frac{•1}{•2})}}}
m^2_{\Delta_{L(\overline{3},3,\frac{1}{3})}}
m^2_{\Delta_{L(\overline{6},3,-\frac{1}{3})}}
}
\{
\left(
 \overline{•u}_{L p}^{\overline{\rho}} d_{R q \overline{\tau}} 
 \right)
 \left[
\left(
 e^T_{L k} C u_{L l \overline{\lambda}}
 \right)
\left(
 u^T_{L m \overline{\rho}} C u_{L n \overline{\chi}}
 \right) 
 +
 \left(
 \nu^T_{L k} C u_{L l \overline{\lambda}} 
 \right)
\left(
 u^T_{L m \overline{\rho}} C d_{L n \overline{\chi}}
 \right)
 \right]
\nonumber\\&
-
\left(
\overline{•d}_{L p}^{\overline{\rho}} d_{R q \overline{\tau}} 
 \right)
 \left[
\left(
e^T_{L k} C u_{L l \overline{\lambda}}
 \right)
\left(
 d^T_{L m \overline{\rho}} C u_{L n \overline{\chi}}
 \right) 
 +
 \left(
 \nu^T_{L k} C u_{L l \overline{\lambda}}
 \right)
\left(
 d^T_{L m \overline{\rho}} C d_{L n \overline{\chi}}
 \right)
 \right]
 \}
\nonumber \\&
 +
 \frac{1}{m^2_{\Sigma_{{(\overline{3},2,-\frac{•1}{•6})}}}
 m^4_{\Delta_{L(\overline{3},3,\frac{1}{3})}}
 }
 \{
 \left(
 \overline{•\nu}_{L p} d_{R q \overline{\tau}} 
 \right)
 \left[
\left(
 \nu^T_{L k} C u_{L l \overline{\lambda}} 
 \right)
 \left(
 e^T_{L k} C u_{L l \overline{\lambda}}
 \right)
 +
 \left(
 \nu^T_{L k} C d_{L l \overline{\lambda}} 
 \right)
 \left(
 \nu^T_{L k} C u_{L l \overline{\lambda}} 
 \right)
 \right]
\nonumber \\&
 +
 \left(
 \overline{•e}_{L p} d_{R q \overline{\tau}} 
 \right)
 \left[
\left(
 e^T_{L k} C u_{L l \overline{\lambda}} 
 \right)
 \left(
 e^T_{L k} C u_{L l \overline{\lambda}} 
 \right)
 +
 \left(
 e^T_{L k} C d_{L l \overline{\lambda}} 
 \right)
 \left(
 \nu^T_{L k} C u_{L l \overline{\lambda}} 
 \right)
 \right]
 \}
\Biggr\} +h.c\;, 
\end{align}

\begin{align}
\mathcal{L}^{(e_4)}_{eff}&= 
(\widetilde{\gamma}_{10L} v_{2}) 
\epsilon^{\overline{•\tau}\overline{•\lambda}
\overline{•\chi}} 
{Y_{15}}_{pq}{Y^{L}_{10}}_{kl}{Y^L_{10}}_{mn}
\; \times
\nonumber \\&
\Biggl\{ 
\frac{1}{m^2_{\Sigma_{{(8,2,-\frac{•1}{•2})}}}
m^2_{\Delta_{L(\overline{3},3,\frac{1}{3})}}
m^2_{\Delta_{L(\overline{6},3,-\frac{1}{3})}}
}
\{
\left(
 \overline{•u}_{L p}^{\overline{\rho}} u_{R q \overline{\tau}}
 \right)
 \left[
\left(
 e^T_{L k} C d_{L l \overline{\lambda}} 
 \right)
\left(
 u^T_{L m \overline{\rho}} C u_{L n \overline{\chi}}
 \right) 
 +
 \left(
 \nu^T_{L k} C d_{L l \overline{\lambda}} 
 \right)
\left(
 u^T_{L m \overline{\rho}} C d_{L n \overline{\chi}}
 \right)
 \right]
\nonumber\\&
-
\left(
 \overline{•d}_{L p}^{\overline{\rho}} u_{R q \overline{\tau}}
 \right)
 \left[
\left(
 e^T_{L k} C d_{L l \overline{\lambda}} 
 \right)
\left(
 d^T_{L m \overline{\rho}} C u_{L n \overline{\chi}}
 \right) 
 +
 \left(
 e^T_{L k} C u_{L l \overline{\lambda}} 
 \right)
\left(
 d^T_{L m \overline{\rho}} C d_{L n \overline{\chi}}
 \right)
 \right]
 \}
\nonumber \\&
 +
 \frac{1}{m^2_{\Sigma_{{(\overline{3},2,-\frac{•7}{•6})}}}
 m^4_{\Delta_{L(\overline{3},3,\frac{1}{3})}}
 }
 \{
 \left(
 \overline{•\nu}_{L p} u_{R q \overline{\tau}} 
 \right)
 \left[
\left(
 \nu^T_{L k} C u_{L l \overline{\lambda}} 
 \right)
 \left(
 e^T_{L k} C d_{L l \overline{\lambda}} 
 \right)
 +
 \left(
 \nu^T_{L k} C d_{L l \overline{\lambda}}
 \right)
 \left(
 \nu^T_{L k} C d_{L l \overline{\lambda}}
 \right)
 \right]
\nonumber \\&
 +
 \left(
 \overline{•e}_{L p} u_{R q \overline{\tau}} 
 \right)
 \left[
\left(
 e^T_{L k} C u_{L l \overline{\lambda}}
 \right)
 \left(
 e^T_{L k} C d_{L l \overline{\lambda}}
 \right)
 +
 \left(
 e^T_{L k} C d_{L l \overline{\lambda}}
 \right)
 \left(
 \nu^T_{L k} C d_{L l \overline{\lambda}}
 \right)
 \right]
 \}
\Biggr\} +h.c\;, 
\end{align}

\noindent  The terms involving color octets mediate neutron decay via the channels $n \rightarrow {\nu_L}^c \pi^0, e^+_L \pi^-$, $\mu^+_L \pi^-, {\nu_L}^c K^0, K^- e^+_L, K^- \mu^+_L$ and proton decay via $p\rightarrow {\nu_L}^c \pi^+, e^+_L \pi^0, {\nu_L}^c K^+, e^+_L K^0, \mu^+_L \pi^0, \mu^+_L K^0$. And the terms where the color triplets replacing color octets, the decay modes are, $n \rightarrow \nu_L {\nu_L}^c {\nu_L}^c$, $e^-_L e^+_L {\nu_L}^c$, $e^-_L \mu^+_L {\nu_L}^c$, $\mu^-_L e^+_L {\nu_L}^c$, $\mu^-_L \mu^+_L {\nu_L}^c$ and $p \rightarrow e_L^+ \nu_L {\nu_L}^c$, $\mu_L^+ \nu_L {\nu_L}^c$, $e_L^- e_L^+ e_L^+$, $\mu_L^- e_L^+ e_L^+$, $\mu_L^- \mu_L^+ e_L^+$, $\mu_L^- \mu_L^+ \mu_L^+$.

Six fermion vertex $d=9$ nucleon decay operators mediate processes like $n\rightarrow$ lepton + meson and $p\rightarrow$ lepton + mesons, whereas   $n\rightarrow$ antilepton + meson and $p\rightarrow$ antilepton + meson processes arise through $d=10$ six fermion vertex operators. $d=10$ operators also induce processes with three lepton final state, which is not the case with $d=9$. The decay width for processes like $n,p\rightarrow$ antilepton + meson is:
\vspace*{-5pt}
\begin{align}
\Gamma^{d=10}_{n,p\to \ell^c+\rm{meson}} \sim \frac{1}{8 \pi} \left|\frac{v_{ew}\; \Lambda^5_{QCD}}{M^6}\right|^2 m_p ,
\end{align}

\noindent and for the three lepton final state processes is:
\vspace*{-5pt}
\begin{align}
\Gamma^{d=10}_{n,p\to \ell \ell^c \ell^c} \sim \frac{1}{256 \pi^3} \left|\frac{v_{ew}\; \Lambda^3_{QCD}}{M^6}\right|^2 m_p^5 .
\end{align}

\noindent  For $n,p\rightarrow$ antilepton + meson to be within the observable range ($\tau \sim 10^{34}$ yrs   \cite{Nishino:2012bnw}), the requirement on the PS scale is $v_R \sim 10^5$ GeV.  The three lepton final state also requires $v_R \sim 10^5$ GeV (here $\tau \sim 10^{33}$ yrs   \cite{Takhistov:2014pfw}). Again, as mentioned above, by some choice of  the quartic couplings involved in these decay rate of these processes can simultaneously satisfy the lower bound of the  PS breaking by making the masses of these scalars somewhat smaller than $v_R$,  but still be in the interesting observable range.  \\

\vspace*{5pt}
\noindent
\textbf{Nucleon decay relative branching fractions} \\
\noindent
By using the formulae as aforementioned  one can compare  the decay widths of the different  modes. A naive estimation of the relative branching fractions reveal 
\vspace*{0pt}
\begin{align}
&\frac{\Gamma^{d=9}_{p\to \ell + mesons}}{\Gamma^{d=9}_{n\to \ell + meson}}\sim \frac{\Lambda^4_{QCD}}{m^4_p}  \sim 10^{-3},
\\&
\frac{\Gamma^{d=10}_{n,p\to \ell^c + meson}}{\Gamma^{d=9}_{n\to \ell + meson}}\sim \frac{v^2_{ew}}{v^2_R}  \sim  10^{-8} ,
\\&
\frac{\Gamma^{d=10}_{n,p\to \ell \ell^c \ell^c }}{\Gamma^{d=9}_{n\to \ell + meson}}\sim \frac{1}{32 \pi^2}\frac{v^2_{ew}}{v^2_R} \frac{m^4_p}{\Lambda^4_{QCD}}  \sim  10^{-7} ,
\\&
\frac{\Gamma^{d=10}_{n,p\to \ell^c + meson}}{\Gamma^{d=9}_{p\to \ell + mesons}}\sim \frac{v^2_{ew}}{v^2_R}  \frac{m^4_p}{\Lambda^4_{QCD}}   \sim 10^{-5} ,
\\&
\frac{\Gamma^{d=10}_{n,p\to \ell \ell^c \ell^c}}{\Gamma^{d=9}_{p\to \ell +mesons}}\sim  \frac{1}{32 \pi^2} \frac{v^2_{ew}}{v^2_R} \frac{m^8_p}{\Lambda^8_{QCD}} \sim 10^{-4},
\\&
\frac{\Gamma^{d=10}_{n,p\to  \ell^c +meson}}{\Gamma^{d=10}_{n,p\to \ell \ell^c \ell^c}}\sim 32 \pi^2
 \frac{\Lambda^4_{QCD}}{m^4_p}   \sim 0.34 .
\end{align}

\noindent Here we have chosen $v_R= 10^6$ GeV. This estimation shows that for the $d=9$ case, neutron decay will be dominating over the proton decay due the the presence of extra pion in the final state for the proton decay modes. Again $d=10$ processes are suppressed compared to the $d=9$ processes due to the extra suppression factor of $v^2_{ew}/v^2_R$. We remind the readers that these results are not general, since  the Higgs boson mass spectrum is expected to be non-degenerate and appropriate  hierarchical pattern can be realized to make these two processes comparable.  Also note that the details of the structure of the Yukawa couplings are ignored for this analysis. Since nucleon decay processes involve more than one quartic coupling, definite predictions about the relative  branching fractions of different decay channels can not be firmly predicted.

\vspace*{5pt}
\noindent
\textbf{Comment on $d=7$ B-violating  operators} \\
\noindent In unified theories, another interesting B-violating  operators involving Higgs bosons that can mediate nucleon decay correspond to the case of $d=7$.  In addition to the leptoquark color triplets present in our theory, if also diquark color triplets exist, then $d=7$ operators can mediate nucleon decay. For example, quartic terms in the Higgs potential involving a triplet leptoquark, a triplet diquark, a Higgs doublet and the neutral component from $\bm\Delta_R$  is responsible for generating nucleon decay processes \cite{Babu:2012vb} when the $B-L$ violating VEV of $\bm\Delta_R$ is inserted. In our minimal model due to the absence of diquark color triplets, $d=7$ operators are not present.

\subsection{\texorpdfstring{$n-\overline{n}$}{TEXT} Oscillation}
\begin{figure}[th!]
\centering
\includegraphics[scale=0.45]{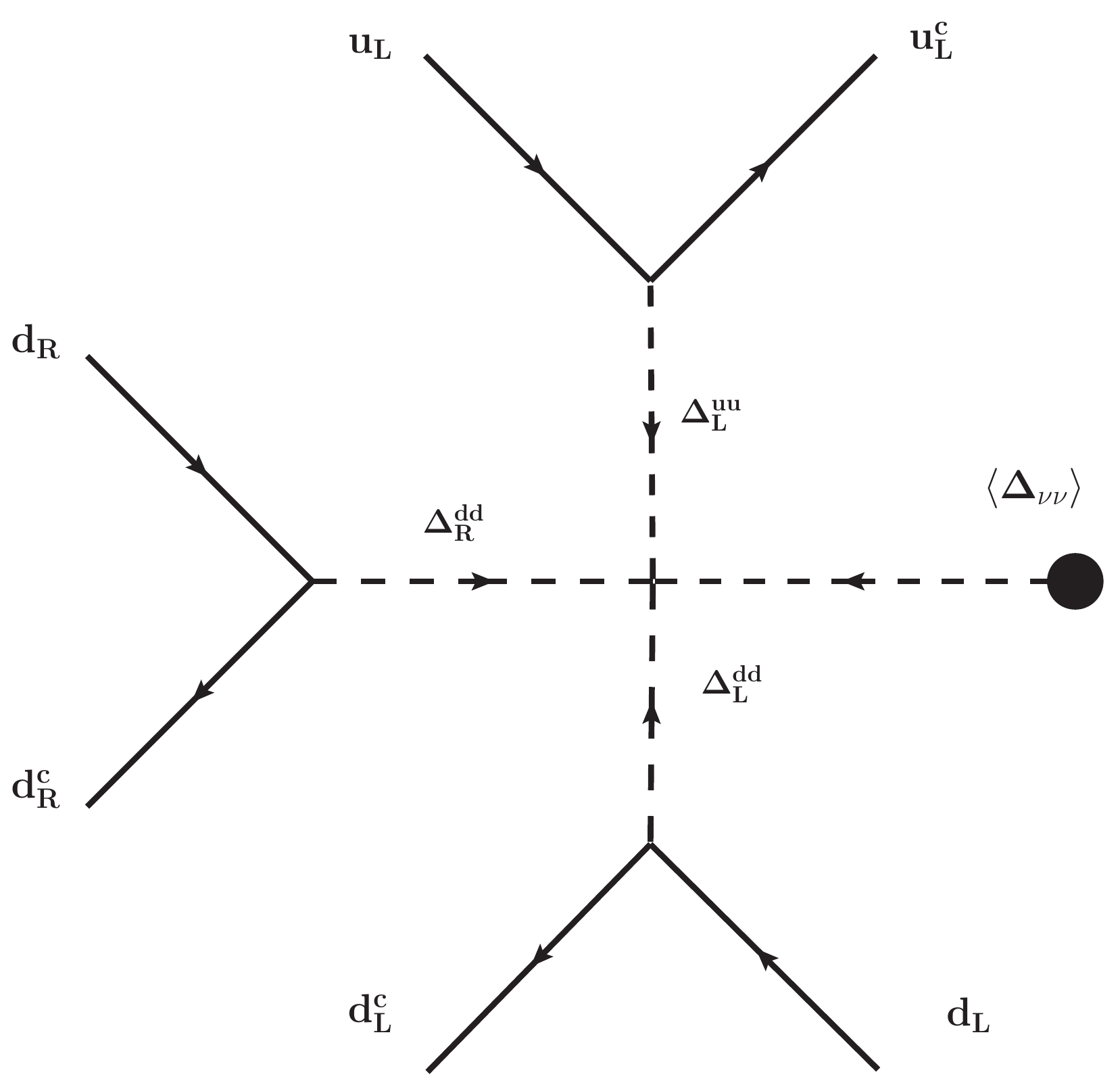}\hspace{10pt}
\includegraphics[scale=0.45]{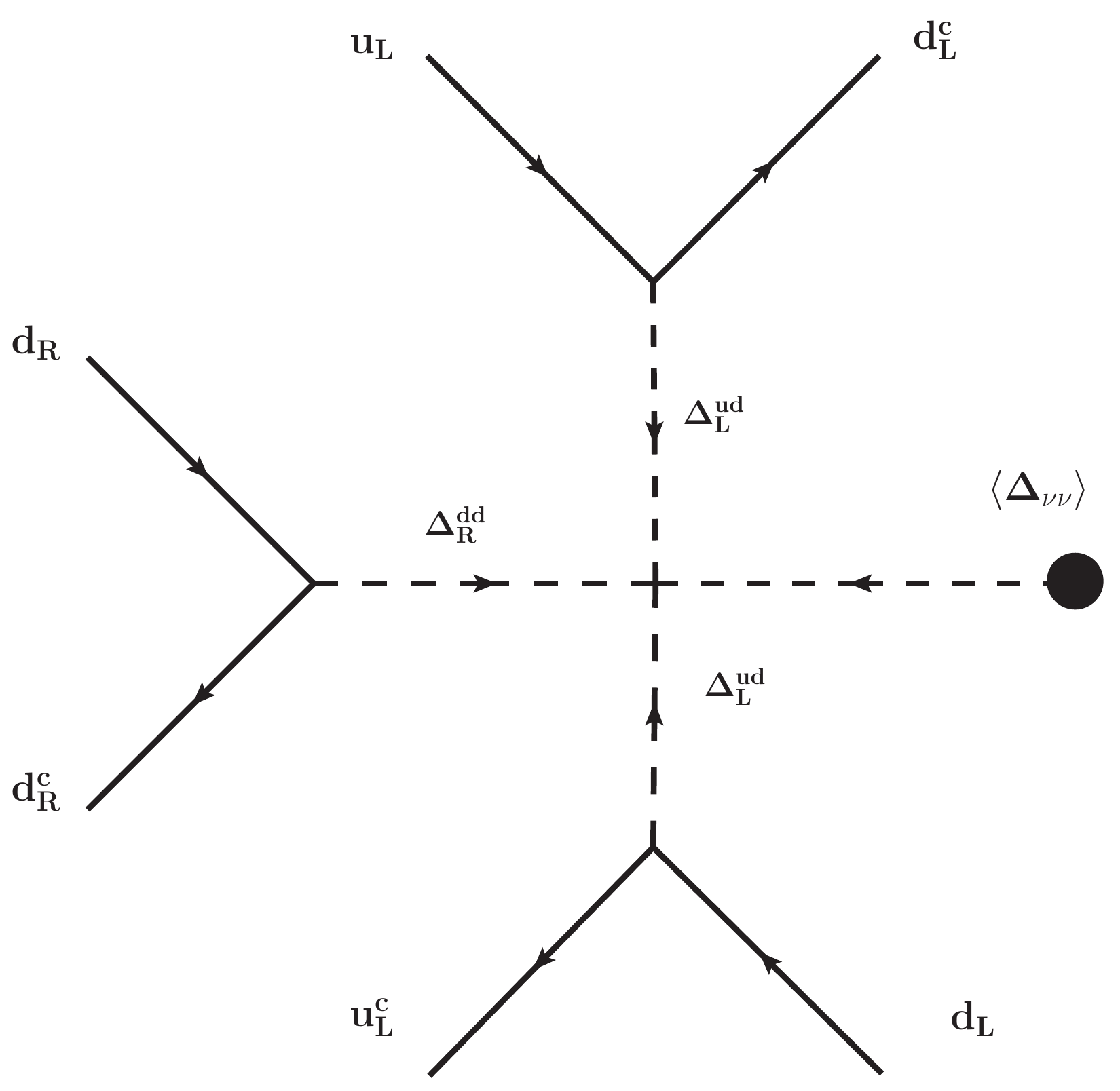}
\caption{Feynman diagrams for $n-\overline{n}$ oscillation. }\label{nnbar}
\end{figure}

Another phenomenologically interesting process that can take place in PS model is the $\Delta B=2$ interactions that can give rise to $n-\overline{n}$ oscillation. A PS model with the presence of only $\Delta_R$(1,3,10)  scalar  can have nucleon transition at the tree level that includes six-fermion $\Delta B=2$ vertex \cite{Mohapatra:1980qe,Phillips:2014fgb}. Such transitions are again led by a specific type of term in the scalar potential and has the  form $\bm\Delta_R^4$. Note that, due to the additional $U(1)_{PQ}$ symmetry, terms of this form are not present in our theory since this field carries non-zero PQ charge. However, there is a quartic term involving both $\Delta_R$ and $\Delta_L$  in our potential which is,

\vspace{-15pt}
\begin{align}\label{nantin}
V_{\Delta} \supset \widetilde{\lambda_{9}}\; \bm\Delta_{R\mu \nu \;\dot{\alpha}}^{\;\;\;\;\;\;\;\dot{\beta}} \bm\Delta_{R\rho \tau \;\dot{\beta}}^{\;\;\;\;\;\;\;\dot{\alpha}} \bm\Delta_{L\lambda \chi \;\alpha}^{\;\;\;\;\;\;\;\beta} \bm\Delta_{L\zeta \omega \;\beta}^{\;\;\;\;\;\;\;\alpha} \epsilon^{\mu \rho \lambda \zeta} \epsilon^{\nu \tau \chi \omega} + h.c.   
\end{align} 
\noindent
Interactions generated by Eq. \eqref{nantin} and Eq. \eqref{yukawa} after the spontaneous PS symmetry breaking by $\langle \Delta_R \rangle$ cause   baryon number violating $n-\overline{n}$ oscillation as shown in Fig. \ref{nnbar}. The existing term is of the form $\bm\Delta_R^2 \bm\Delta_L^2$, which indicates that if  $\bm\Delta_L$ field is not present, $n-\overline{n}$ transition is forbidden in this set-up due to the added $U(1)_{PQ}$ symmetry. 

Expanding in terms of the SM multiplets, this term gives,
\begin{align}
2 \tilde \lambda_9^{\ast} v_R \epsilon_3 \epsilon_3\; \Delta^{\ast}_{R}(\overline 6,1,2/3) \;\Delta^{\ast}_{L}(\overline 6,3,-1/3) \;\Delta^{\ast}_{L}(\overline 6,3,-1/3) + h.c. \;.
\end{align}

\noindent From this, the  effective Lagrangian describing the six-fermion vertex ($d=9$ operators) that corresponds to $n-\overline{n}$ oscillation can be written down,
 \vspace{-2pt}
 \begin{align}
\mathcal{L}^{n-\overline{n}}_{eff}&= 
- \widetilde{\lambda}_9 \; (2\;v_R)\; \epsilon^{\overline{•\rho}\overline{•\lambda}\overline{•\zeta}} 
\epsilon^{\overline{•\tau}\overline{•\chi}\overline{•\omega}}
{Y^R_{10}}_{kl}{Y^L_{10}}_{mn}{Y^L_{10}}_{pq}
\;\;\;
\frac{\left(
 d^T_{R k \overline{\rho}} C d_{R l \overline{\tau}}
 \right)}{m^2_{\Delta_{R(\overline{6},1,\frac{2}{3})}}
m^4_{\Delta_{L(\overline{6},3,-\frac{1}{3})}}
}
\nonumber \\ & \times
\Biggl\{ 
\left(
d^T_{L m \overline{\lambda}} C u_{L n \overline{\chi}}
 \right)
 \left(
u^T_{L p \overline{\zeta}} C d_{L q \overline{\omega}}
 \right)
 -  2
 \left(
 d^T_{L m \overline{\lambda}} C d_{L n \overline{\chi}}
 \right)
 \left(
 u^T_{L p \overline{\zeta}} C u_{L q \overline{\omega}}
 \right)
\Biggr\}, 
 \end{align}

\noindent here $k, l, m, n, p, q$ are the generation indices.

Again on the dimensional ground, the  $n-\overline{n}$ oscillation transition time can be computed as 
\vspace{0pt}
\begin{align}
\tau_{n-\overline{n}} = \frac{M^6}{v_R\; \Lambda^6_{QCD}}.
\end{align} 

\noindent The present limit on this transition time is constraint by the matter disintegration, which is $\tau_{n-\overline{n}} \geq 2\times 10^8$ sec. \cite{Abe:2011ky}. A slightly weaker bound but with less uncertainty is obtained from the free neutron oscillation search, $\tau_{n-\overline{n}} \geq 10^8$ sec. \cite{BaldoCeolin:1994jz}. By taking $\tau_{n-\overline{n}} = 10^8$ sec. one can find the lower bound on the scale $v_R \sim 3.2\times 10^5$ GeV (like before $M= v_R = 10^3$ TeV is assumed).  So if the scalar fields responsible for this oscillation have masses somewhat smaller than $10^3$ TeV, which is certainly possible  with some choice of the quartic couplings,     $n-\overline{n}$ transition time can be within the interesting observable range.

\section{Conclusion} \label{chapter07}
In this work, we have presented a minimal renormalizable  model based on the Pati-Salam gauge group, $SU(2)_L\times SU(2)_R\times SU(4)_C$ that unifies quark and leptons by treating leptons as the fourth color. We extend the symmetry of our theory by imposing a global $U(1)_{PQ}$ Peccei-Quinn symmetry, that automatically solves the strong CP problem and provides axion as a dark matter candidate. The minimal Higgs sector consists of (2,2,1), (2,2,15) and (1,3,10) multiplets under the Pati-Salam group, is fixed by the requirement of  consistent symmetry breaking pattern and by the phenomenological constraints of reproducing the observed fermion spectrum. Due to the imposition of the global $U(1)_{PQ}$ symmetry, several terms in the Higgs potential and few terms in the Yukawa Lagrangian are forbidden.    This  theory is highly predictive and with only 14 parameters in the Yukawa sector a good fit to the charged fermion masses and mixings are obtained. The origin of the baryon asymmetry of the universe is linked to the seesaw mechanism that is also responsible for the observed neutrino oscillations. Detailed search of the parameter space for  successful generation of matter-antimatter asymmetry  is carried out that makes connection between the low scale and high scale parameters of the theory. With a  minimal scalar content,  comprehensive analysis in the Higgs sector is carried out by  constructing the  complete Higgs potential  and then by computing the   mass spectrum of the Higgs fields. We also studied two baryon number violating processes present in our theory that are nucleon decay and neutron-antineutron oscillation. Possible nucleon decay modes arising from dimension 9 and dimension 10 operators are discussed and branching fractions of different channels are computed with certain approximations. Neutron-antineutron oscillation via dimension 9 operators in this framework is also analyzed. Both the nucleon lifetime and neutron-antineutron transition time can be within the interesting observable range if the Pati-Salam symmetry breaking scale is low.

\section*{Acknowledgments}
The author would like to thank Dr. Kaladi Babu for useful discussions. This work has been partially supported in part by the U.S. Department
of Energy Grant No. de-sc0010108.   For this project, S.S has also been partially supported by the OSU Graduate College's ``Robberson Summer Research Fellowships'' 2016 award. The author would like to thank the Fermilab theory group for the hospitality during the summer visit where part of the work was done.

\newpage
\begin{appendices}
\appendixpageoff

\section{\large Parameter space for successful Leptogenesis for the case with \texorpdfstring{$m_1=0.8$}{TEXT} meV and \texorpdfstring{$m_1=4$}{TEXT} meV}\label{A}
In this appendix we present the plots of the parameter space for successful generation of the baryon asymmetry for the cases with $m_{1}=0.8$ and 4 meV.

\FloatBarrier
\begin{figure}[ht]
\begin{subfigure}{.3\textwidth}
  \centering
    \caption*{$m_1=0.8$ meV (Flavored)}
  \includegraphics[scale=0.28]{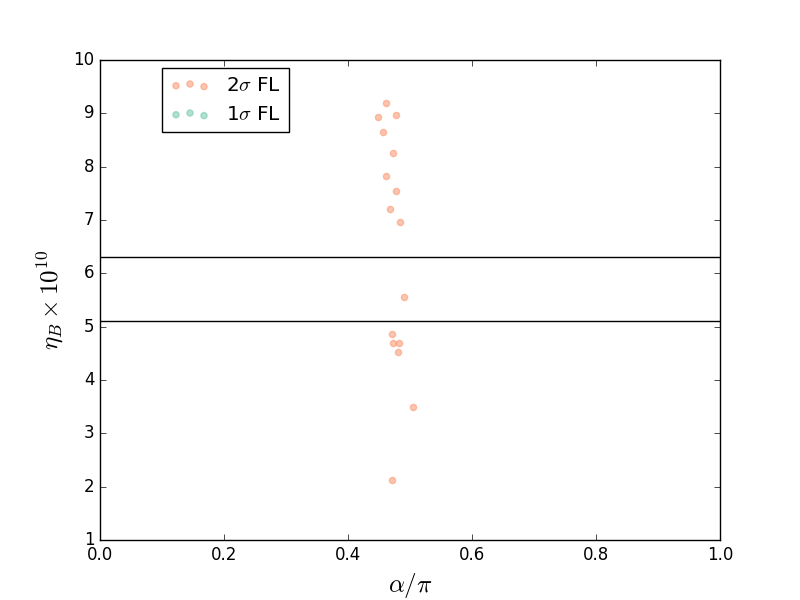}
\end{subfigure}
 \hspace{20pt}
\begin{subfigure}{.3\textwidth}
  \centering
  \caption*{$m_1=4$ meV (Flavored)}
  \includegraphics[scale=0.28]{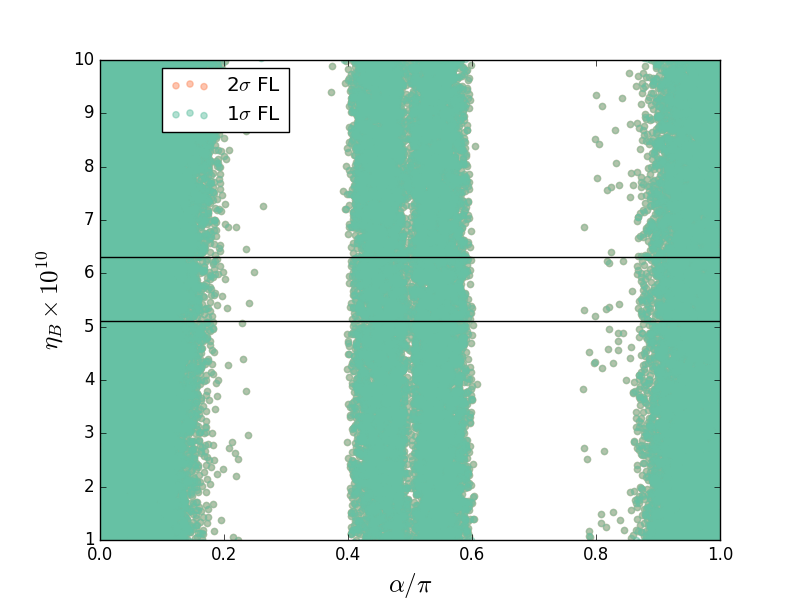}
\end{subfigure}
 \hspace{20pt}
\begin{subfigure}{.3\textwidth}
  \centering
   \caption*{$m_1=4$ meV (Unflavored)}
  \includegraphics[scale=0.28]{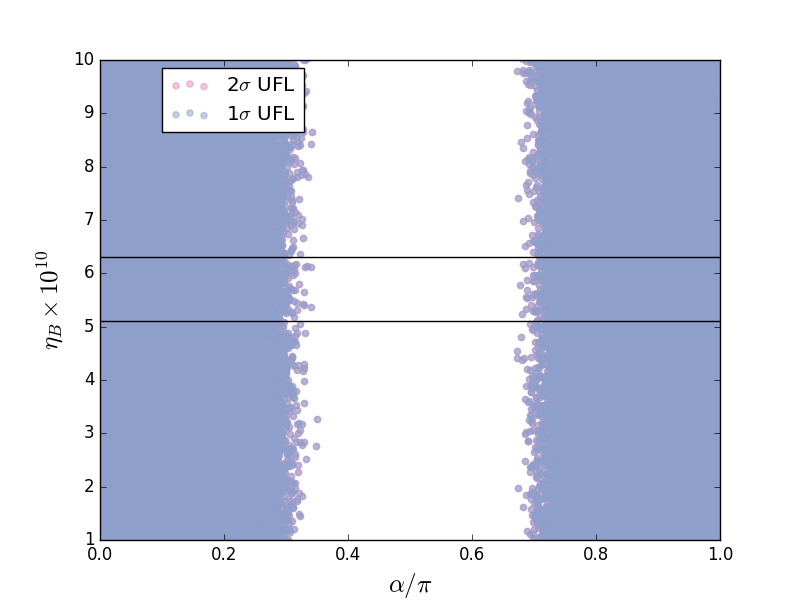}
\end{subfigure} 
\begin{subfigure}{.3\textwidth}
  \centering
  \includegraphics[scale=0.28]{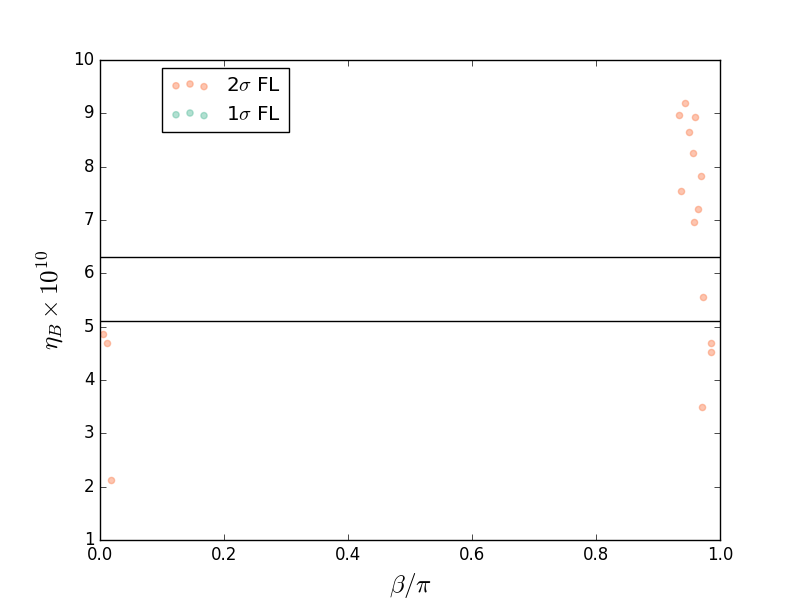}
\end{subfigure}
 \hspace{20pt}
\begin{subfigure}{.3\textwidth}
  \centering
  \includegraphics[scale=0.28]{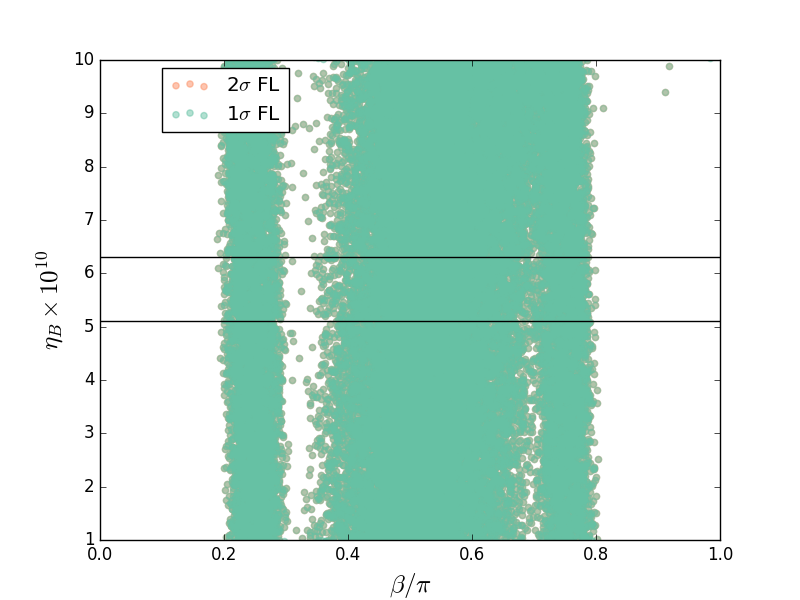}
\end{subfigure}
 \hspace{20pt}
\begin{subfigure}{.3\textwidth}
  \centering
  \includegraphics[scale=0.28]{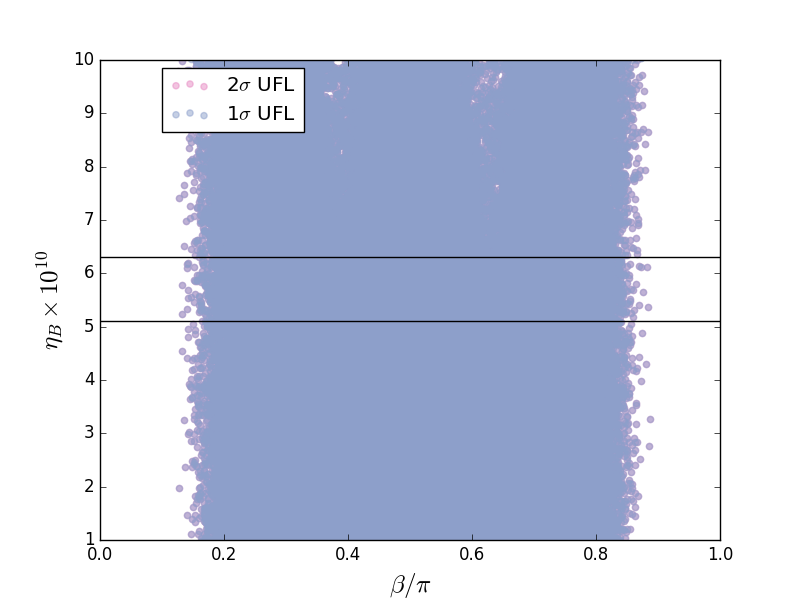}
\end{subfigure} 
\begin{subfigure}{.3\textwidth}
  \centering
  \includegraphics[scale=0.28]{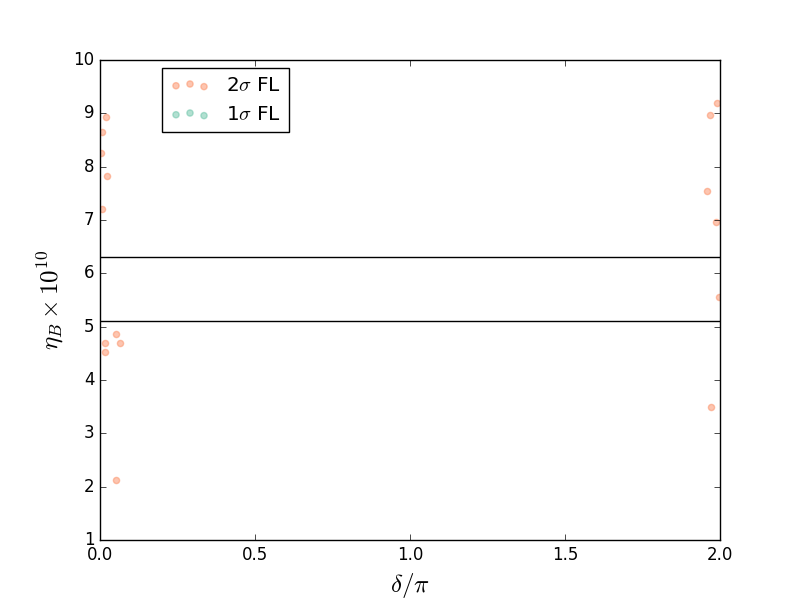}
\end{subfigure}
 \hspace{20pt}
\begin{subfigure}{.3\textwidth}
  \centering
  \includegraphics[scale=0.28]{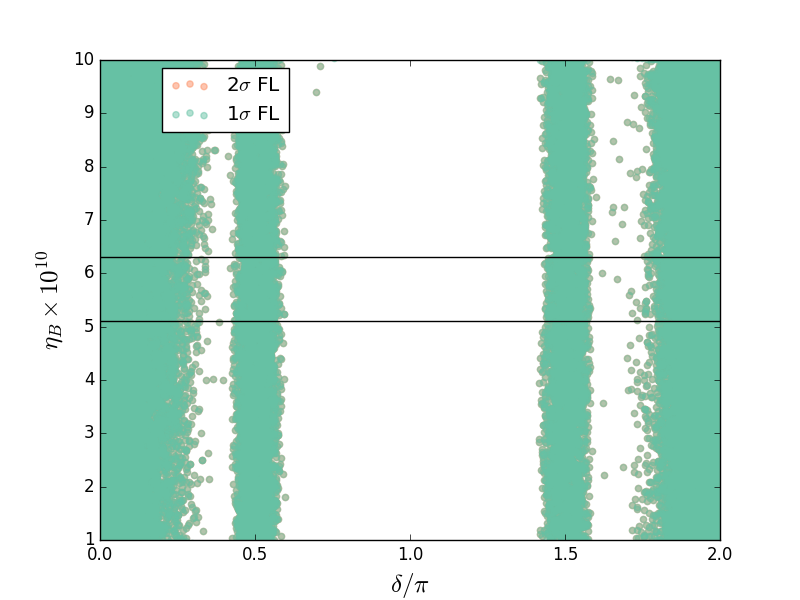}
\end{subfigure}
 \hspace{20pt}
\begin{subfigure}{.3\textwidth}
  \centering
  \includegraphics[scale=0.28]{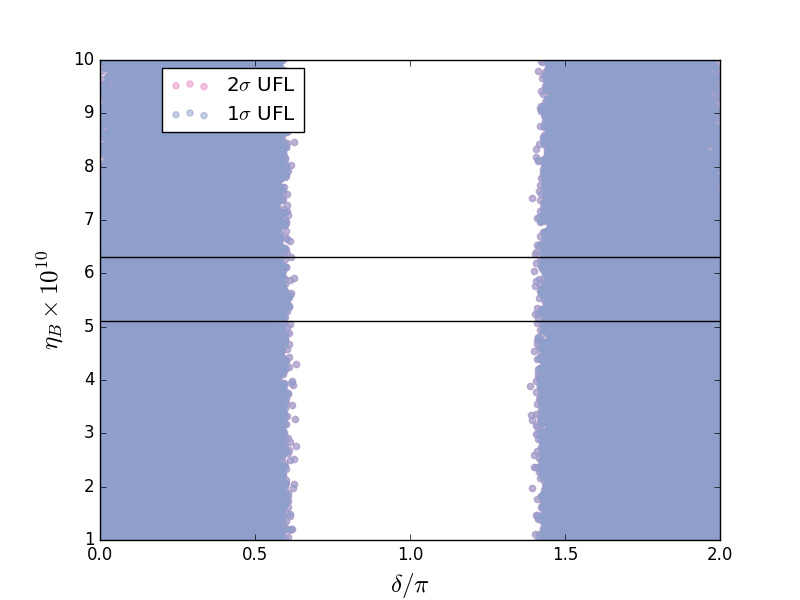}
\end{subfigure} 
\caption{ Allowed parameter space for the unknown quantities $\alpha, \beta, \delta$  to generate successful baryon asymmetry for  two different values of  $m_{1}=0.8, 4$ meV are presented here. As mentioned in the text, these case with $m_{1}=0.8$ meV does not permit any solution.   Color code is the same as in Fig. \ref{fig:001}. }
\label{fig:84}
\end{figure}

\newpage
\section{\large Allowed right-handed neutrino mass spectrum for successful leptogenesis}\label{B}

\begin{figure}[ht]
\begin{subfigure}{.3\textwidth}
  \centering
    \caption*{$m_1=1$ meV (Flavored)}
  \includegraphics[scale=0.3]{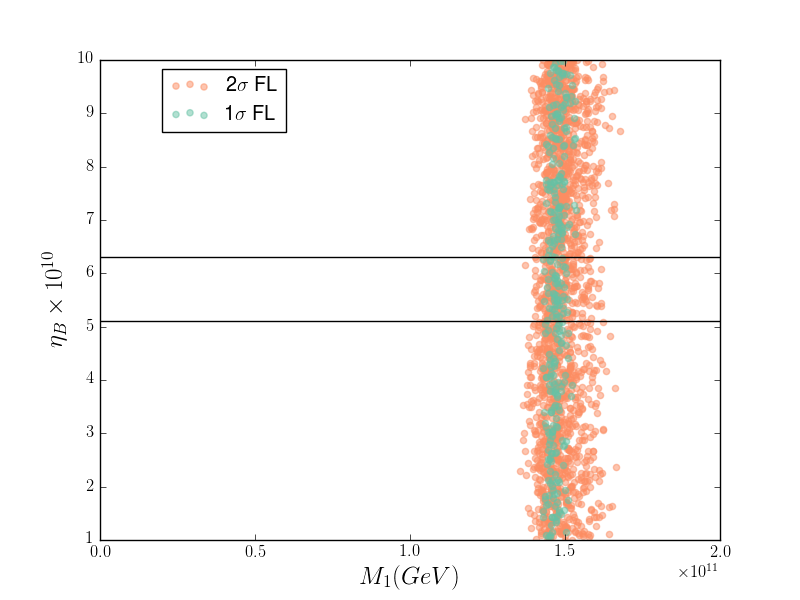}
\end{subfigure}
 \hspace{20pt}
\begin{subfigure}{.3\textwidth}
  \centering
  \caption*{$m_1=2$ meV (Flavored)}
  \includegraphics[scale=0.3]{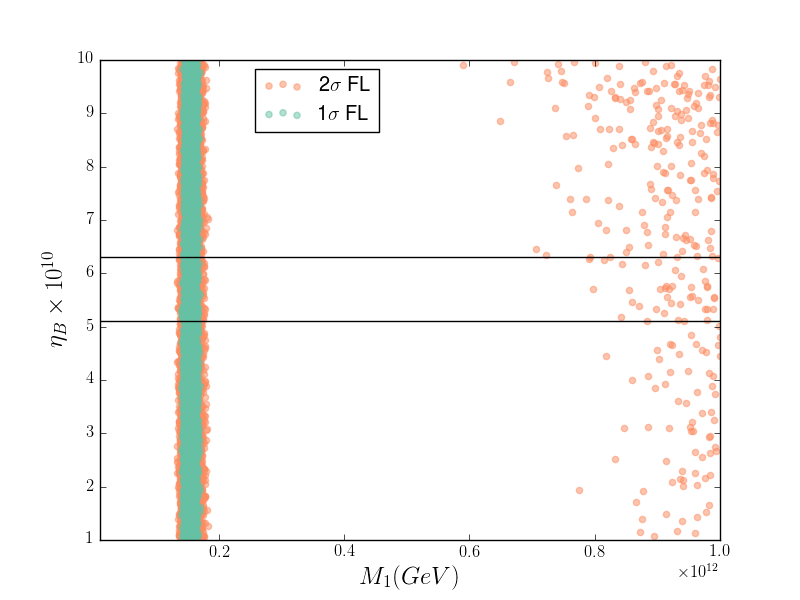}
\end{subfigure}
 \hspace{20pt}
\begin{subfigure}{.3\textwidth}
  \centering
   \caption*{$m_1=2$ meV (Unflavored)}
  \includegraphics[scale=0.3]{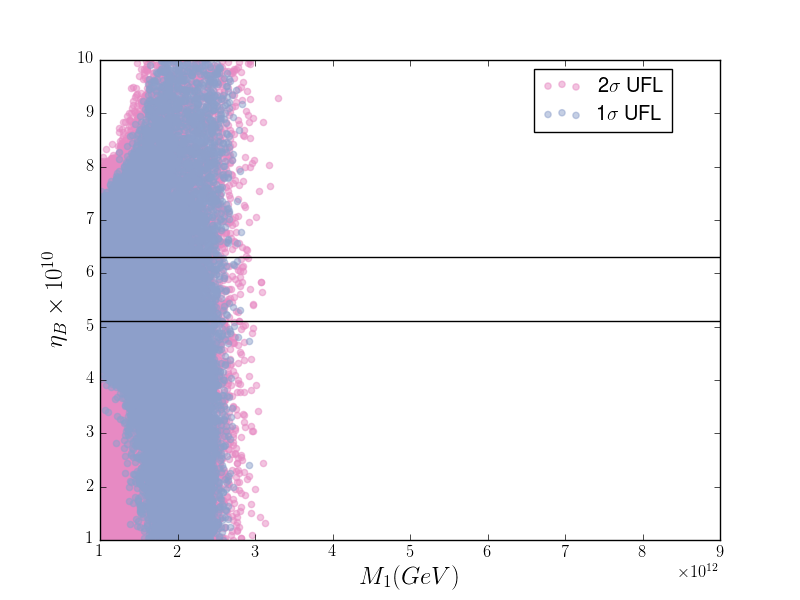}
\end{subfigure} 
\begin{subfigure}{.3\textwidth}
  \centering
  \includegraphics[scale=0.3]{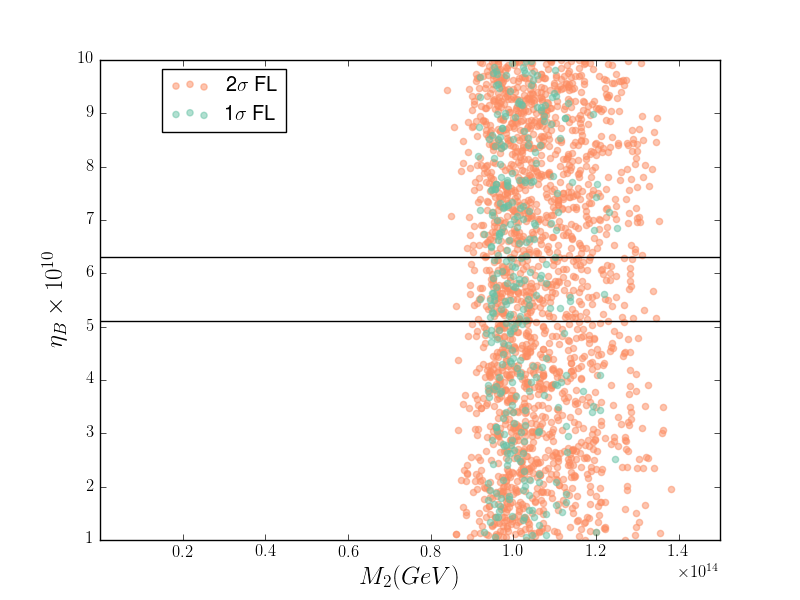}
\end{subfigure}
 \hspace{20pt}
\begin{subfigure}{.3\textwidth}
  \centering
  \includegraphics[scale=0.3]{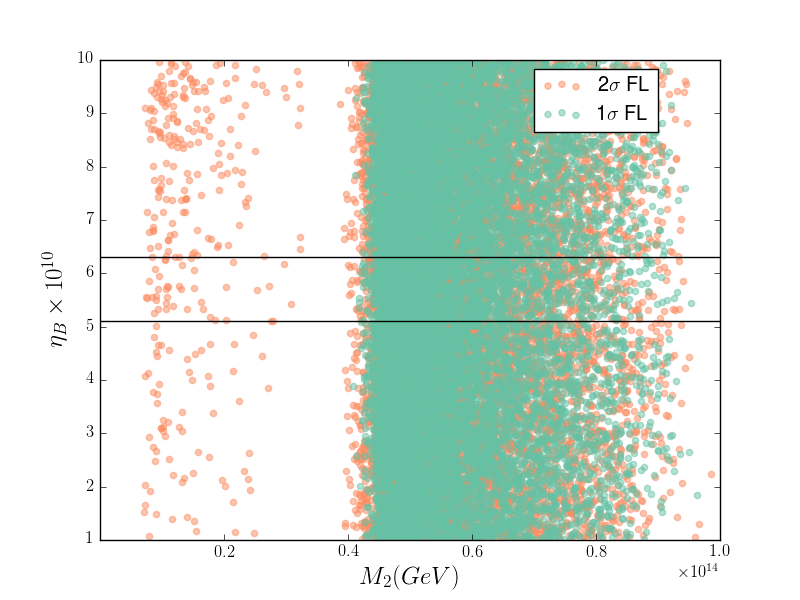}
\end{subfigure}
 \hspace{20pt}
\begin{subfigure}{.3\textwidth}
  \centering
  \includegraphics[scale=0.3]{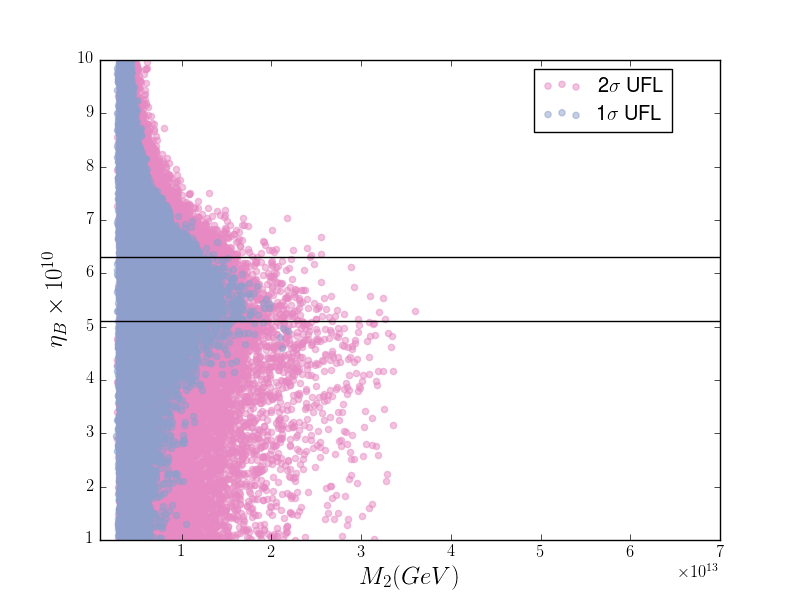}
\end{subfigure} 
\begin{subfigure}{.3\textwidth}
  \centering
  \includegraphics[scale=0.3]{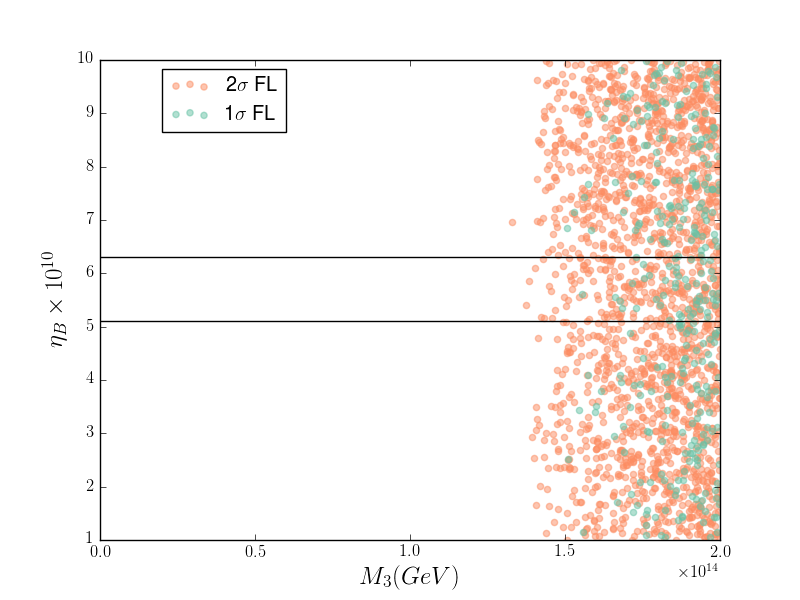}
\end{subfigure}
 \hspace{20pt}
\begin{subfigure}{.3\textwidth}
  \centering
  \includegraphics[scale=0.3]{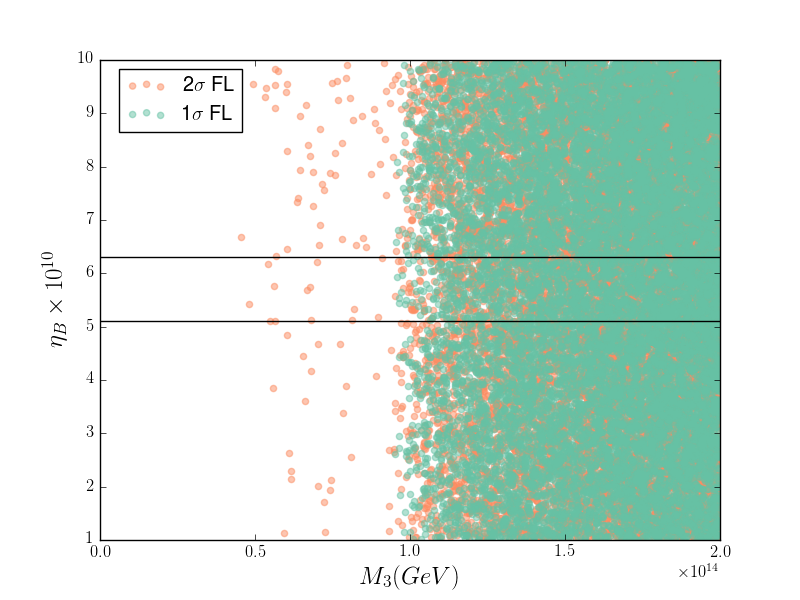}
\end{subfigure}
 \hspace{20pt}
\begin{subfigure}{.3\textwidth}
  \centering
  \includegraphics[scale=0.3]{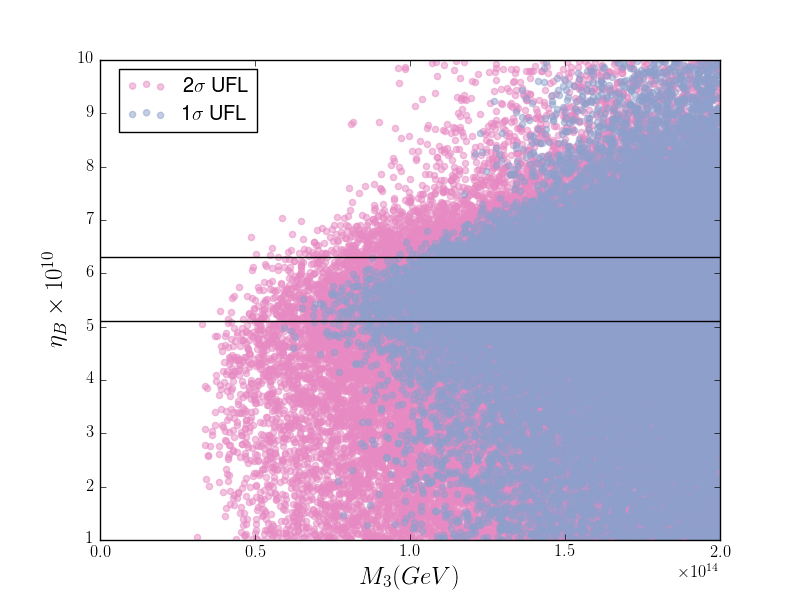}
\end{subfigure} 
\caption{The correspondence between the baryon asymmetry and the heavy right-handed neutrino mass spectrum $M_{i}$ are plotted. Color code is the same as in Fig. \ref{fig:001}.}
\label{fig:004}
\end{figure}

The correspondence between the baryon asymmetry and the heavy right-handed neutrino mass spectrum $M_{i}$ are  shown in Fig. \ref{fig:004} for two fixed values of the lightest neutrino mass,  $m_1=1$ meV and $m_1=2$ meV.

\FloatBarrier

\newpage
\section{\large Relation between baryon asymmetric parameters and \texorpdfstring{$m_{\beta}$}{TEXT} and \texorpdfstring{$m_{\beta \beta}$}{TEXT} for the case with \texorpdfstring{$m_1=1$}{TEXT} and 2 meV}\label{C}

\begin{figure}[ht]
\begin{subfigure}{.3\textwidth}
  \centering
    \caption*{$m_1=1$ meV (Flavored)}
  \includegraphics[scale=0.3]{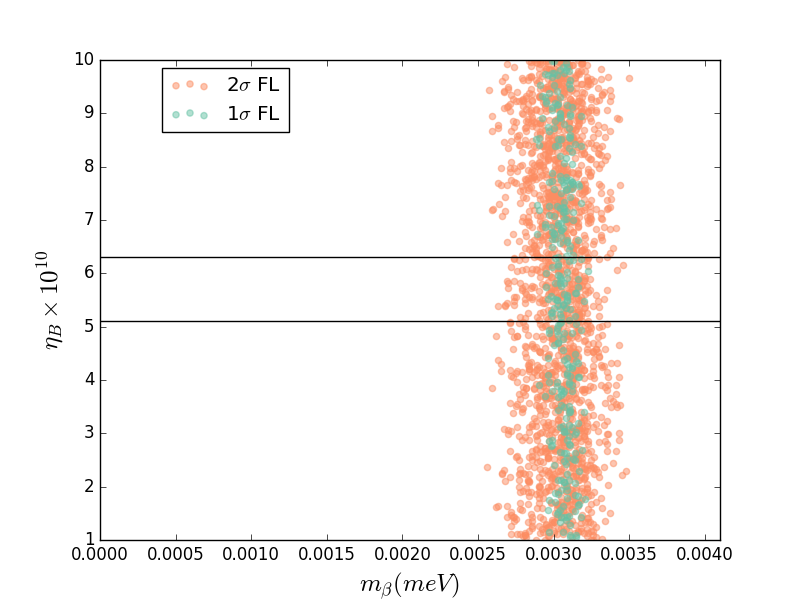}
\end{subfigure}
 \hspace{20pt}
\begin{subfigure}{.3\textwidth}
  \centering
  \caption*{$m_1=2$ meV (Flavored)}
  \includegraphics[scale=0.3]{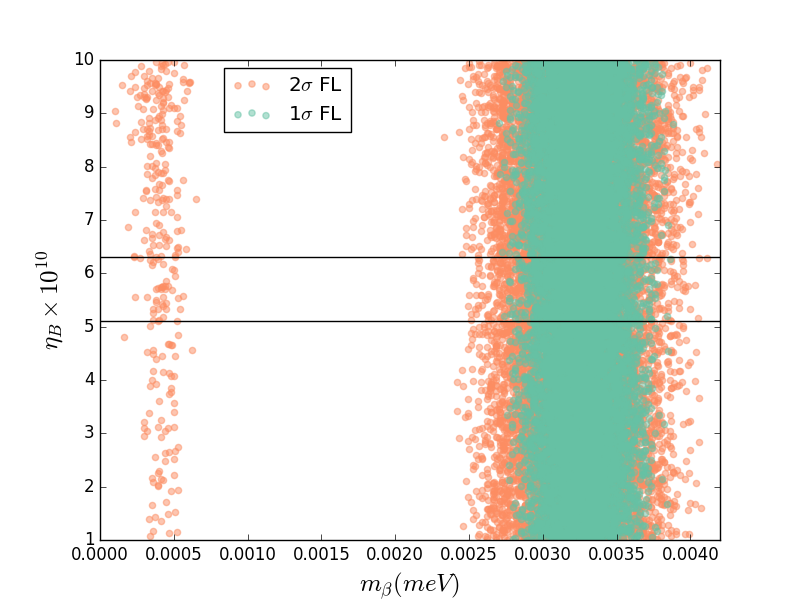}
\end{subfigure}
 \hspace{20pt}
\begin{subfigure}{.3\textwidth}
  \centering
   \caption*{$m_1=2$ meV (Unflavored)}
  \includegraphics[scale=0.3]{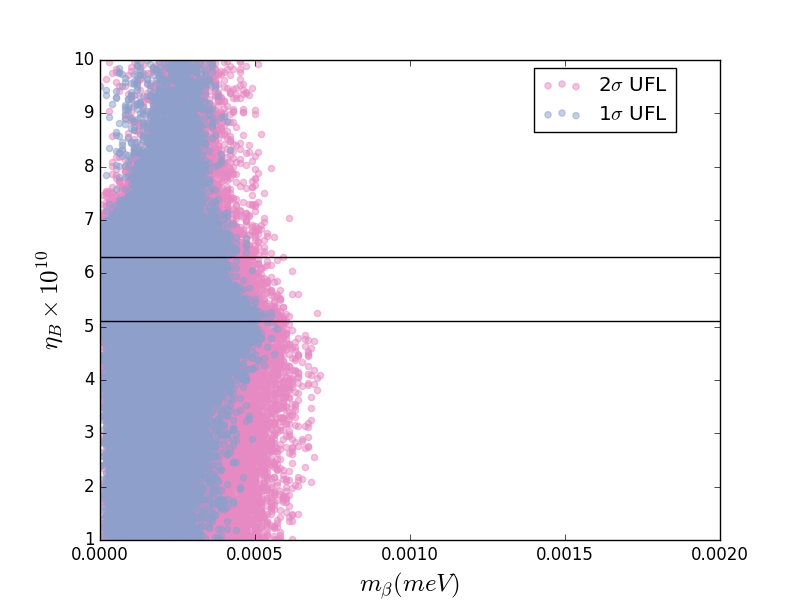}
\end{subfigure} 
\begin{subfigure}{.3\textwidth}
  \centering
  \includegraphics[scale=0.3]{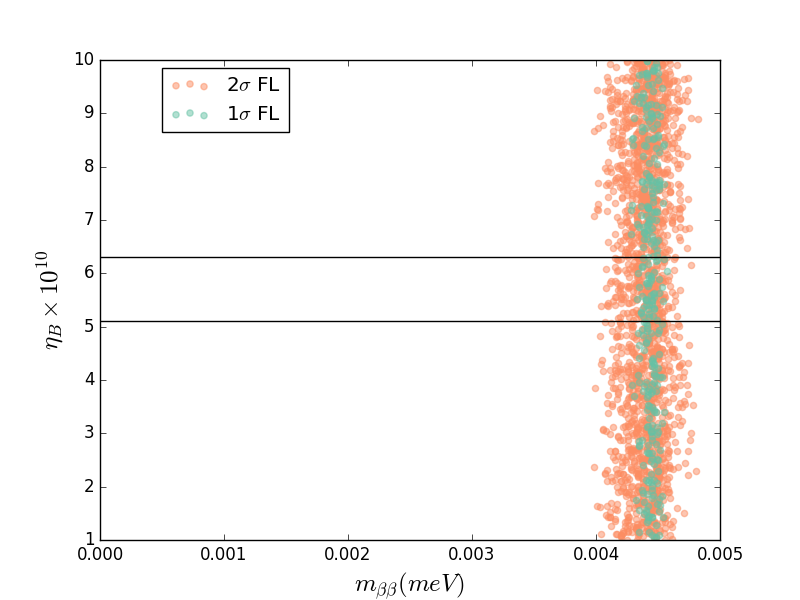}
\end{subfigure}
 \hspace{20pt}
\begin{subfigure}{.3\textwidth}
  \centering
  \includegraphics[scale=0.3]{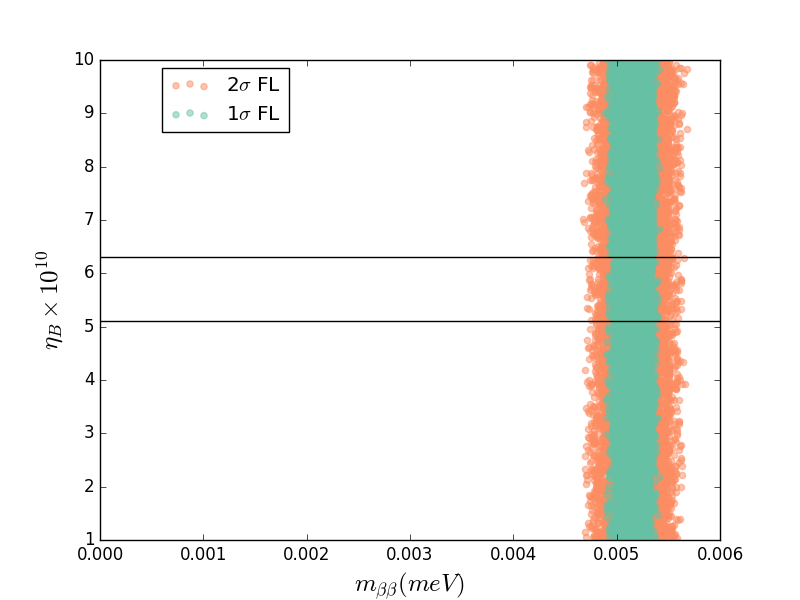}
\end{subfigure}
 \hspace{20pt}
\begin{subfigure}{.3\textwidth}
  \centering
  \includegraphics[scale=0.3]{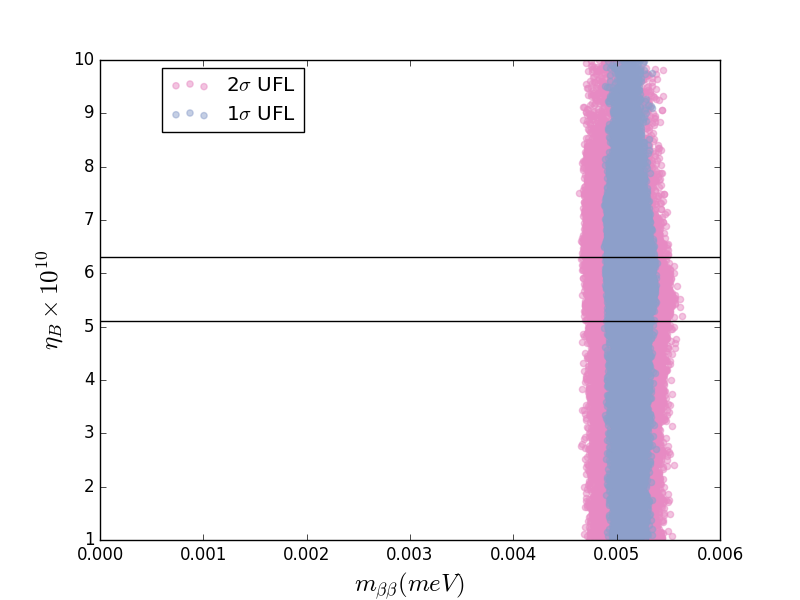}
\end{subfigure} 
\caption{
The correspondence between the baryon asymmetry and $m_{\beta, \beta \beta}$ are plotted, where $m_{\beta}=\sum_{i} |U_{\nu \;\;e i}|^{2} m_{i}$ is the effective mass parameter for the beta-decay and $m_{\beta \beta}= | \sum_{i} U_{\nu \;\;e i}^{2} m_{i} |$ is the effective mass parameter for neutrinoless double beta decay. Color code is the same as in Fig. \ref{fig:001}.}
\label{fig:002}
\end{figure}

In this appendix we present the permitted region  for $m_{\beta}$ and $m_{\beta \beta}$ to have successful leptogenesis in Fig. \ref{fig:002}.

\FloatBarrier

\newpage
\section{\large The correlation between the CP-violating Dirac phase \texorpdfstring{$\delta$}{TEXT} and \texorpdfstring{$\theta_{13}$}{TEXT} }\label{D}

\begin{figure}[ht]
\begin{subfigure}{.3\textwidth}
  \centering
    \caption*{$m_1=1$ meV (Flavored)}
  \includegraphics[scale=0.3]{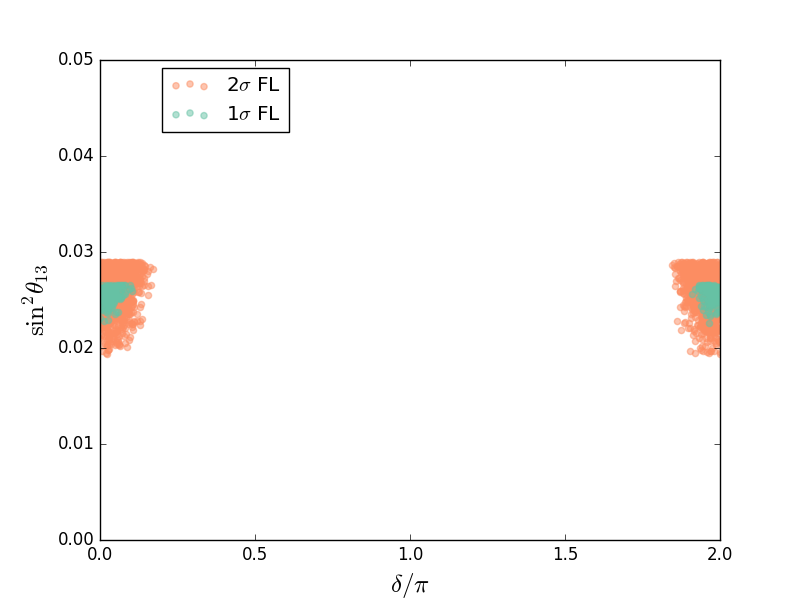}
\end{subfigure}
 \hspace{20pt}
\begin{subfigure}{.3\textwidth}
  \centering
  \caption*{$m_1=2$ meV (Flavored)}
  \includegraphics[scale=0.3]{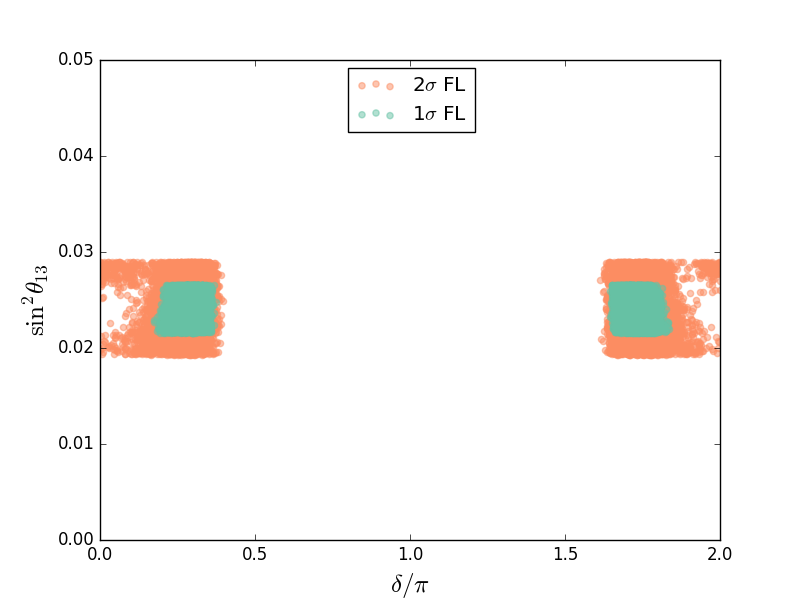}
\end{subfigure}
 \hspace{20pt}
\begin{subfigure}{.3\textwidth}
  \centering
   \caption*{$m_1=2$ meV (Unflavored)}
  \includegraphics[scale=0.3]{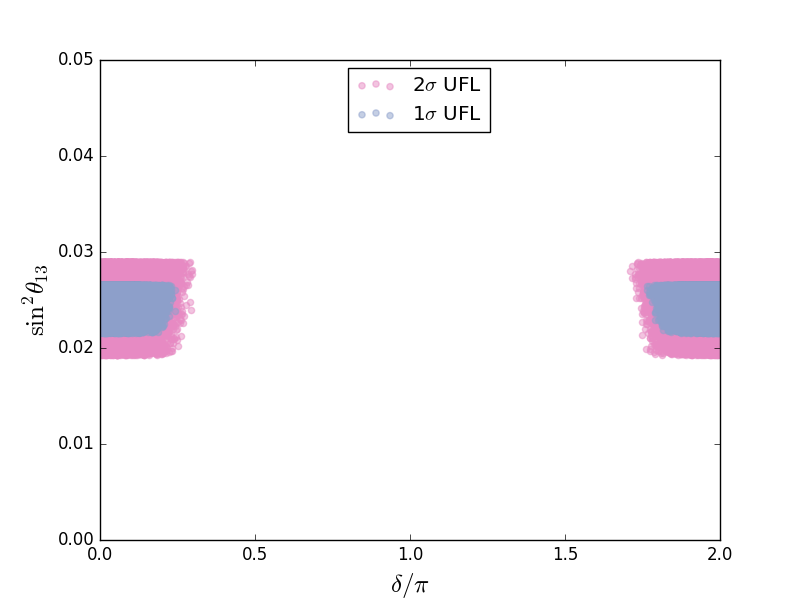}
\end{subfigure} 
\caption{Correlation between the quantities $\delta$ and $\sin^{2}_{\theta^{PMNS}_{13}}$ is plotted for three different values of $m_{1}=0.8, 1,2$ meV. Color code is the same as in Fig. \ref{fig:001}.}
\label{fig:003}
\end{figure}

The correlations between the  phase $\delta$ and the  angle $\theta_{13}$ permitted by reproducing correct baryon asymmetry is presented in Fig. \ref{fig:003}.

\FloatBarrier

\end{appendices}

\FloatBarrier

\end{document}